\documentclass[iop]{emulateapj}
\usepackage{epstopdf}
\usepackage{nicefrac}
\usepackage{amsmath}
\newcommand{\myemail}{iarcavi@lcogt.net}
\def\gtorder{\mathrel{\raise.3ex\hbox{$>$}\mkern-14mu
             \lower0.6ex\hbox{$\sim$}}} 
\def\ltorder{\mathrel{\raise.3ex\hbox{$<$}\mkern-14mu
             \lower0.6ex\hbox{$\sim$}}}
\def\ltsima{$\; \buildrel < \over \sim \;$}
\def\simlt{\lower.5ex\hbox{\ltsima}}
\def\gtsima{$\; \buildrel > \over \sim \;$}
\def\simgt{\lower.5ex\hbox{\gtsima}} 

\shorttitle{Rapidly Rising Luminous Transients}
\shortauthors{Arcavi et al.}

\begin{document} 


\title{Rapidly Rising Transients in the Supernova - Superluminous Supernova Gap}


\author{Iair~Arcavi\altaffilmark{1,2},
William~M.~Wolf\altaffilmark{3},
D.~Andrew~Howell\altaffilmark{1,3},
Lars~Bildsten\altaffilmark{2,3},
Giorgos~Leloudas\altaffilmark{4,5},
Delphine~Hardin\altaffilmark{6},
Szymon~Prajs\altaffilmark{7},
Daniel~A.~Perley\altaffilmark{5},
Gilad~Svirski\altaffilmark{8},
Avishay~Gal-Yam\altaffilmark{4},
Boaz~Katz\altaffilmark{4},
Curtis~McCully\altaffilmark{2,3},
S.~Bradley~Cenko\altaffilmark{9,10},
Chris~Lidman\altaffilmark{11},
Mark~Sullivan\altaffilmark{7},
Stefano~Valenti\altaffilmark{2,3},
Pierre~Astier\altaffilmark{6},
Cristophe~Balland\altaffilmark{6},
Ray~G.~Carlberg\altaffilmark{13},
Alex~Conley\altaffilmark{14},
Dominique~Fouchez\altaffilmark{15},
Julien~Guy\altaffilmark{6},
Reynald~Pain\altaffilmark{6},
Nathalie~Palanque-Delabrouille\altaffilmark{16},
Kathy~Perrett\altaffilmark{17},
Chris~J.~Pritchet\altaffilmark{18},
Nicolas~Regnault\altaffilmark{6},
James~Rich\altaffilmark{16}
and
Vanina~Ruhlmann-Kleider\altaffilmark{16}
}

\affil{\altaffilmark{1}Las Cumbres Observatory Global Telescope, 6740 Cortona Dr, Suite 102, Goleta, CA 93111, USA \myemail}
\affil{\altaffilmark{2}Kavli Institute for Theoretical Physics, University of California, Santa Barbara, CA 93106, USA}
\affil{\altaffilmark{3}Department of Physics, University of California, Santa Barbara, CA 93106, USA}
\affil{\altaffilmark{4}Department of Particle Physics and Astrophysics, The Weizmann Institute of Science, Rehovot, 76100, Israel}
\affil{\altaffilmark{5}Dark Cosmology Centre, Niels Bohr Institute, University of Copenhagen, Juliane Maries Vej 30, DK-2100 Copenhagen, Denmark}
\affil{\altaffilmark{6}LPNHE, CNRS-IN2P3 and University of Paris VI \& VII, F-75005 Paris, France}
\affil{\altaffilmark{7}School of Physics and Astronomy, University of Southampton, Southampton, SO17 1BJ, UK}
\affil{\altaffilmark{8}Racah Institute for Physics, The Hebrew University, Jerusalem 91904, Israel}
\affil{\altaffilmark{9}Astrophysics Science Division, NASA Goddard Space Flight Center, Mail Code 661, Greenbelt, MD 20771, USA}
\affil{\altaffilmark{10}Joint Space-Science Institute, University of Maryland, College Park, MD 20742, USA}
\affil{\altaffilmark{11}Australian Astronomical Observatory, PO Box 915, North Ryde, NSW 1670, Australia}
\affil{\altaffilmark{12}University of Paris-Sud, Orsay, F-91405, France}
\affil{\altaffilmark{13}Department of Astronomy and Astrophysics, University of Toronto, 50 St. George Street, Toronto, ON M5S 3H8, Canada}
\affil{\altaffilmark{14}Center for Astrophysics and Space Astronomy, University of Colorado, 389 UCB, Boulder, CO 80309-389, USA}
\affil{\altaffilmark{15}Aix Marseille Universit\'{e}, CNRS/IN2P3, CPPM UMR 7346, 13288, Marseille, France}
\affil{\altaffilmark{16}DSM/IRFU/SPP, CEA-Saclay, F-91191 Gif-sur-Yvette, France}
\affil{\altaffilmark{17}DRDC Ottawa, 3701 Carling Avenue, Ottawa, ON K1A 0Z4, Canada}
\affil{\altaffilmark{18}Department of Physics and Astronomy, University of Victoria, P.O. Box 3055, Victoria, BC V8W 3P6, Canada}

%\altaffiltext{$\star$}{Hubble Fellow}
%\altaffiltext{$\dagger$}{Carnegie-Princeton Fellow}


%\altaffiltext{1}{Hubble Fellow.} 

\newpage

\begin{abstract} 

We present observations of four rapidly rising ($t_{rise}\approx10$\,d) transients with peak luminosities between those of supernovae (SNe) and superluminous SNe ($M_{peak}\approx-20$) - one discovered and followed by the Palomar Transient Factory (PTF) and three by the Supernova Legacy Survey (SNLS). The light curves resemble those of SN\,2011kl, recently shown to be associated with an ultra-long-duration gamma ray burst (GRB), though no GRB was seen to accompany our SNe. The rapid rise to a luminous peak places these events in a unique part of SN phase space, challenging standard SN emission mechanisms. Spectra of the PTF event formally classify it as a Type~II SN due to broad H$\alpha$ emission, but an unusual absorption feature, which can be interpreted as either high velocity H$\alpha$ (though deeper than in previously known cases) or Si II (as seen in Type~Ia SNe), is also observed. We find that existing models of white dwarf detonations, CSM interaction, shock breakout in a wind (or steeper CSM) and magnetar spindown can not readily explain the observations. We consider the possibility that a ``Type 1.5 SN'' scenario could be the origin of our events. More detailed models for these kinds of transients and more constraining observations of future such events should help better determine their nature.

\end{abstract} 




\section{Introduction} 

Supernovae (SNe), the explosive deaths of stars, are observed to occur in a variety of types and a spread of luminosities. Most notable is the division into core collapse SNe (Types~Ib/c and II; see Filippenko 1997 for a review), which are associated with the deaths of massive ($M\gtrsim8M_{\odot}$) stars, and Type~Ia SNe, associated with the thermonuclear disruptions of white dwarfs (Hoyle \& Fowler 1960; Hansen \& Wheeler 1969, Nomoto 1982a,b; Nomoto, Thielemann \& Yokoi 1984; Branch et al. 1985; see Nugent et al. 2011 for the most direct observational evidence of this association). Recently, a third class of explosions has been identified, superluminous SNe (SLSNe), characterized by their high luminosity at peak (e.g. Quimby et al. 2007; Smith et al. 2007; Ofek et al. 2007; Gal-Yam et al. 2009; Pastorello et al. 2010; Quimby et al. 2011; see Gal-Yam 2012 for a review). These events likely originate in massive stars, but clear progenitor scenarios for SLSNe have yet to be determined.

\begin{figure}
\includegraphics[trim=0 0 00 0 clip=true,width=\columnwidth]{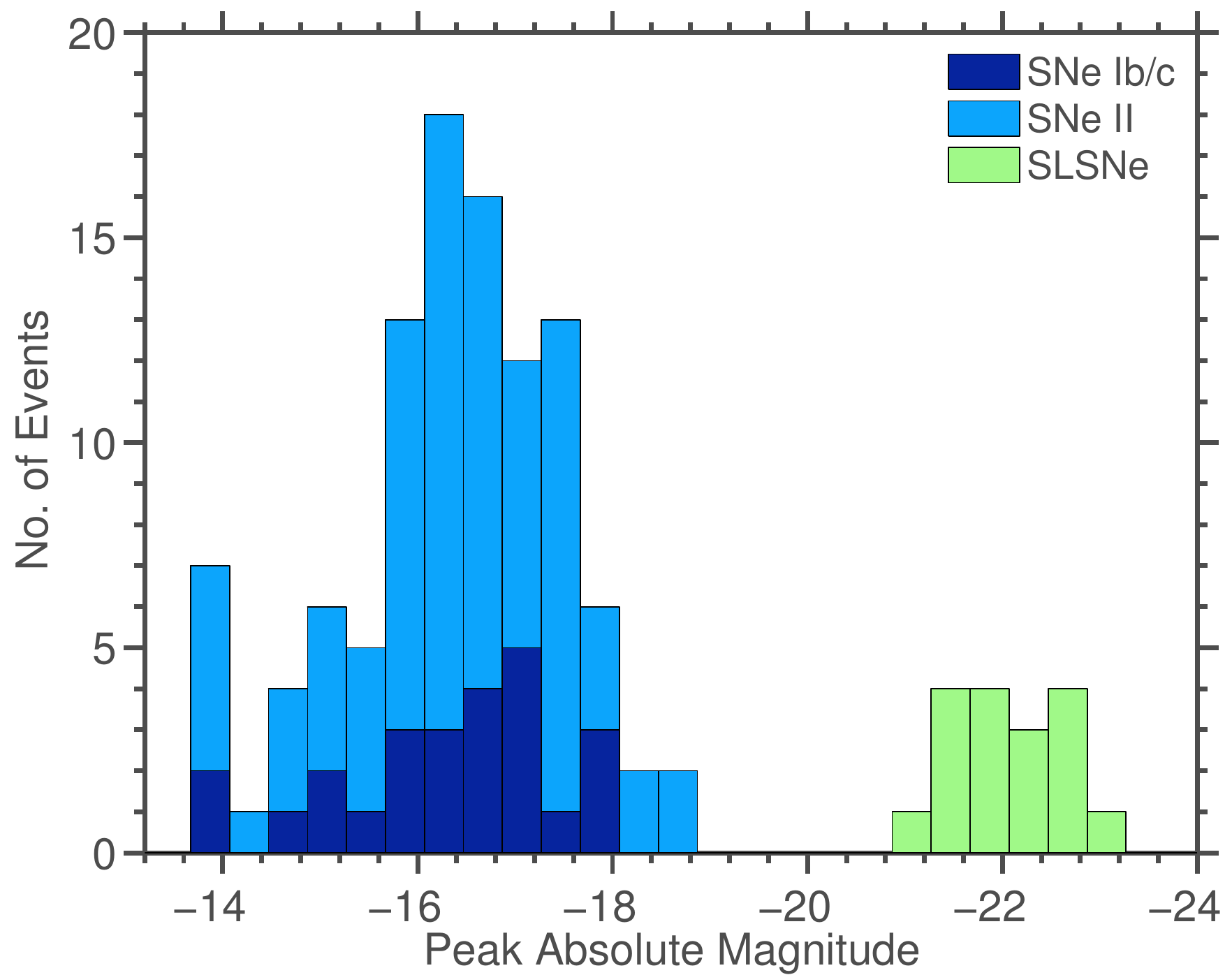}
\caption{\label{fig:peakhist}Peak magnitudes of core collapse SNe (Li et al. 2011; see also Bazin et al. 2009 and Taylor et al. 2014) and super luminous SNe (SLSNe; Gal-Yam 2012). Here all strongly interacting (Type~IIn) SNe are excluded. A gap between core collapse SNe and SLSNe is apparent.}
\end{figure}

An open question related to the possible connection between core collapse SNe and SLSNe is whether the apparent lack of ``intermediate'' events (i.e. SNe with peak absolute magnitudes in the range $-19$ to $-21$; Fig. \ref{fig:peakhist}) is real or just a selection effect. Arcavi et al. (2014) searched for such events in the spectroscopically confirmed H-rich core collapse sample from the Palomar Transient Factory (PTF; Rau et al. 2009; Law et al. 2009). Three events were found in the centers of non-starforming galaxies, and turned out to be tidal disruptions of stars by supermassive black holes (Arcavi et al. 2014). A fourth event, PTF10iam, was shown to be significantly offset from its host center, was in a starforming galaxy and displayed a faster rise to peak magnitude.

Here we investigate the nature of PTF10iam as a SN with peak luminosity in the SN-SLSN ``gap'', but with a surprisingly rapid ($\approx10$ day) rise to peak. We present three additional events with similarly short rise times and peak luminosities discovered and followed by the Supernova Legacy Survey (SNLS\footnote{Not to be confused with the acronym for superluminous supernova - SLSN}; Astier et al. 2006). These events were identified as part of a broader search for SLSNe in the SNLS dataset, the details of which will be discussed in Wolf et al. (in prep). In brief, moderately luminous events ($M_{peak}\gtrsim-19$) with available redshifts that were not initially identified as Type~Ia SNe were all checked by eye. The three studied here, SNLS04D4ec, SNLS05D2bk, and SNLS06D1hc all exhibited short rise times ($\lesssim10$ days) and luminous peak magnitudes ($\approx-20$).

Various SNe with rapidly evolving light curves, such as PTF09uj (Ofek et al. 2010), SN 2002bj (Poznanski et al. 2011), SN 2010X (Kasliwal et al. 2011), OGLE-2013-SN-079 (Inserra et al. 2015) and a sample of events discovered by Pan-STARRS 1 (Drout et al. 2014), have been studied in the past. Taddia et al. (2015) show that ``normal'' stripped envelope SNe Ib/c can also have a rapid rise to peak. However, none of the previously studied rapid SNe were as luminous at peak as the sample presented here, with two exceptions. One is the transient ``Dougie'' (Vink{\'o} et al. 2015). Dougie was discovered by the ROTSE-IIIb survey and displays a $\approx7$ day rise to an extremely luminous peak magnitude of $\approx-23$ in $R$-band. Being much more luminous than our events, it may be of a completely different nature. The second known rapidly rising luminous event is SN\,2011kl (Greiner et al. 2015). SN\, 2011kl was recently recovered from the optical afterglow of the ultra-long-duration gamma ray burst (GRB) 111209A (Gendre et al. 2013; Stratta et al. 2013; Levan et al. 2014). Both the GRB properties and the SN properties are not similar to any previous GRB-SN event, but the SN light curves are similar to those of our sample.

We present the observations of our events (and a new host galaxy spectrum of Dougie, confirming its redshift and thus peak absolute magnitude) in \S2 and analyze the light curves and spectra of our events in \S3, comparing to those of SN\,2011kl. We discuss possible physical mechanisms for creating the transients in our sample in \S4 and summarize in \S5.

\section{Observations}

PTF10iam was discovered by the Palomar 48-inch Oschin Schmidt telescope (P48) as part of the PTF survey. It was classified as a Type~II SN following spectra showing a blue continuum and later broad H${\alpha}$ emission. The three SNLS events (SNLS04D4ec, SNLS05D2bk and SNLS06D1hc) were discovered by the deep survey of the Canada France Hawaii Telescope Legacy Survey (CFHTLS 2002)\footnote{http://cfht.hawaii.edu/Science/CFHTLS/}, using the CFHT 3.6-meter telescope. The events were marked as non-Ia SNe based on their light curves (Sullivan et al. 2006). The discovery information for all four events is presented in Table \ref{tab:events}. 

\renewcommand{\arraystretch}{1.4}
\begin{table*}
\begin{center}
{\caption{\label{tab:events}Discovery details of our events. Error values denote $1{\sigma}$ uncertainties.}}

\begin{tabular*}{0.62\textwidth}{lccccc}
\hline
\hline
{Name} & {RA} & {Dec} & {Redshift} & {Discovery} & {Discovery} \tabularnewline
{} & {(J2000)} & {(J2000)} & {} & {Date} & {Mag} \tabularnewline
\hline
{PTF10iam} & {15:45:30.85} & {$+$54:02:33.0} & {$0.109$} & {2010 May 22} & {$19.14\pm0.11$} \tabularnewline
{SNLS04D4ec} & {22:16:29.29} & {$-$18:11:04.1} & {$0.593$} & {2004 Jul 9} & {$22.70\pm0.06$} \tabularnewline
{SNLS05D2bk} & {10:02:13.96} & {$+$02:05:55.2} & {$0.699$} & {2005 Jan 15} & {$23.33\pm0.06$} \tabularnewline
{SNLS06D1hc} & {02:24:48.25} & {$-$04:56:03.6} & {$0.555$} & {2006 Nov 14} & {$22.55\pm0.04$} \tabularnewline
\hline

\end{tabular*}
\end{center}
\end{table*}
\renewcommand{\arraystretch}{1}

\subsection{Photometry} \label{sec:photometry}

The photometry of PTF10iam was released in Arcavi et al. (2014). Here we recover an additional pre-explosion non-detection upper limit. For the SNLS events, images in the $g$, $r$, $i$ and $z$ bands were obtained as part of the SNLS rolling survey. The SN flux was measured by removing a modeled host contamination, which is taken from reference images after PSF matching (see Astier et al. 2006 and references therein for more details).

The photometry for PTF10iam is presented in the AB system, and for the SNLS events in the Vega system, in Table \ref{tab:phot}. In Figure \ref{fig:lcs} we present the light curves in observed filters after correcting for Galactic extinction using the Schlafly \& Finkbeiner (2011) maps, extracted via the NASA Extragalctic Database (NED\footnote{http://ned.ipac.caltech.edu/}). 

The PTF10iam spectra presented in Section \ref{sec:specs} do not display noticable Na I D absorption, indicating that any host extinction for PTF10iam is likely small. Additionally, we fit the first spectrum of PTF10iam with a variety of temperatures and extinction values. Assuming no extinction, our best blackbody fit gives a temperature $T$ of $11000$\,K (see Setion \ref{sec:bb}). We find equally good fits up to $T=13200$\,K and $\textrm{E(B-V)}=0.19$, but the fits significantly worsen with higher extinction values regardless of temperature. The best fit is found for $T=11000$\,K with $\textrm{E(B-V)}=0.09$, which corresponds to an extinction of $A=0.2$\,mag in the PTF $R$-band (assuming a Cardelli et al. 1989 extinction law). All of our events show similar blackbody temperatures (Setion \ref{sec:bb}), suggesting that they all suffer comparably low host extinction. We thus neglect host extinction for all of our events. 

Distance moduli are calculated from spectroscopic redshifts of the host galaxies, determined from narrow spectral features (Fig. \ref{fig:spec_hosts}). A cosmological model with $H_0 = 70\,\textrm{km\,s}^{-1}\,\textrm{Mpc}^{-1}$, $\Omega_{m}=0.3$ and $\Omega_{\Lambda}=0.7$ is assumed throughout.

\renewcommand{\arraystretch}{1.4}
\begin{table}
{\caption{\label{tab:phot}Photometric observations (upper limits mark $3\sigma$ non-detections). The PTF10iam data were presented also in Arcavi et al. (2014), but are given here again for completeness with the addition of a new pre-explosion upper limit. This table is published in its entirety in the electronic version. A portion is shown here for guidance regarding its form and content.}}
\begin{tabular}{lccccc}
\hline
\hline
{Object} & {Telescope} & {Filter} & {MJD} & {Mag} & {Error}\tabularnewline
\hline
{PTF10iam} & {P48} & {$R$} & {$55323.462$} & {$>21.222$} & {}\tabularnewline
{PTF10iam} & {P48} & {$R$} & {$55324.263$} & {$>21.134$} & {}\tabularnewline
{PTF10iam} & {P48} & {$R$} & {$55324.306$} & {$>21.216$} & {}\tabularnewline
{PTF10iam} & {P48} & {$R$} & {$55330.491$} & {$>20.914$} & {}\tabularnewline
{PTF10iam} & {P48} & {$R$} & {$55338.779$} & {$>20.994$} & {}\tabularnewline
{PTF10iam} & {P48} & {$R$} & {$55345.472$} & {$19.142$} & { $0.114$ }\tabularnewline
{PTF10iam} & {P48} & {$R$} & {$55346.267$} & {$18.973$} & { $0.046$ }\tabularnewline
{PTF10iam} & {P48} & {$R$} & {$55346.311$} & {$18.905$} & { $0.048$ }\tabularnewline
{PTF10iam} & {P48} & {$R$} & {$55351.3$} & {$18.451$} & { $0.025$ }\tabularnewline
\hline

\end{tabular}
\end{table}
\renewcommand{\arraystretch}{1}

\newlength{\leftimagewidth}
\newlength{\rightimagewidth}
\settowidth{\leftimagewidth}{\includegraphics[height=6.7cm]{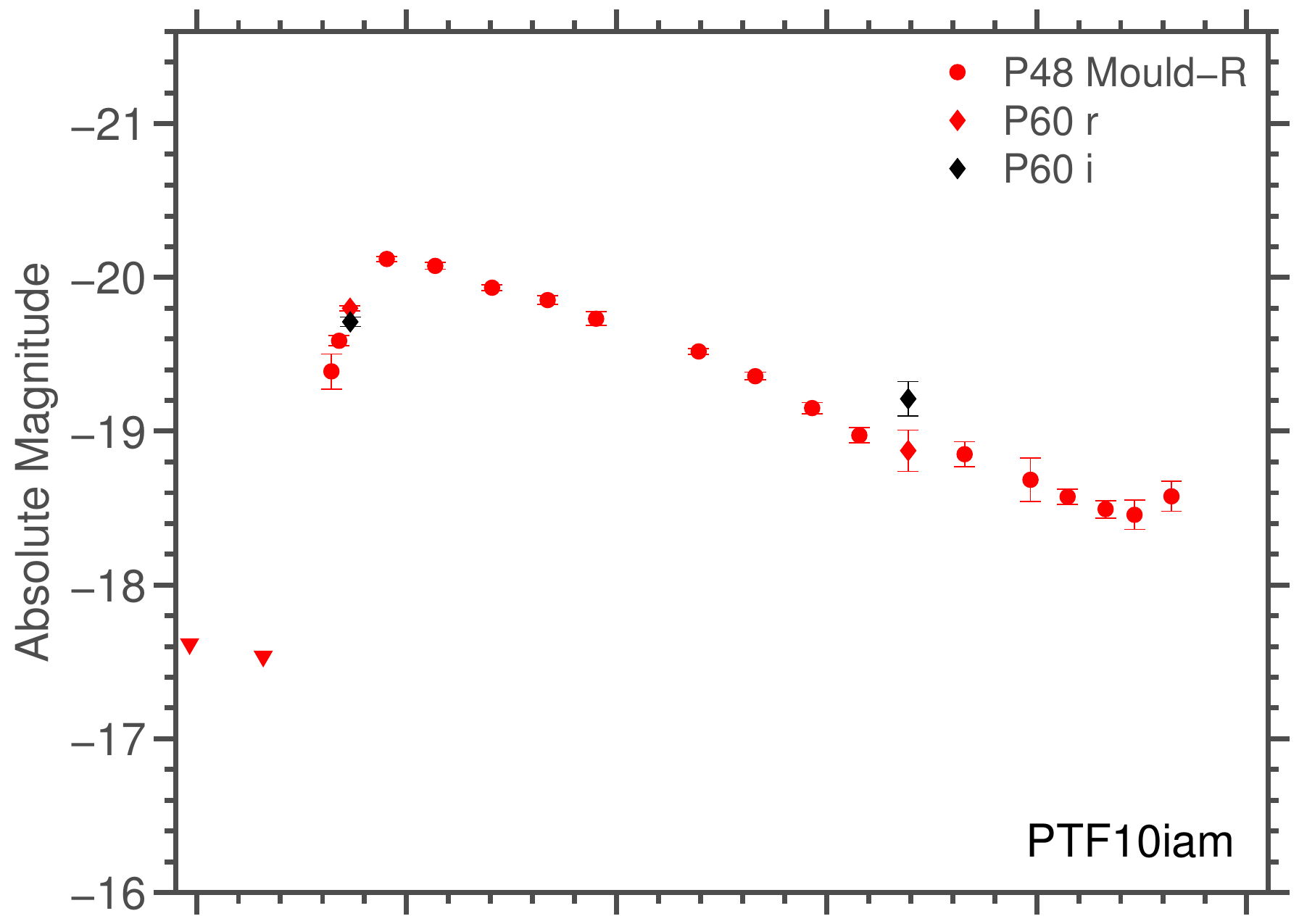}}
\settowidth{\rightimagewidth}{\includegraphics[height=6.7cm]{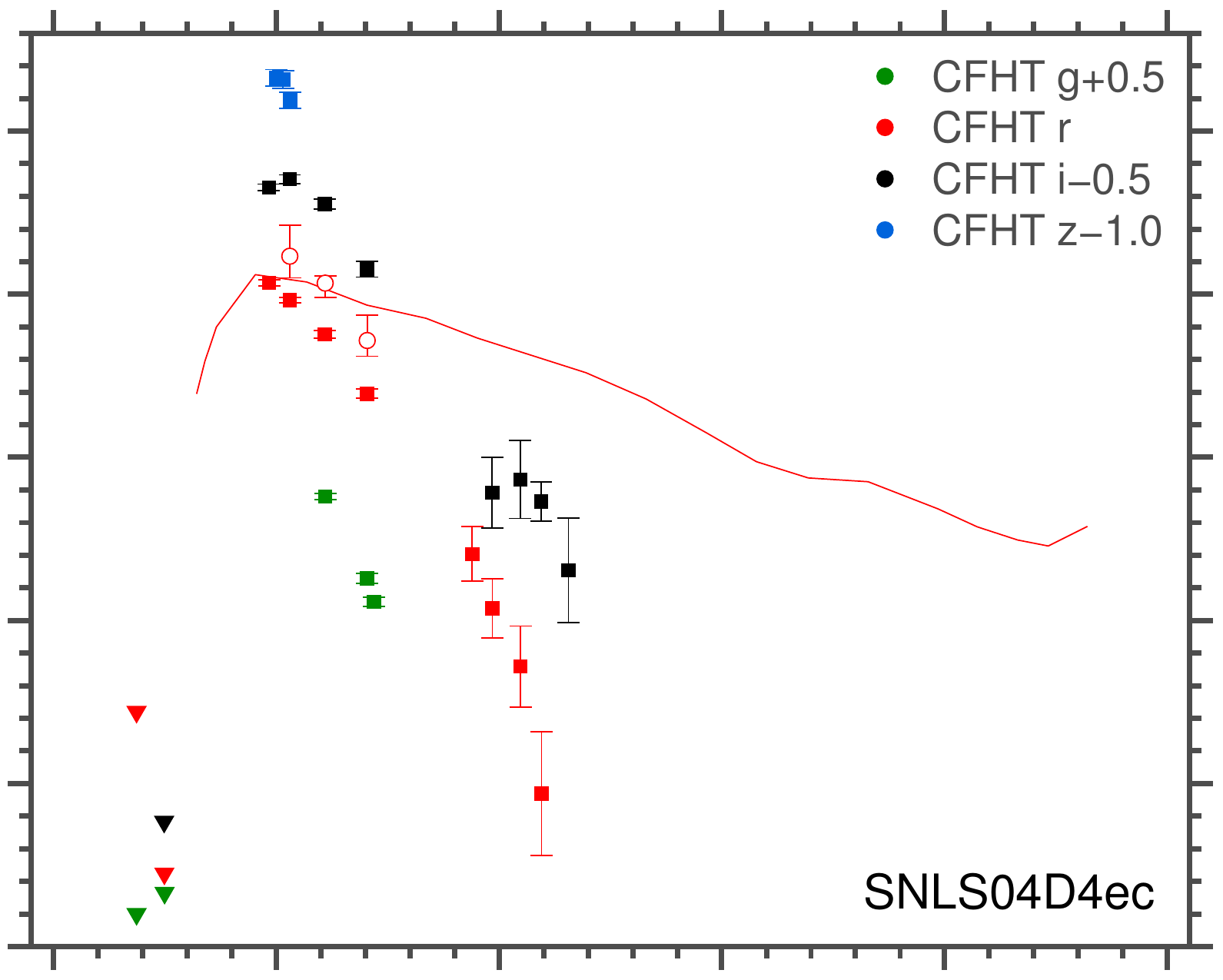}}

\begin{figure*}[t]
\includegraphics[height=6.7cm]{lc_ptf10iam.eps}\,\,\includegraphics[height=6.7cm]{lc_04d4ec.eps}
\includegraphics[width=\leftimagewidth]{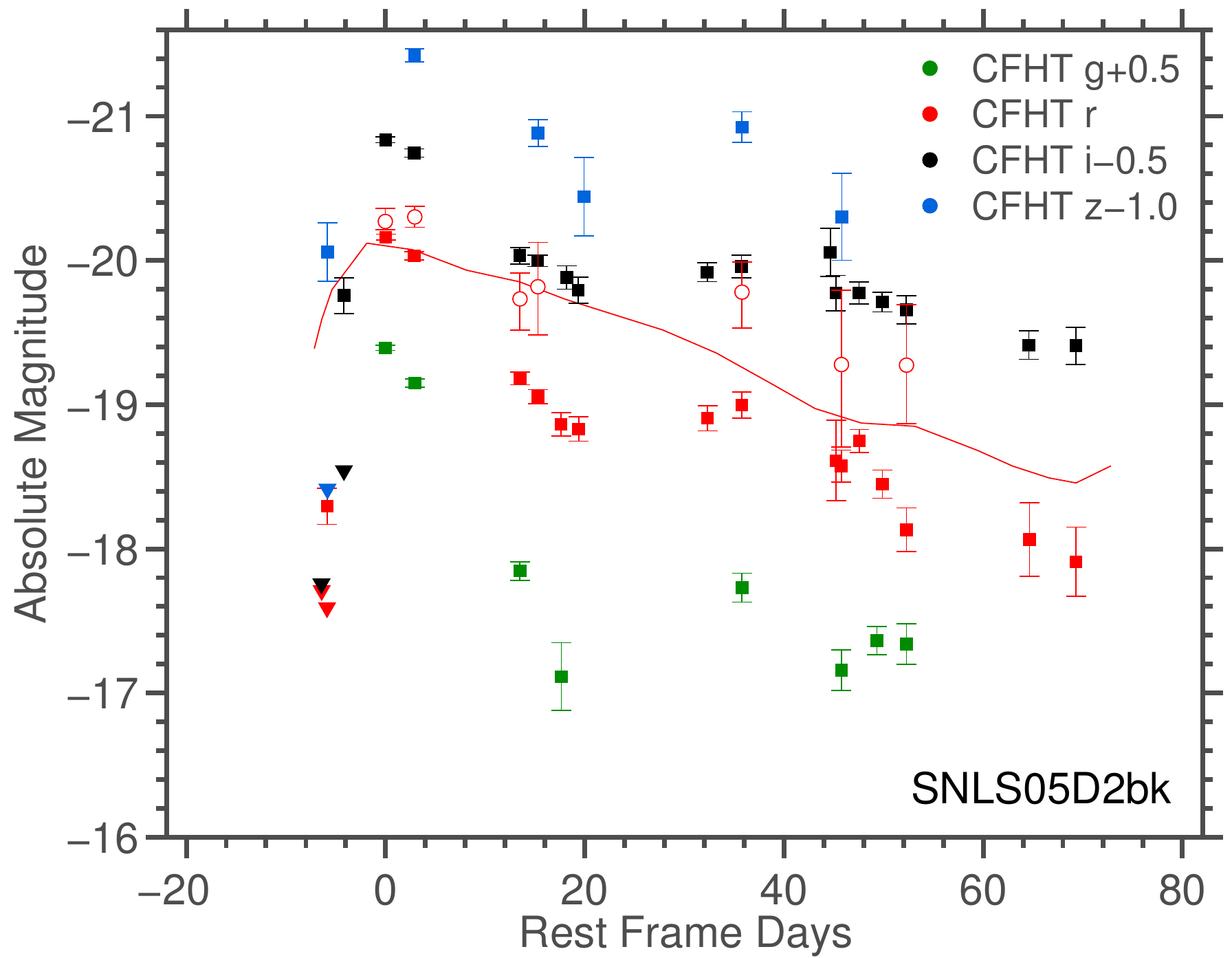}\,\,\includegraphics[width=\rightimagewidth]{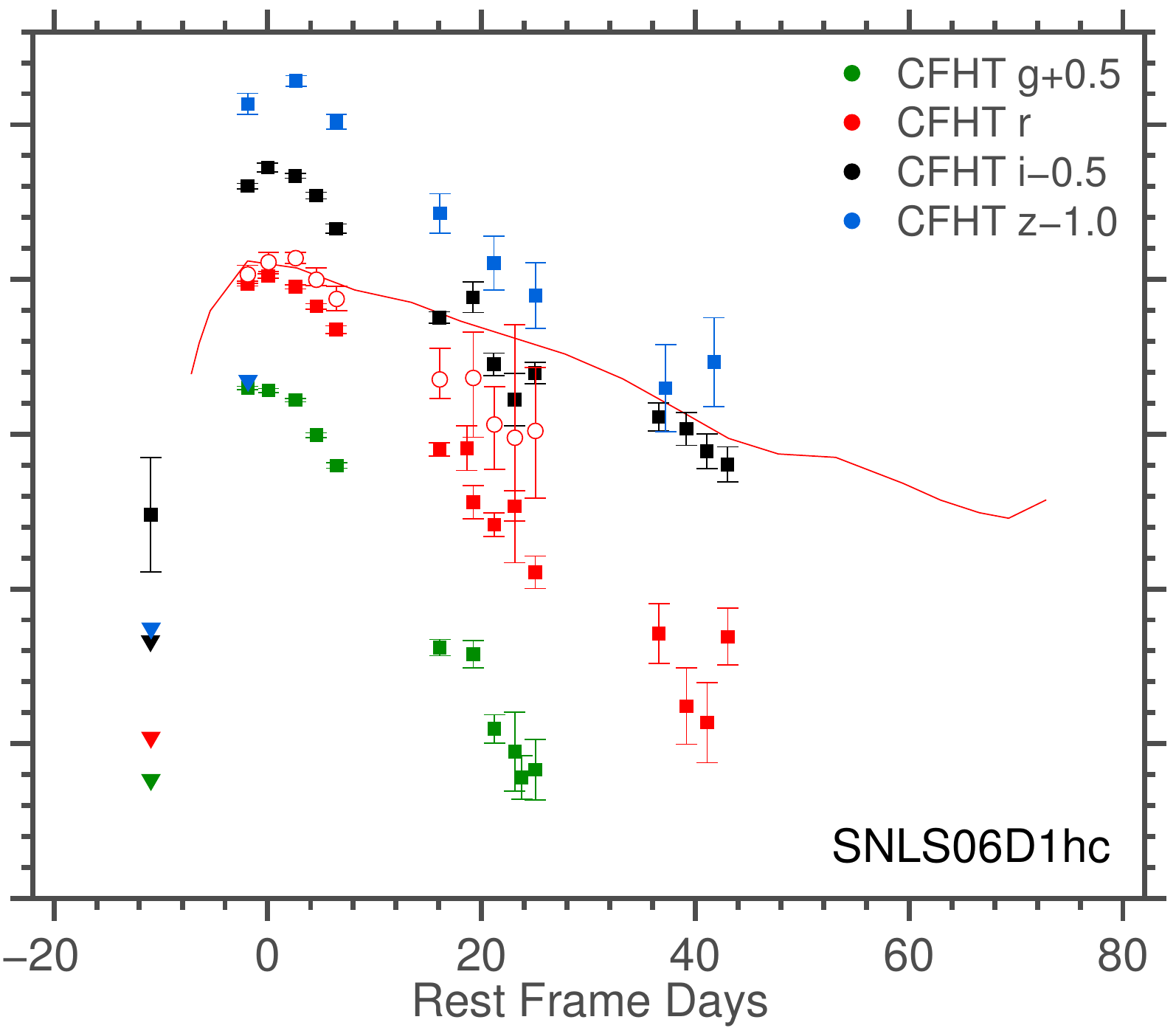}
\caption{\label{fig:lcs}Light curves (in observed filters) of our four rapidly rising luminous transients. Triangles denote $3{\sigma}$ non-detection upper limits. The solid line represents the PTF10iam light curve for comparison to the SNLS events. Empty circles in the SNLS light curves are the expected magnitudes after K-correction to the PTF Mould-$R$ filter in the PTF10iam rest frame (i.e. these points are the ones that can be directly compared to the solid line). The K-corrections are based on the blackbody fits performed for epochs with $3$ or more bands observed within $0.5$ days (see text for details). All four events exhibit a rapid $\lesssim10$ day rise to a luminous ($\approx-20$ mag) peak. SNLS05d2bk shows a second peak approximately $40$ days after the main peak.}
\end{figure*}

\begin{figure}
\includegraphics[width=\columnwidth]{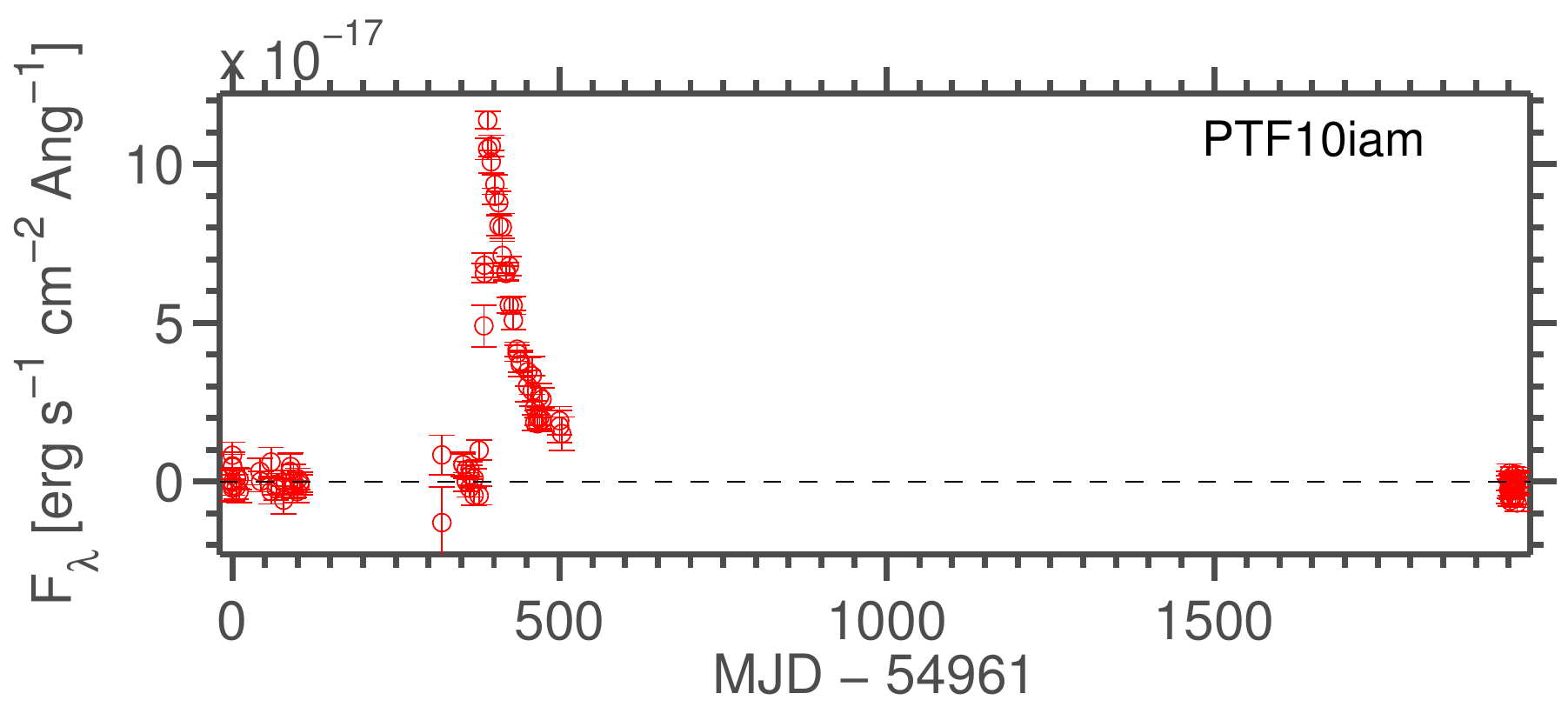}
\includegraphics[width=\columnwidth]{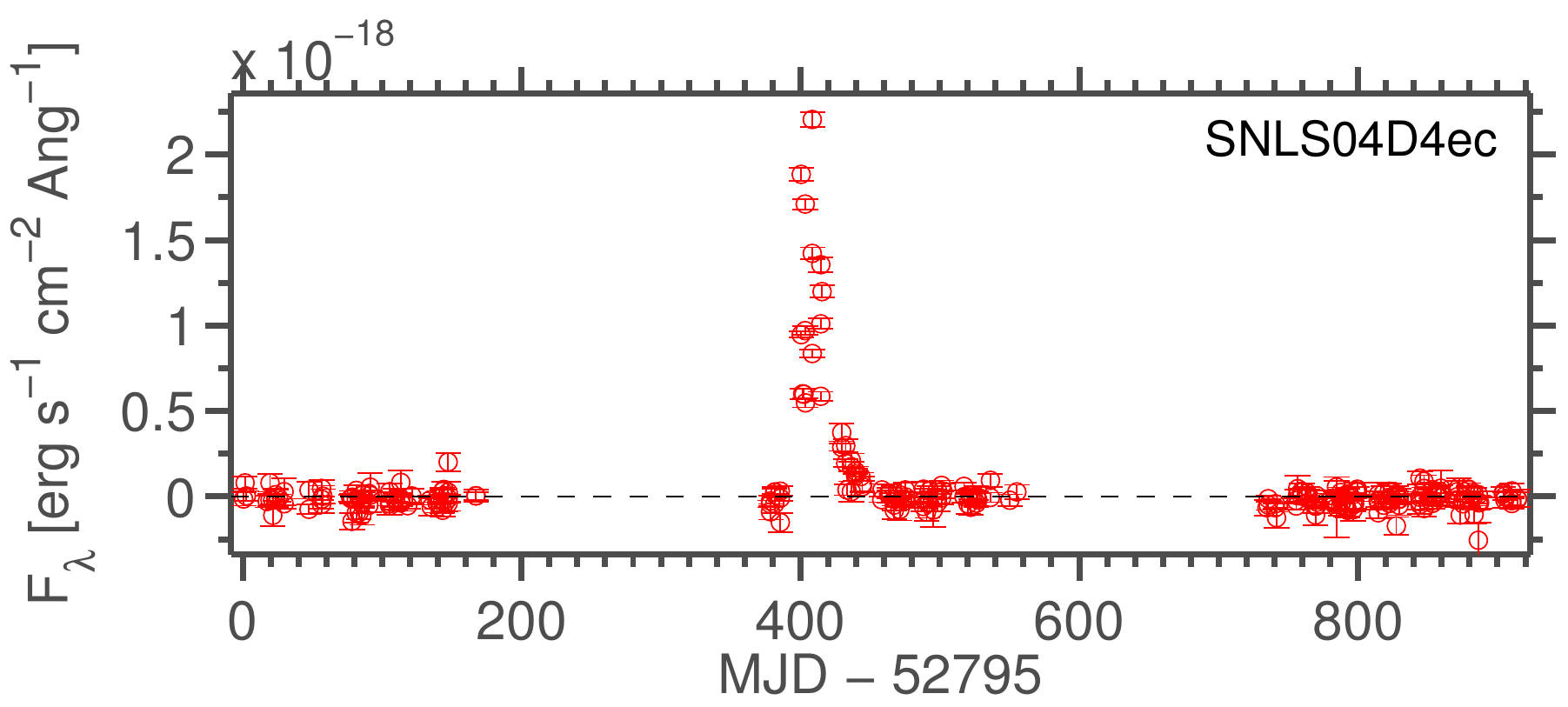}
\includegraphics[width=\columnwidth]{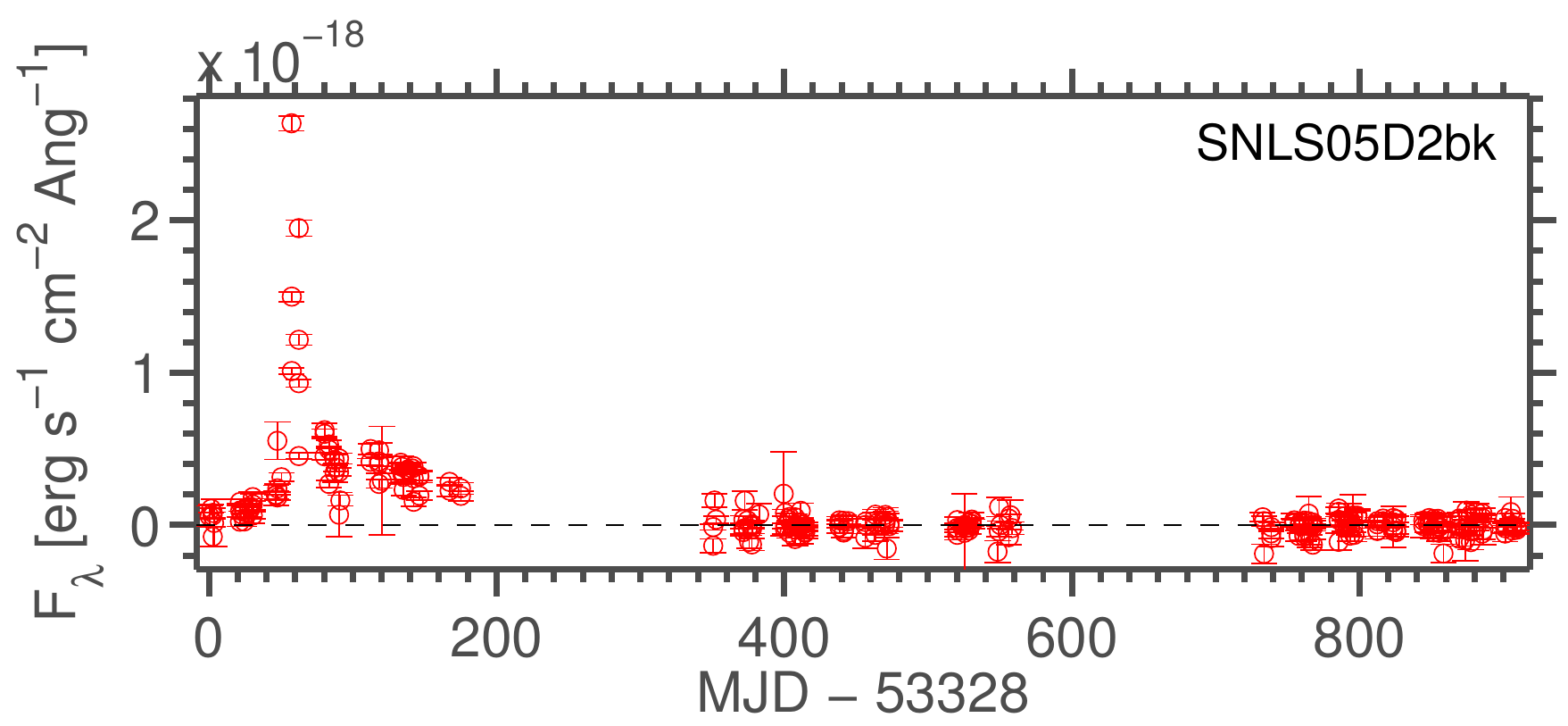}
\includegraphics[width=\columnwidth]{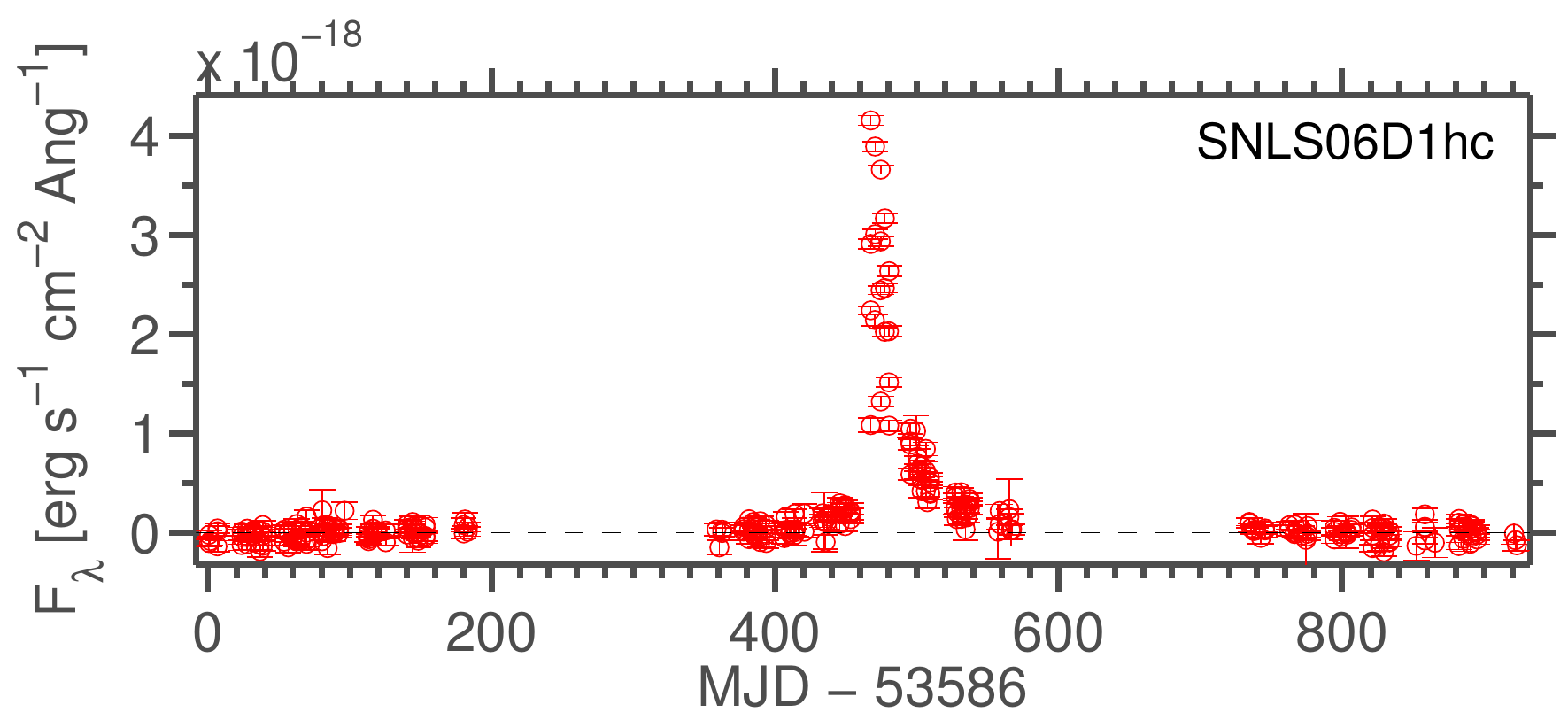}
\caption{\label{fig:longterm_lcs}Long-term light curves of our events from the PTF and SNLS surveys. No activity is detected outside the main outburst for each event.}
\end{figure}

We present long-term light curves in observed flux for our events in Figure \ref{fig:longterm_lcs}. There is no evidence for additional activity other than the main eruptions.

The SNLS host-galaxy magnitudes were obtained from the SNLS 5-year imaging data set (Hardin et al, in prep), following the general method described in Kronborg et al. (2010). In short, photometry was performed on deep image stacks in the $ugriz$ Megacam filters. The deep stacks are constructed by selecting 60\% of the best quality images. Transmission and seeing cuts (${\rm FWHM} < 1.1$'') were applied. Because we have fewer exposures in the $u$-band than in the other bands, less stringent quality cuts are applied to these images. For the Deep-D2 field (relevant for SNLS objects with ``D2'' in their name), we use the Terapix T0006 D2-$u$ stack\footnote{ http://terapix.iap.fr/}, as it incorporates COSMOS (Scoville 2007, Koekemoer 2007) CFHT-Megacam-$u$ data that partially overlap with the D2 field and were not processed in the SNLS pipeline.

The selected images were co-added using the {\sc swarp v2.17.1} package\footnote{http://terapix.iap.fr/soft/swarp/} to produce a large contiguous $1$ square degree ``season''-stack (a season corresponds to the 6 consecutive months during which the field was observed). These ``season'' frames are further co-added excluding the season during which the supernova exploded.

The source detection and photometry is performed using SExtractor V2.4.4 (Bertin \& Arnouts, 1996) in double image mode. The detection is made in the $i$ band. Zero points are computed using aperture photometry on a tertiary star catalog described in Regnault et al. (2009).

For PTF10iam and Dougie, host-galaxy magnitudes are taken from the Sloan Digital Sky Survey (SDSS)\footnote{http://www.sdss.org} via DR10 (Ahn et al. 2014). All host-galaxy magnitudes are presented in Table \ref{tab:hostmags}.

\renewcommand{\arraystretch}{1.4}
\begin{table*}
\begin{center}
{\caption{\label{tab:hostmags}Host galaxy magnitudes for our events and for Dougie. The data for the host galaxies of PTF10iam and Dougie are taken from SDSS. For the SNLS events the magnitudes are taken from deep co-add pre-explosion SNLS images and are given in the Vega system.}}
\begin{tabular}{lccccc}
\hline
\hline
{Object} & {$u$} & {$g$} & {$r$} & {$i$} &{$z$}\tabularnewline
\hline
{PTF10iam Host} & {$18.76\pm0.025$} & {$17.67\pm0.006$} & {$17.22\pm0.005$} & {$16.90\pm0.006$} & {$16.80\pm0.015$} \tabularnewline
{SNLS04D4ec Host} & {$22.54\pm0.046$} & {$22.90\pm0.027$} & {$21.98\pm0.022$} & {$21.36\pm0.020$} & {$21.18\pm0.038$} \tabularnewline
{SNLS05D2bk Host} & {$22.85\pm0.057$} & {$22.89\pm0.033$} & {$22.03\pm0.026$} & {$21.17\pm0.021$} & {$20.86\pm0.033$} \tabularnewline
{SNLS06D1hc Host} & {$23.56\pm0.122$} & {$23.91\pm0.067$} & {$22.89\pm0.045$} & {$22.30\pm0.042$} & {$22.07\pm0.078$} \tabularnewline
\hline
{Dougie Host} & {$22.90\pm0.481$} & {$21.17\pm0.048$} & {$19.88\pm0.023$} & {$19.45\pm0.024$} & {$19.12\pm0.058$} \tabularnewline
\hline

\end{tabular}
\end{center}
\end{table*}
\renewcommand{\arraystretch}{1}

\subsection{Spectroscopy} \label{sec:specs}

\begin{figure*}
\includegraphics[width=\textwidth]{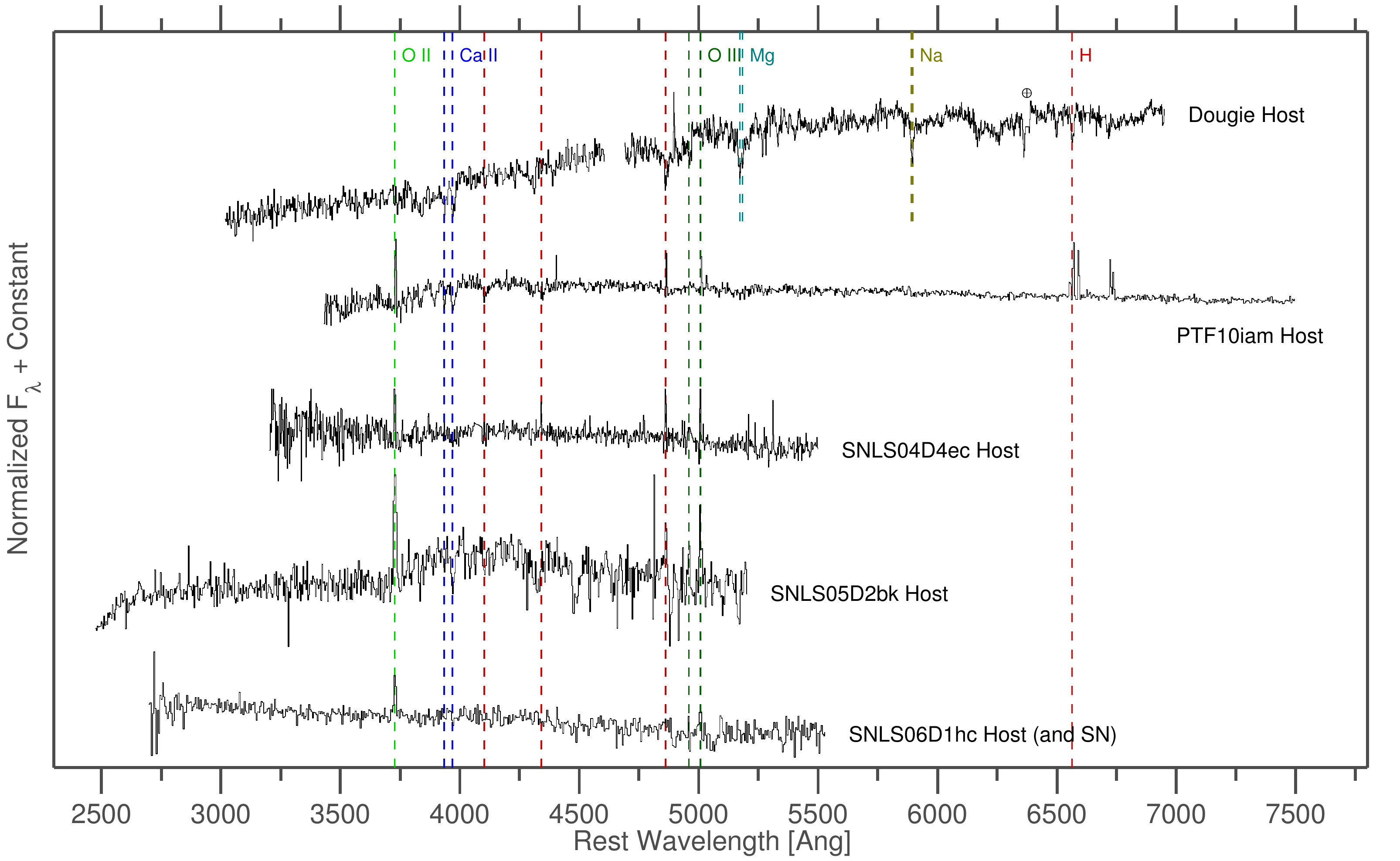}
\caption{\label{fig:spec_hosts}Spectra of the host galaxies of our events and of Dougie. Several narrow emission and/or absorption features are used to securely determine the redshift of each galaxy. The host galaxies of our events all exhibit strong emission features, indicative of ongoing star formation. Dougie's host, in contrast, shows only absorption features. It is therefore less likely that Dougie was the explosion of a massive star.}
\end{figure*}

\begin{figure}
\includegraphics[trim=0 0 0 0 clip=true,width=\columnwidth]{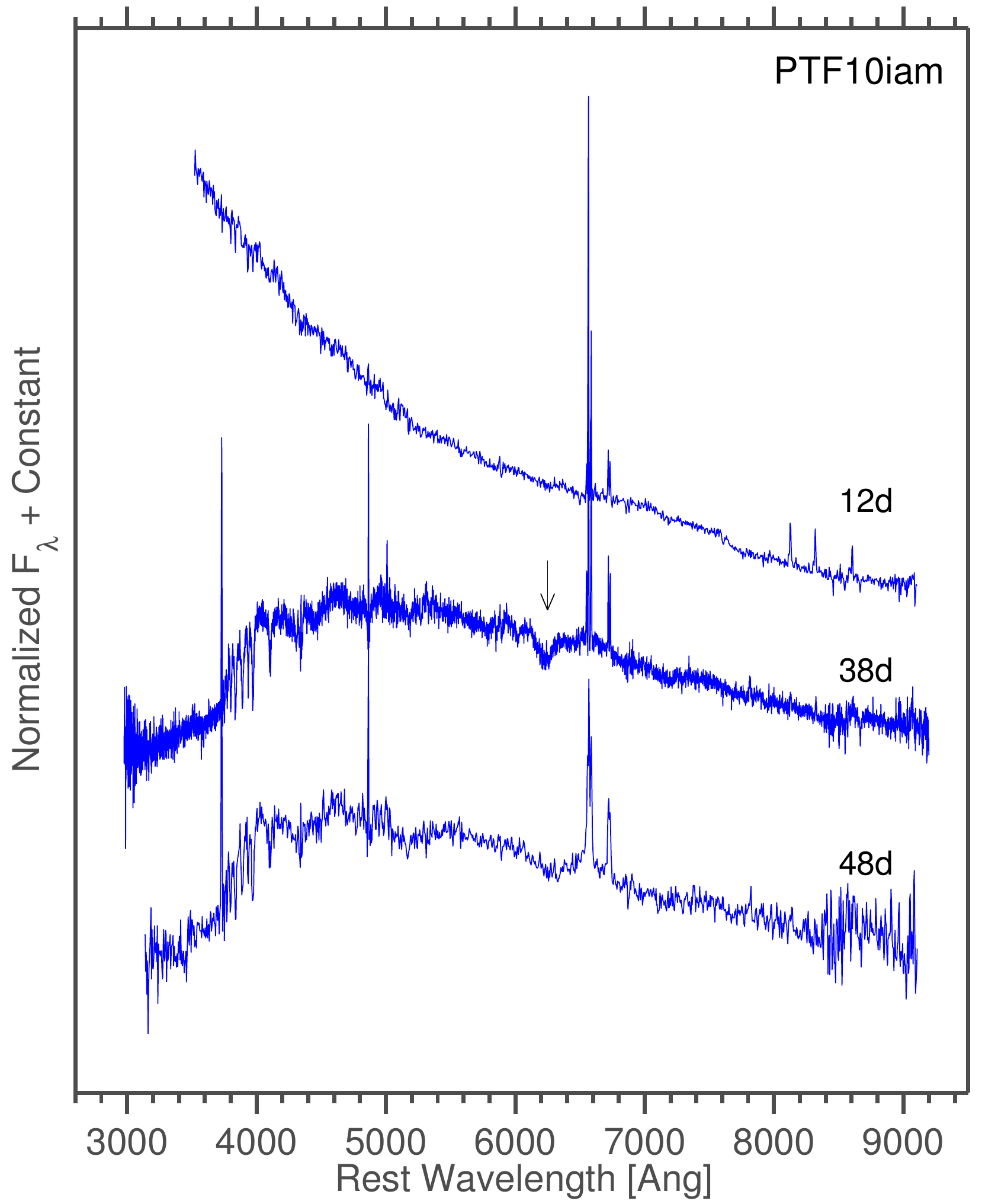}
\caption{\label{fig:spec_10iam}Spectra of PTF10iam. Phases are shown in rest-frame days from explosion. The first spectrum is mostly a blue continuum, the second spectrum displays broad H$\alpha$ in emission as well as a notable absorption feature bluewards of H${\alpha}$ (marked with an arrow). The last spectrum is heavily host-galaxy contaminated. The narrow emission and absorption features are from the host galaxy.}
\end{figure}

Spectra of PTF10iam were released in Arcavi et al. (2014), and are presented here in Figure \ref{fig:spec_10iam}. A spectrum of the host galaxy of PTF10iam was obtained by SDSS in 2002 and downloaded via DR10. This spectrum is presented in Figure \ref{fig:spec_hosts}.

No spectra were obtained of SNLS04D4ec and SNLS05D2bk during outburst, but spectra of their host galaxies were taken after the transients faded significantly. A spectrum of SNLS06D1hc was obtained about three weeks after explosion, but no discernible SN features (other than a possible blue continuum) can be seen in the spectrum. All SNLS spectra were taken with the FOcal Reducer and Spectrograph (FORS1 and FORS2; Appenzeller et al. 1998) mounted on the Very Large Telescope (VLT). The FORS data were reduced using a mixture of standard IRAF\footnote{IRAF, the Image Reduction and Analysis Facility, is a general purpose software system for the reduction and analysis of astronomical data. IRAF is written and supported by the National Optical Astronomy Observatories (NOAO) in Tucson, Arizona} tasks and our own routines that were specifically written to process MOS data from FORS1 and FORS2. For the SNe, we also derive an error spectrum, which is computed from regions in the 2D sky-subtracted spectrum that are free of objects. The spectra are presented in Figure \ref{fig:spec_hosts}.

To verify the redshift (and thus peak luminosity) of Dougie reported by Vink{\'o} et al. (2015), we observed its host galaxy with the Low Resolution Imaging Spectrometer (LRIS; Oke et al. 1995) mounted on the Keck I 10-meter telescope, and with the Gemini Multi-Object Spectrograph (GMOS; Hook et al. 2004) mounted on the Gemini-North 8.1-meter telescope. The LRIS data were reduced using standard IRAF and IDL routines. The Gemini data were reduced using the Gemini IRAF package in addition to custom Python spectral reduction scripts. The highest signal to noise was obtained for the blue part of the LRIS spectrum, and for the red part of the GMOS spectrum, and we present these in Figure \ref{fig:spec_hosts}. We identify several narrow features in the spectrum and use them to determine a redshift of $0.194$ (only slightly different than the value of $0.191$ measured by Vink{\'o} et al. 2015 using cross correlation with galaxy spectral templates). 

We thus confirm the peak absolute magnitude of Dougie ($M_R\approx-23$) reported by Vink{\'o} et al. (2015), placing it amongst the brightest SLSNe. It does not fit in the SN-SLSN luminosity gap which is the focus of our interest here. In addition, our observations of Dougie's host galaxy reveal it to be absorption-feature dominated, unlike the emission-rich hosts of our sample (Fig. \ref{fig:spec_hosts}). Its redder color also indicates a much lower star formation rate (SFR) compared to the other hosts (Table \ref{tab:host_params}). With its extreme luminosity and passive host galaxy, Dougie is likely a different type of event compared to those in our sample and we do not discuss it further in this work.

Our full spectral log is presented in Table \ref{tab:spec}. Digital versions of our spectra are available online through the Weizmann Interactive Supernova data REPository (WISeREP\footnote{http://wiserep.weizmann.ac.il}; Yaron \& Gal-Yam 2012).

\renewcommand{\arraystretch}{1.4}
\begin{table}
{\caption{\label{tab:spec}Spectroscopic observations. Phases are denoted in rest-frame days relative to peak luminosity. The spectra of PTF10iam were presented also in Arcavi et al. (2014), but are noted here for completeness.}}
\begin{tabular}{lccc}
\hline
\hline
{Object} & {UT Date} & {Phase} & {Telescope}\tabularnewline
{} & {} & {[days]} & {(Instrument)}\tabularnewline
\hline
{PTF10iam} & {2010 June 8} & {2} & {Keck I (LRIS)}\tabularnewline
{PTF10iam} & {2010 July 7} & {28} & {Keck I (LRIS)}\tabularnewline
{PTF10iam} & {2010 July 18} & {38} & {P200 (DBSP)}\tabularnewline
{SNLS04D4ec Host} & {2007 Jun 19} & {} & {VLT (FORS1)}\tabularnewline
{SNLS05D2bk Host} & {2007 Jan 28} & {} & {VLT (FORS2)}\tabularnewline
{SNLS06D1hc} & {2006 Nov 25} & {5} & {VLT (FORS1)}\tabularnewline
\hline
{Dougie Host} & {2014 Dec 18} & {} & {Keck I (LRIS)}\tabularnewline
{Dougie Host} & {2015 Feb 19} & {} & {Gemini N (GMOS)}\tabularnewline
\hline

\end{tabular}
\end{table}
\renewcommand{\arraystretch}{1}

\section{Analysis}

\subsection{Blackbody Fits and Bolometric Light Curves} \label{sec:bb}

\begin{figure}
\includegraphics[trim=0 10 0 0,clip,width=\columnwidth]{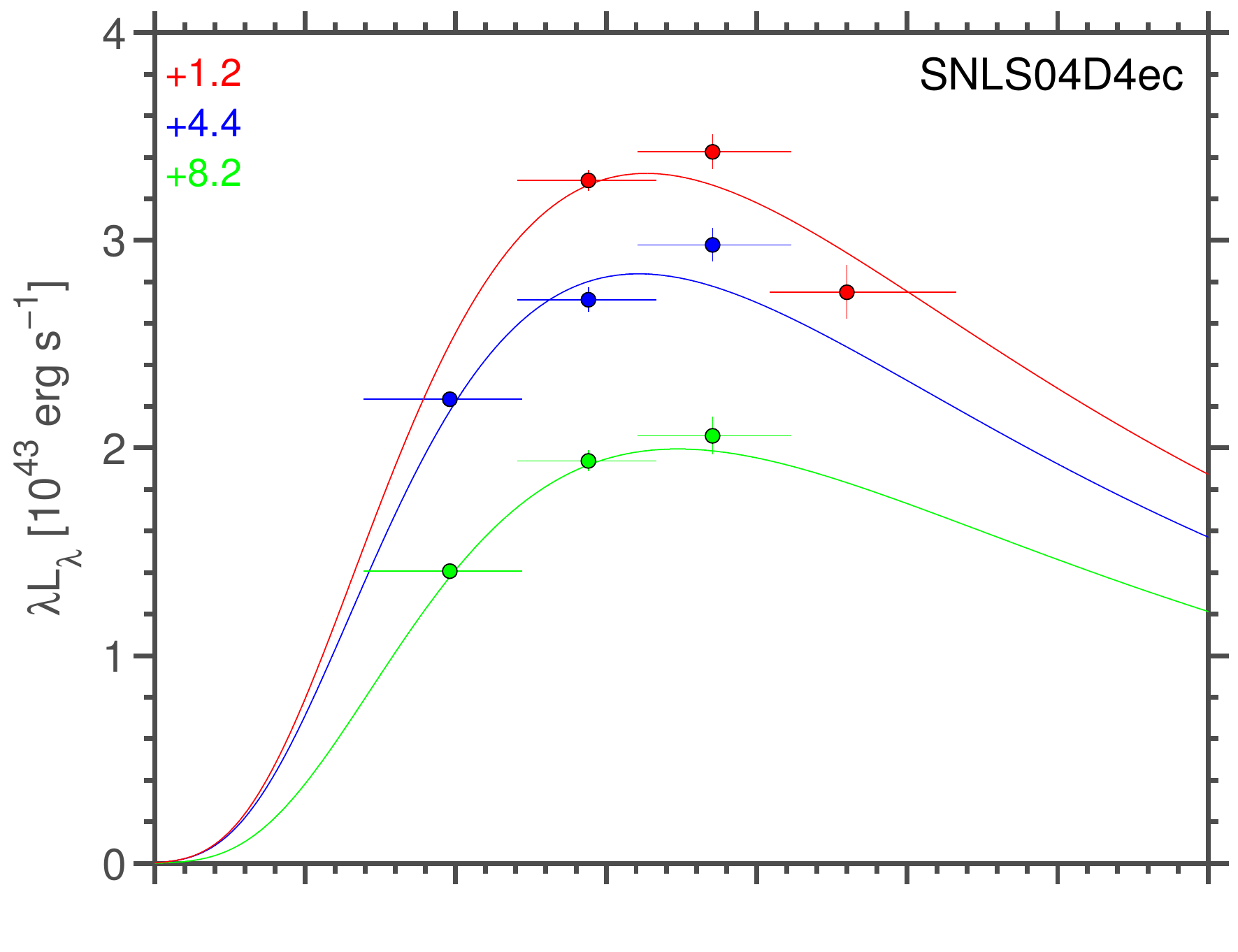}
\includegraphics[trim=0 10 0 0,clip,width=\columnwidth]{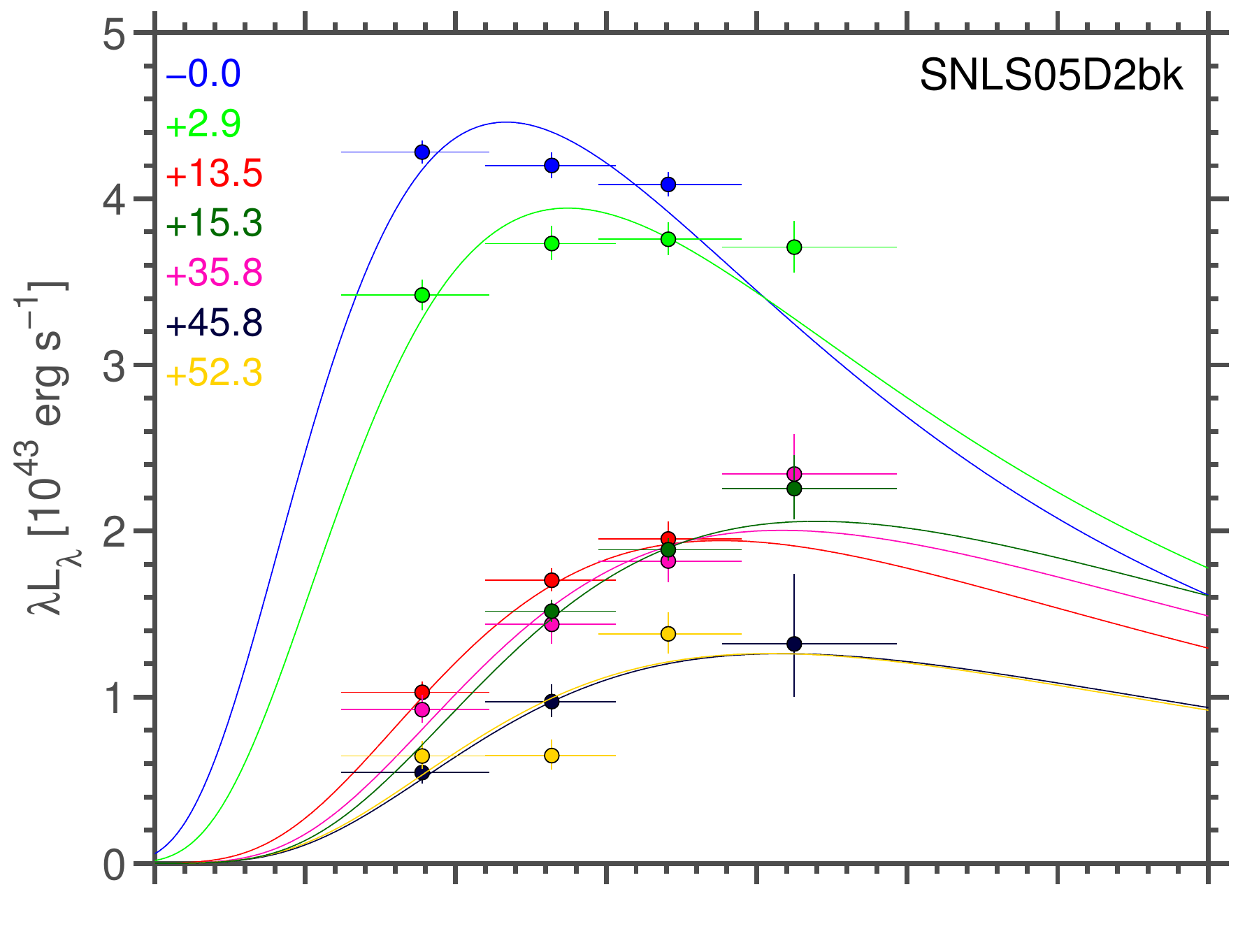}
\includegraphics[trim=0 0 12 0,clip,width=\columnwidth]{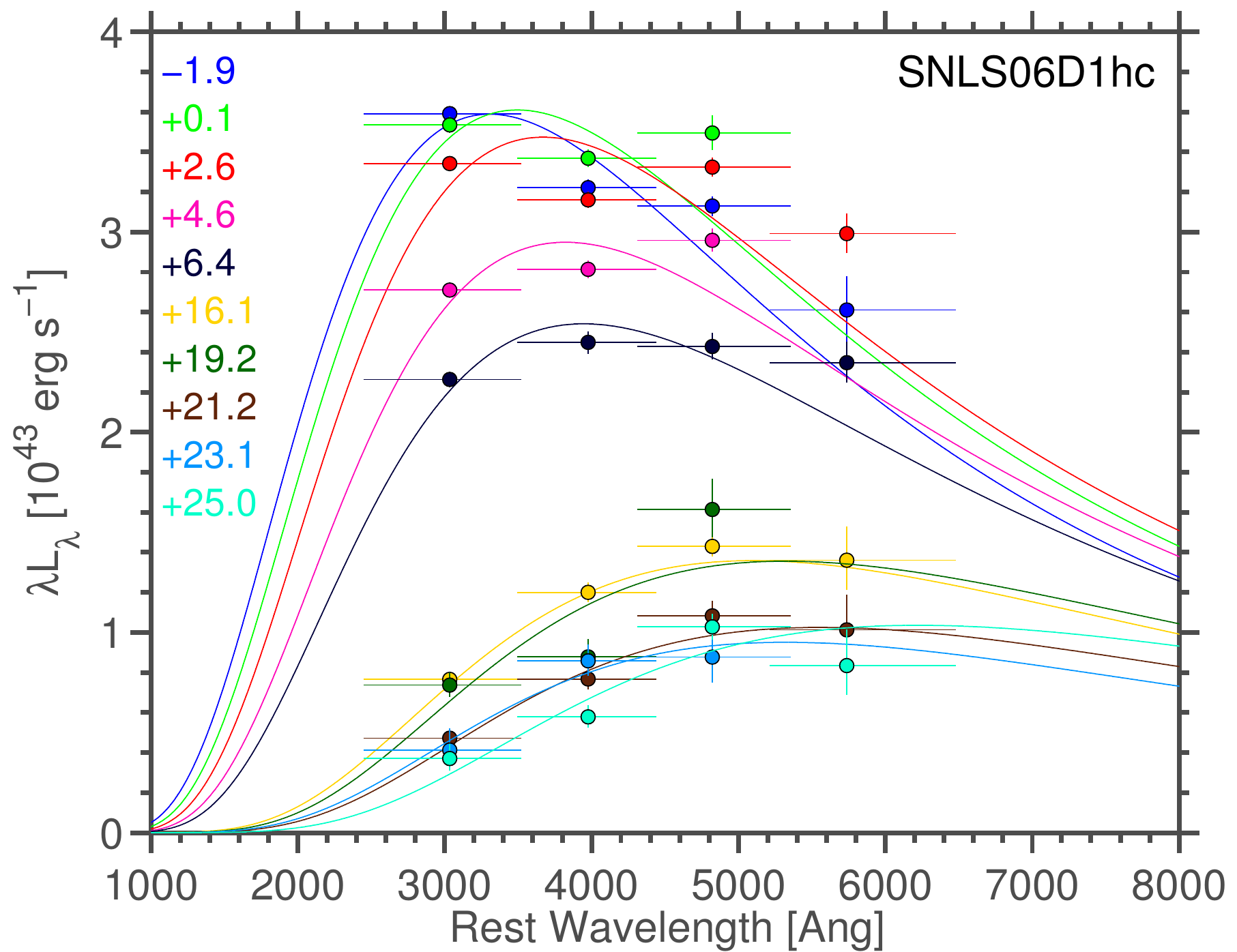}
\caption{\label{fig:seds}Blackbody fits to the SEDs of the SNLS events (for epochs when at least three filters were observed within half a day). Epochs are shown in rest frame days relative to peak. The best-fit temperatures and radii are presented in Table \ref{tab:bbfits} and plotted in Figure \ref{fig:blackbody}.}
\end{figure}

\begin{figure}
\includegraphics[trim=0 10 0 0,clip,width=\columnwidth]{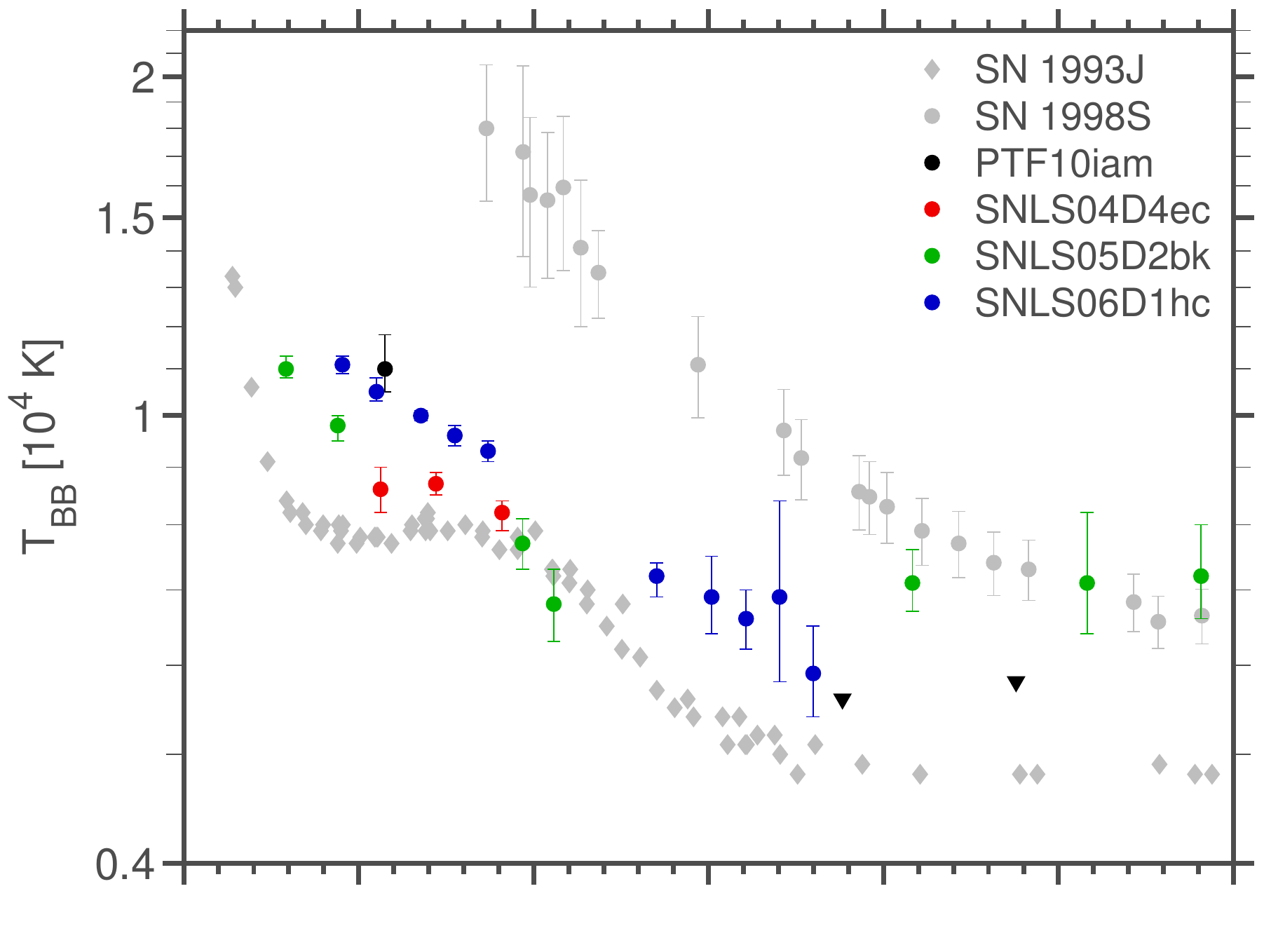}
\includegraphics[trim=0 10 0 0,clip,width=\columnwidth]{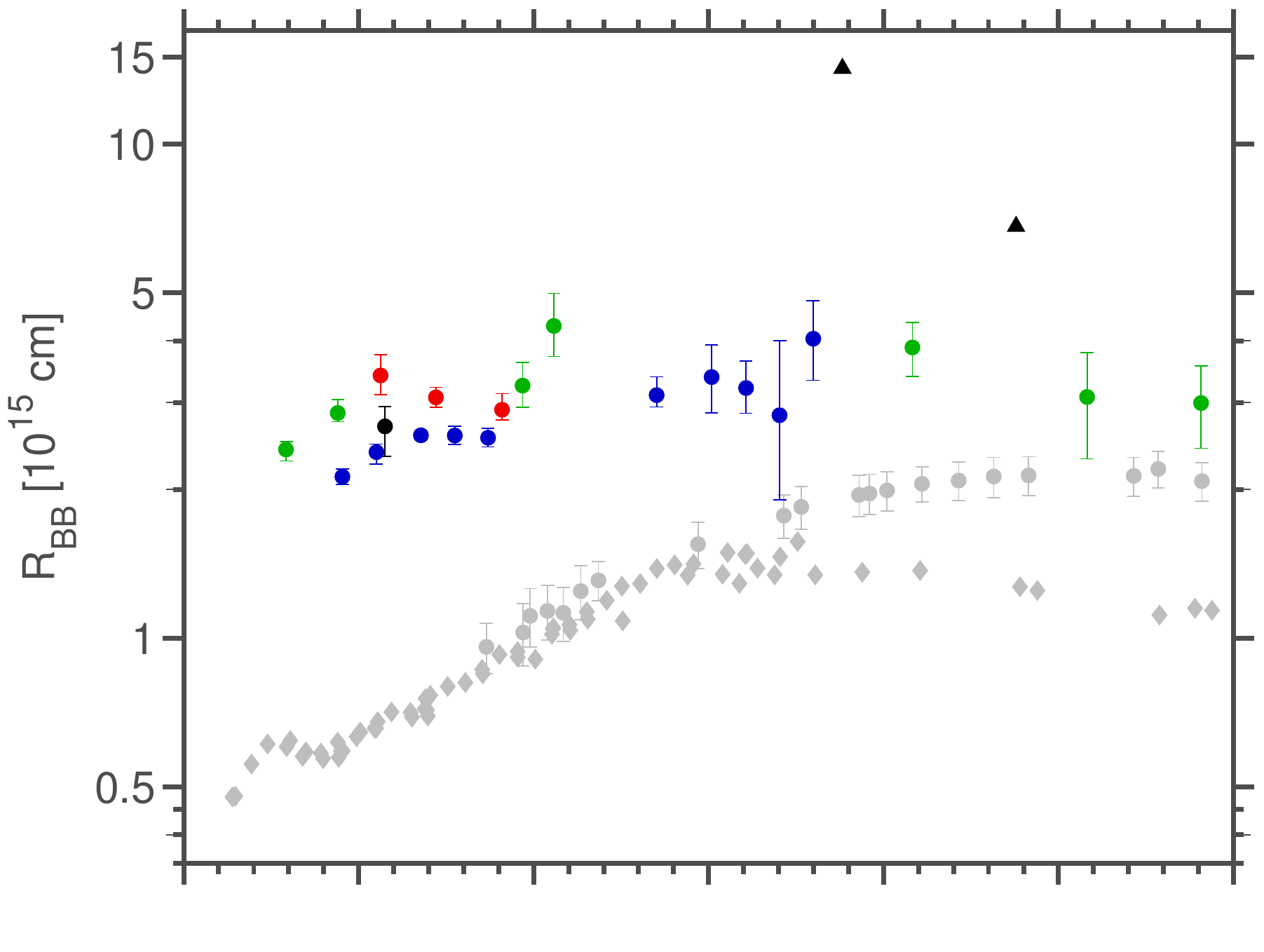}
\includegraphics[trim=0 0 2 0,clip,width=\columnwidth]{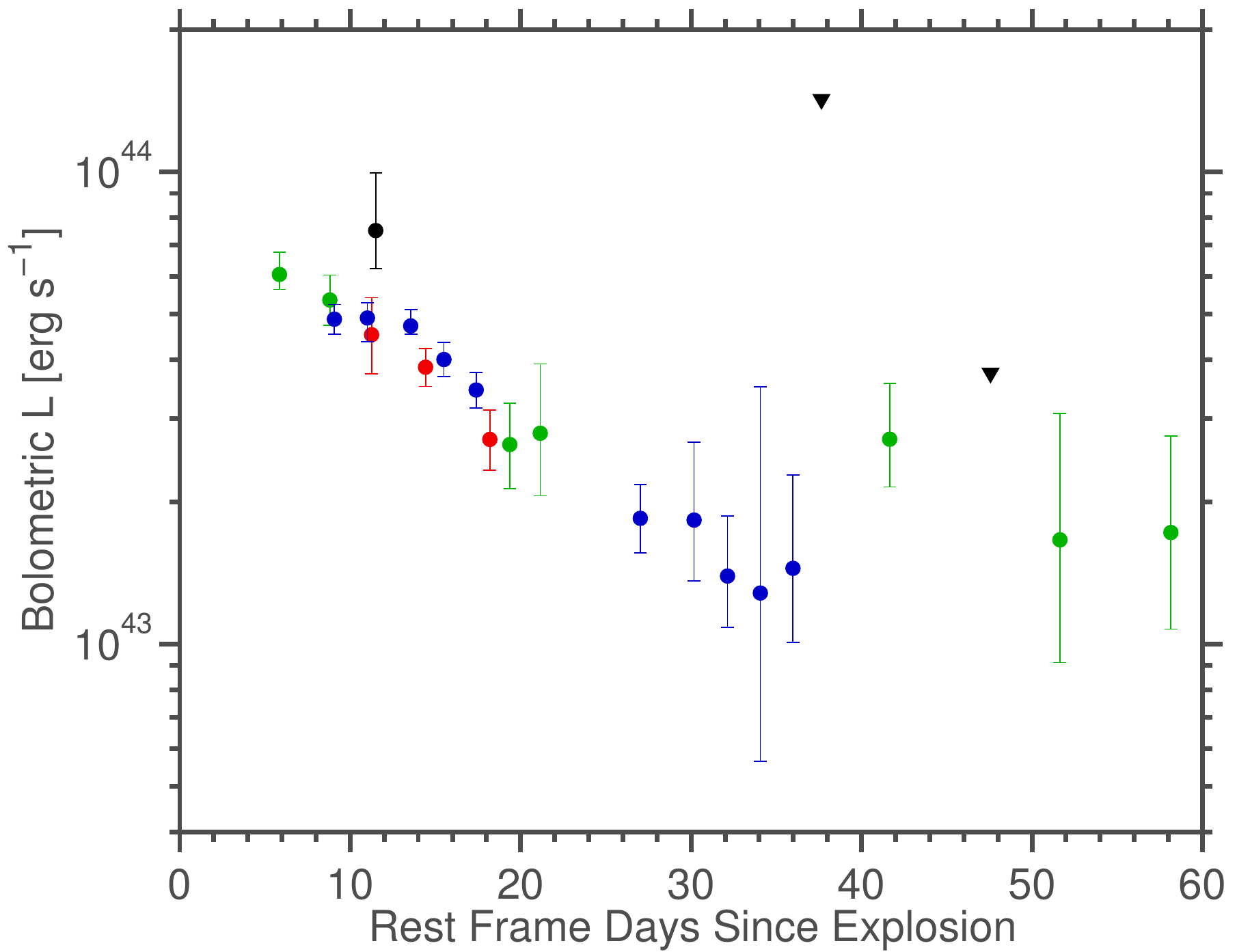}
\caption{\label{fig:blackbody}Best fit blackbody temperatures (top) and radii (middle) for our events, and for the Type~II SNe 1993J (from Richmond et al. 1994; explosion date from Filippenko et al. 1993) and 1998S (from Fassia et al. 2000; explosion date from Chugai 2001) for comparison. The resulting blackbody bolometric luminosities are shown in the bottom panel. Our events show similar blackbody evolution amongst themselves. They have temperatures between those of the partially hydrogen-stripped IIb SN 1993J and the hydrogen-rich SN 1998S, but show more extended blackbody radii than both comparison events.}
\end{figure}

We compare each spectrum of PTF10iam to the sum of a blackbody and a scaled host galaxy spectrum, and fit for the blackbody parameters and host scaling factor simultaneously. For the SNLS events we use the SED from epochs when photometric data is available for at least three different bands within half a day (while using each photometric data point only once). We present the SED fits in Figure \ref{fig:seds}. Our best fit blackbody temperatures and radii for all events are presented in the top and middle panels of Figure \ref{fig:blackbody} and in Table \ref{tab:bbfits}. All events show very similar temperatures and radii and evolve at similar rates. The temperatures are between those of the Type IIb SN 1993J (data from Richmond et al. 1994) and the Type II SN 1998S (data from Fassia et al. 2000), but the radii of our events are larger. PTF10iam appears to display higher blackbody radii than the SNLS events, but this may be due to the fact that it is the only object for which a spectrum was used in the fit, and not calibrated photometry. The accurate flux calibration of the PTF10iam spectra can not be tested since no multi-color photometry are available for that object.

We use the blackbody radius and temperatures to calculate bolometric light curves, and present these light curves in the bottom panel of Figure \ref{fig:blackbody} and in Table \ref{tab:bbfits}. We take the brightest bolometric point to be the peak bolometric luminosity for each event.

\renewcommand{\arraystretch}{1.4}
\begin{table}
{\caption{\label{tab:bbfits}Best fit blackbody temperatures and radii, and resulting bolometric luminosities, to the spectra of PTF10iam and to the multi-band photometry of the SNLS events. Phases are listed in rest frame days from peak. Errors denote $1\sigma$ confidence intervals for the blackbody fits, and are propagated to the calculated luminosities.}}
\begin{tabular}{lcccc}

\hline
\hline
{Object} & {Phase} & {$T$} & {$R$} & {$L_{bol}$}\tabularnewline
{} & {[d]} & {[K]} & {[$10^{15}$\,cm]} & {[$10^{43}$\,erg\,s$^{-1}$]}\tabularnewline
\hline
{PTF10iam} & {$1.6$} & {$11000_{-500}^{+800}$} & {$2.68_{-0.35}^{+0.26}$} & {$7.51_{-1.27}^{+2.43}$}\tabularnewline
{PTF10iam} & {$27.8$} & {$<5600$} & {{$>14.24$}} & {{$<14.22$}}\tabularnewline
{PTF10iam} & {$37.7$} & {$<5800$} & {{$>6.82$}} & {{$<3.75$}}\tabularnewline
\hline
{SNLS04D4ec} & {$1.2$} & {$8600_{-400}^{+400}$} & {$3.40_{-0.30}^{+0.34}$} & {$4.52_{-0.78}^{+0.90}$}\tabularnewline
{SNLS04D4ec} & {$4.4$} & {$8700_{-200}^{+200}$} & {$3.07_{-0.14}^{+0.15}$} & {$3.86_{-0.34}^{+0.37}$}\tabularnewline
{SNLS04D4ec} & {$8.2$} & {$8200_{-300}^{+300}$} & {$2.90_{-0.20}^{+0.22}$} & {$2.71_{-0.38}^{+0.42}$}\tabularnewline
\hline
{SNLS05D2bk} & {$0.0$} & {$11000_{-200}^{+300}$} & {$2.41_{-0.13}^{+0.09}$} & {$6.06_{-0.43}^{+0.69}$}\tabularnewline
{SNLS05D2bk} & {$2.9$} & {$9800_{-300}^{+300}$} & {$2.85_{-0.17}^{+0.18}$} & {$5.35_{-0.63}^{+0.69}$}\tabularnewline
{SNLS05D2bk} & {$13.5$} & {$7700_{-400}^{+400}$} & {$3.25_{-0.31}^{+0.37}$} & {$2.64_{-0.51}^{+0.59}$}\tabularnewline
{SNLS05D2bk} & {$15.3$} & {$6800_{-500}^{+600}$} & {$4.28_{-0.67}^{+0.71}$} & {$2.80_{-0.74}^{+1.13}$}\tabularnewline
{SNLS05D2bk} & {$35.8$} & {$7100_{-400}^{+500}$} & {$3.87_{-0.49}^{+0.48}$} & {$2.71_{-0.56}^{+0.85}$}\tabularnewline
{SNLS05D2bk} & {$45.8$} & {$7200_{-1000}^{+1200}$} & {$2.95_{-0.78}^{+1.03}$} & {$1.66_{-0.75}^{+1.42}$}\tabularnewline
{SNLS05D2bk} & {$52.3$} & {$7200_{-800}^{+900}$} & {$3.00_{-0.63}^{+0.80}$} & {$1.72_{-0.65}^{+1.04}$}\tabularnewline
\hline
{SNLS06D1hc} & {$-1.9$} & {$11100_{-200}^{+200}$} & {$2.12_{-0.07}^{+0.08}$} & {$4.87_{-0.34}^{+0.36}$}\tabularnewline
{SNLS06D1hc} & {$0.1$} & {$10600_{-300}^{+200}$} & {$2.33_{-0.09}^{+0.14}$} & {$4.90_{-0.53}^{+0.38}$}\tabularnewline
{SNLS06D1hc} & {$2.6$} & {$10000_{-100}^{+200}$} & {$2.57_{-0.10}^{+0.05}$} & {$4.72_{-0.19}^{+0.39}$}\tabularnewline
{SNLS06D1hc} & {$4.6$} & {$9600_{-200}^{+200}$} & {$2.57_{-0.10}^{+0.11}$} & {$4.01_{-0.32}^{+0.34}$}\tabularnewline
{SNLS06D1hc} & {$6.4$} & {$9300_{-200}^{+200}$} & {$2.54_{-0.11}^{+0.11}$} & {$3.45_{-0.29}^{+0.31}$}\tabularnewline
{SNLS06D1hc} & {$16.1$} & {$7200_{-300}^{+300}$} & {$3.11_{-0.24}^{+0.28}$} & {$1.85_{-0.29}^{+0.33}$}\tabularnewline
{SNLS06D1hc} & {$19.2$} & {$7000_{-500}^{+700}$} & {$3.27_{-0.57}^{+0.52}$} & {$1.83_{-0.47}^{+0.85}$}\tabularnewline
{SNLS06D1hc} & {$21.2$} & {$6600_{-400}^{+500}$} & {$3.21_{-0.44}^{+0.43}$} & {$1.39_{-0.31}^{+0.47}$}\tabularnewline
{SNLS06D1hc} & {$23.1$} & {$7000_{-1300}^{+2000}$} & {$2.74_{-1.08}^{+1.39}$} & {$1.28_{-0.72}^{+2.22}$}\tabularnewline
{SNLS06D1hc} & {$25.0$} & {$5800_{-500}^{+700}$} & {$4.24_{-0.86}^{+0.84}$} & {$1.45_{-0.44}^{+0.84}$}\tabularnewline
\hline

\end{tabular}
\end{table}
\renewcommand{\arraystretch}{1}

We use the SNLS blackbody fits to K-correct the SNLS light curves to the PTF10iam Mould $R$-band observed frame, and present the results as empty red circles in Figure \ref{fig:lcs} (only for epochs in the SNLS light curves when blackbody fits are available). These magnitudes can then be compared to the observed Mould-$R$ photometry of PTF10iam (plotted as a solid red line in each SNLS light curve plot). The SNLS and PTF10iam light curves are very similar at the epochs where a comparison is possible. 

SNLS05D2bk, however, displays a second peak approximately $40$ rest-frame days after the main peak. The second peak is broader and fainter than the main peak but is clearly seen in all filters. Double peaks are common in the rest-frame IR light curves of Type~Ia SNe, and have also been seen in the optical bands of Type~IIb SNe (intermediate cases between H-rich SNe II and He-dominated SNe Ib) and in at least one energetic Type~Ic (H- and He-stripped) SN. For all cases, however, the second peak is never seen as late as $40$ days after the first one (Richmond et al. 1994; Kasen 2006; Arcavi et al. 2011; Kumar et al. 2013; Bufano et al. 2014; Morales-Garoffolo 2014; Nakar et al. 2015; Arcavi et al. in prep).

\subsection{Rise Times and Peak Magnitudes} \label{sec:rise_times}

For PTF10iam we calculate the time and magnitude of peak brightness ($t_{peak}$ and $M_{peak}$ respectively) by fitting a $2^{nd}$ order polynomial to the $R$-band light curve (in flux space) around peak. The explosion time ($t_{exp}$) is taken as the time of zero flux from the polynomial fit. For the SNLS events, which have less densely sampled rises and peaks, but tighter non-detection constraints, we conservatively take the explosion time to be that of the last non-detection before the first detection. We take the peak time and magnitude to be that of the brightest measured photometric point in $z$-band for SNLS04D4ec, and in $i$-band for the other SNLS events. The measured parameters are presented in Table \ref{tab:events_params}.

\begin{figure*}[t]
\includegraphics[width=1.0\textwidth]{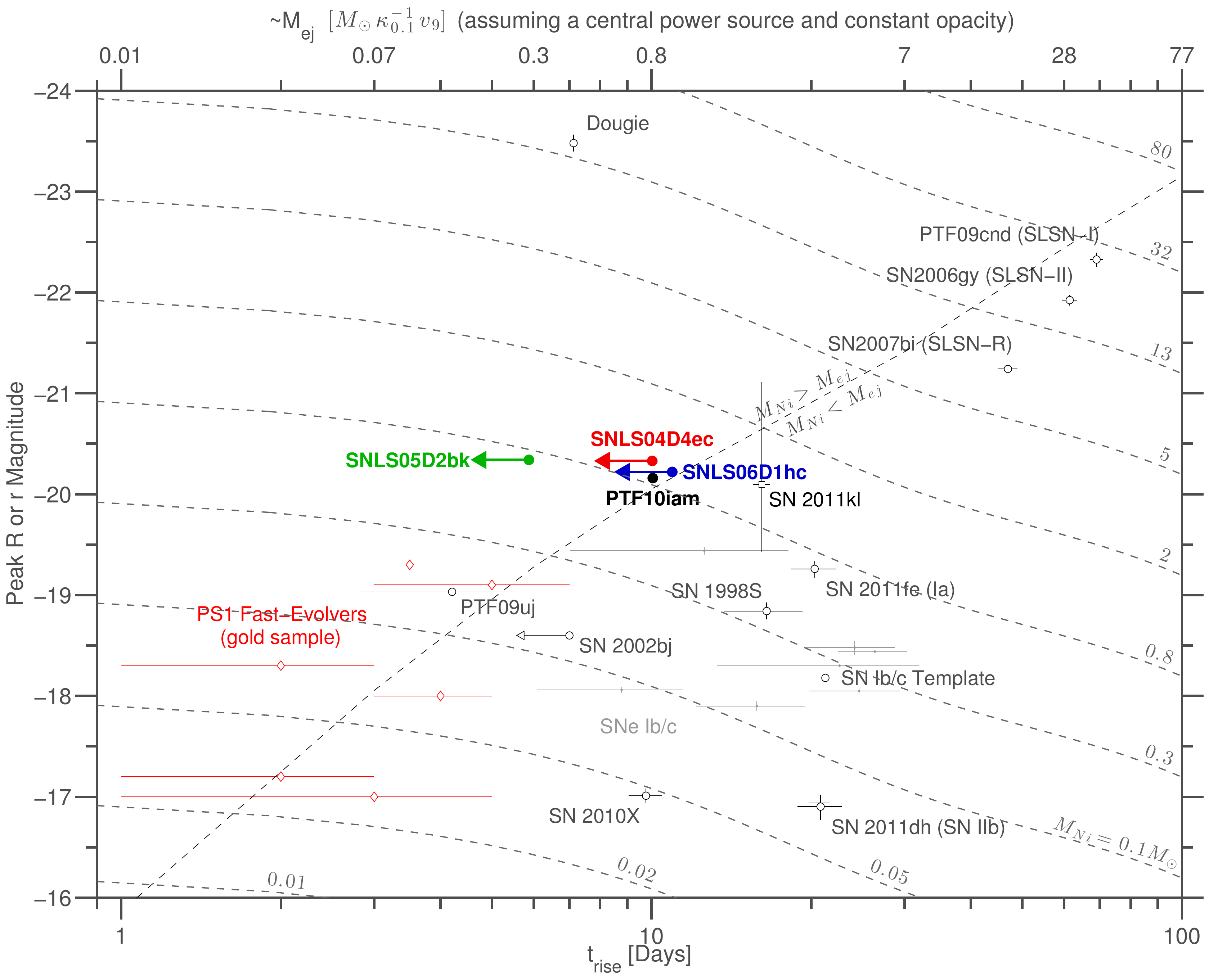}
\caption{\label{fig:peak_vs_trise}Peak magnitude vs. rise time of our events (upper limits for the SNLS rise times) compared to other SNe (see text for references). All comparison data peak magnitudes and rise times are in the observed $R$ or $r$-band. Rise times are in the rest frame of each event. Ejecta mass estimates are normalized to an expansion velocity of $10,000$\,km\,s$^{-1}$ (see text for details, also regarding the calculated nickel masses), and should only be considered approximate. Our events have shorter rise times compared to most SNe, and are more luminous than all similarly-rapid events (except for Dougie, which is a clear outlier in this context). The only similar event to ours is SN\,2011kl, which was accompanied by an ultra-long-duration GRB (Greiner et al. 2015). The positions of our events in this phase space require either a very high nickel to ejecta mass ratio or an alternative dominant power source to nickel decay.}
\end{figure*}

The rise time of each event ($t_{rise}$) is taken to be the rest-frame difference between the time of (main) peak and the explosion time (for the SNLS events this is interpreted as an upper limit on the rise time). We plot the resulting rise times and peak magnitudes in Figure \ref{fig:peak_vs_trise} together with comparison SNe\footnote{The following comparison data is used in Figure \ref{fig:peak_vs_trise}: the SLSNe PTF09cnd from Quimby et al. (2011), SN\,2006gy from Smith et al. (2007) and Ofek et al. (2007) and SN\,2007bi from Gal-Yam et al. (2009), the normal type Ia SN\,2011fe from Vink{\'o} et al. (2012), the average type Ib/c light curve from Drout et al. (2011) and the Type Ib/c $r$-band sample of Taddia et al. (2015; grey points), the Type~IIb SN\,2011dh from Arcavi et al. (2011), the peculiar Type~IIn SN\,1998S from Li et al. (1998), Nakamura et al. (1998), Bignotti et al. (1998), Leonard et al. (2000), Liu et al. (2000) and Fassia et al. (2000), PTF09uj from Ofek et al. (2010), the rapidly evolving SNe 2002bj from Poznanski et al. (2011) and 2010X from Kasliwal et al. (2011), the Pan-STARRS rapidly evolving ``gold'' sample from Drout et al. (2014; red diamonds), Dougie from Vink{\'o} et al. (2015) and SN\,2011kl from Greiner et al. (2015)}.

The rise time and peak magnitudes of the comparison sample were taken from the referenced sources, if stated there explicitly, otherwise they were extracted in the same way as for PTF10iam. Our events have shorter rise times compared to the ``standard'' SNe shown and similar rise times as the rapidly evolving SNe 2010X, 2002bj, PTF09uj and those in the Drout et al. (2014) sample. Our events, however, are much more luminous and are the only ones located in the SN - SLSN gap, aside from SN\,2011kl. We exclude from this plot strongly interacting Type~IIn SNe, identified by narrow emission lines in their spectra (a comparison to such events is shown in Figure \ref{fig:peak_vs_te}).

\renewcommand{\arraystretch}{1.4}
\begin{table*}
\begin{center}
{\caption{\label{tab:events_params}Light curve parameters for our events (the explosion and peak dates are in the observed frame, while rise times are in the rest frame). Peak magnitudes refer to $R$-band for PTF10iam, $z$-band for SNSL04D4ec and $i$-band for SNLS05D2bk and SNLS06D1hc. Bolometric luminosities are based on blackbody fits (to a spectrum of PTF10iam and to multi-band photometry of the SNLS events). Errors and confidence bounds denote $1{\sigma}$ uncertainties.}}

\begin{tabular*}{0.82\textwidth}{lcccccc}
\hline
\hline
{Object} & {$t_{exp}$} & {$t_{peak}$} & {$M_{peak}$} & {$t_{rise}$} & {$t_e$} & {Peak $L_{bol}$} \tabularnewline
{} & {[MJD]} & {[MJD]} & {} & {[days]} & {[days]} & {[$10^{43}$ erg\,s$^{-1}$]} \tabularnewline
\hline
{PTF10iam} & {$55342.24\pm0.14$} & {$55353.38\pm0.06$} & {$-20.16\pm0.01$} & {$10.05\pm0.15$} & {$2.53\pm0.38$} & {$7.51_{-1.28}^{+2.43}$} \tabularnewline
{SNLS04D4ec} & {$>53180.60$} & {$53196.58$} & {$-20.33\pm0.06$} & {$<10.03$} & {$1.95\pm0.88$} & {$4.52_{-0.78}^{+0.90}$} \tabularnewline
{SNLS05D2bk} & {$>53375.58$} & {$53385.55$} & {$-20.34\pm0.02$} & {$<5.87$} & {$2.76\pm1.79$} & {$6.06_{-0.43}^{+0.69}$} \tabularnewline
{SNLS06D1hc} & {$>54039.34$} & {$54056.36$} & {$-20.22\pm0.03$} & {$<10.95$} & {$3.62\pm2.12$} & {$4.90_{-0.53}^{+0.38}$} \tabularnewline
\hline

\end{tabular*}
\end{center}
\end{table*}
\renewcommand{\arraystretch}{1}

We estimate the ejecta mass corresponding to a particular light curve rise time using the following expression from Wheeler et al. (2014), which follows from Arnett (1982) and assumes central deposition of the power source:
\begin{equation} \label{eq:ejecta_mass}
M_{ej}\approx0.77M_{\odot}\left(\frac{\kappa}{0.1\,\textrm{cm}^{2}\,\textrm{g}^{-1}}\right)^{-1}\left(\frac{v_{ph}}{10^{9}\,\textrm{cm}\,\textrm{s}^{-1}}\right)\left(\frac{t_{rise}}{10\,\textrm{d}}\right)^{2}
\end{equation}
Here, $v_{ph}$ is the velocity at the photosphere (sometimes expressed in terms of the kinetic energy of the ejecta). This scaling, normalized to $v_{ph}=10^9$\,cm\,s$^{-1}$ is depicted in the upper axis of Figure \ref{fig:peak_vs_trise}.

We translate peak luminosity and rise time to nickel mass (for fully nickel-powered light curves) following Stritzinger \& Leibundgut (2005):
\begin{multline} \label{eq:ni_mass}
\frac{M_{Ni}}{M_{\odot}}=\frac{L_{peak}}{\textrm{erg\,s}^{-1}}/ \\ \left[6.45\cdot10^{43}\exp\left(-\frac{t_{rise}}{8.8\textrm{\,d}}\right)+1.45\cdot10^{43}\exp\left(-\frac{t_{rise}}{111.3\textrm{\,d}}\right)\right]
\end{multline}
This relation is depicted in the gray dashed lines in Figure \ref{fig:peak_vs_trise} for a few selected nickel masses.

We note that equations \eqref{eq:ejecta_mass} and \eqref{eq:ni_mass} make several simplifying assumptions, such as the opacity being constant and the nickel being concentrated in the center, as well as the photospheric velocity $v_{ph}$ being indicative of the scaling velocity in the model, and the rise to peak being indicative of the effective timescale (i.e. the geometric mean of the diffusion time and the hydrodynamical time) in the model. 

Nevertheless, we use these ejecta and nickel mass estimates to roughly sketch out the phase space in Figure \ref{fig:peak_vs_trise} for which nickel decay can not be the only power source (i.e. the required nickel mass would be larger than the entire ejecta mass). This is the area of the plot to the top and left of the dashed black line (denoted $M_{Ni}>M_{ej}$). Since the estimated masses are approximate, this line should be considered a fuzzy limit rather than a strict one. Our events (as well as PTF09uj and the Drout et al. 2014 sample) require very high ratios of nickel mass to total ejected mass (much higher than seen for nickel powered normal Type Ia SNe). Such ratios are seen in models of pure detonations (Sim et al. 2010) and double detonations (Kromer et al. 2010) of high-mass white dwarfs. We consider these white dwarf detonations in more detail and investigate other possible power sources in Section \ref{sec:powersources}.

\subsection{Comparison to SN\,2011kl}

To the best of our knowledge, SN\,2011kl is the only event which has a similar rise time and peak magnitude as our events (Fig. \ref{fig:peak_vs_trise}). We compare the full light curves of our events to those of SN\,2011kl (after removal of its afterglow and host galaxy components; Greiner et al. 2015). SN\,2011kl is at a similar redshift ($z=0.677$) as our SNLS events, so the same bands can be roughly compared to each other. We plot this comparison in Figure \ref{fig:comp_to_11kl}. All events show similar peak magnitudes and post-peak decline rates, and some also have comparable rise times as SN\,2011kl. Here we allow the explosion time (set as time zero in Fig. \ref{fig:comp_to_11kl}) to shift by $2$ days for SNLS04D4ec and by $3$ days for SNLS06D1hc compared to the values derived earlier in order to improve the match with SN\,2011kl. These shifts remain consistent with the observed upper limits. The resemblance between SNLS06D1hc and SN\,2011kl is the most striking.

The similarities in light curve shapes between our events and SN\,2011kl indicate that they may all belong the same class. Given the association between SN\,2011kl and the ultra-long duration GRB\,111209A (Greiner et al. 2015), we searched for high-energy outbursts consistent with the locations and inferred explosions times for our four events. We considered tabulated catalogs for the all-sky InterPlanetary Network (IPN; Hurley et al. 2010), the 8.8\,sr Gamma-Ray Burst Monitor on-board the \textit{Fermi} satellite (GBM; Meegan et al. 2009), and the 2\,sr Burst Alert Telescope (BAT; Barthelmy et al. 2005) on-board the \textit{Swift} satellite (Gehrels et al. 2004). No potential counterparts for any event were reported by the \textit{Swift}-BAT (for which the precise localizations make chance spatial coincidence highly unlikely). While a search of the IPN database revealed several temporal coincidences, the lack of localization provided for most IPN events makes a firm association impossible.

While we find no affirmative evidence for high-energy emission with any of the SNe presented here, we can not rule out an association. The \textit{Swift}-BAT only observes a modest fraction of the sky at any given moment, so the likelihood of ``missing'' an associated GRB is quite high. For the IPN, which effectively provides all-sky coverage with a 100\% duty cycle, the low sensitivity to very long duration ($\Delta t > 10^{3}$\,s) transients does not guarantee a detection due to the nature of the triggering algorithm (e.g., Levan et al. 2014). We conclude that present data do not allow us to confirm or refute an association between the SNe presented here and GRB-like transients.

\settowidth{\leftimagewidth}{\includegraphics[height=6.7cm]{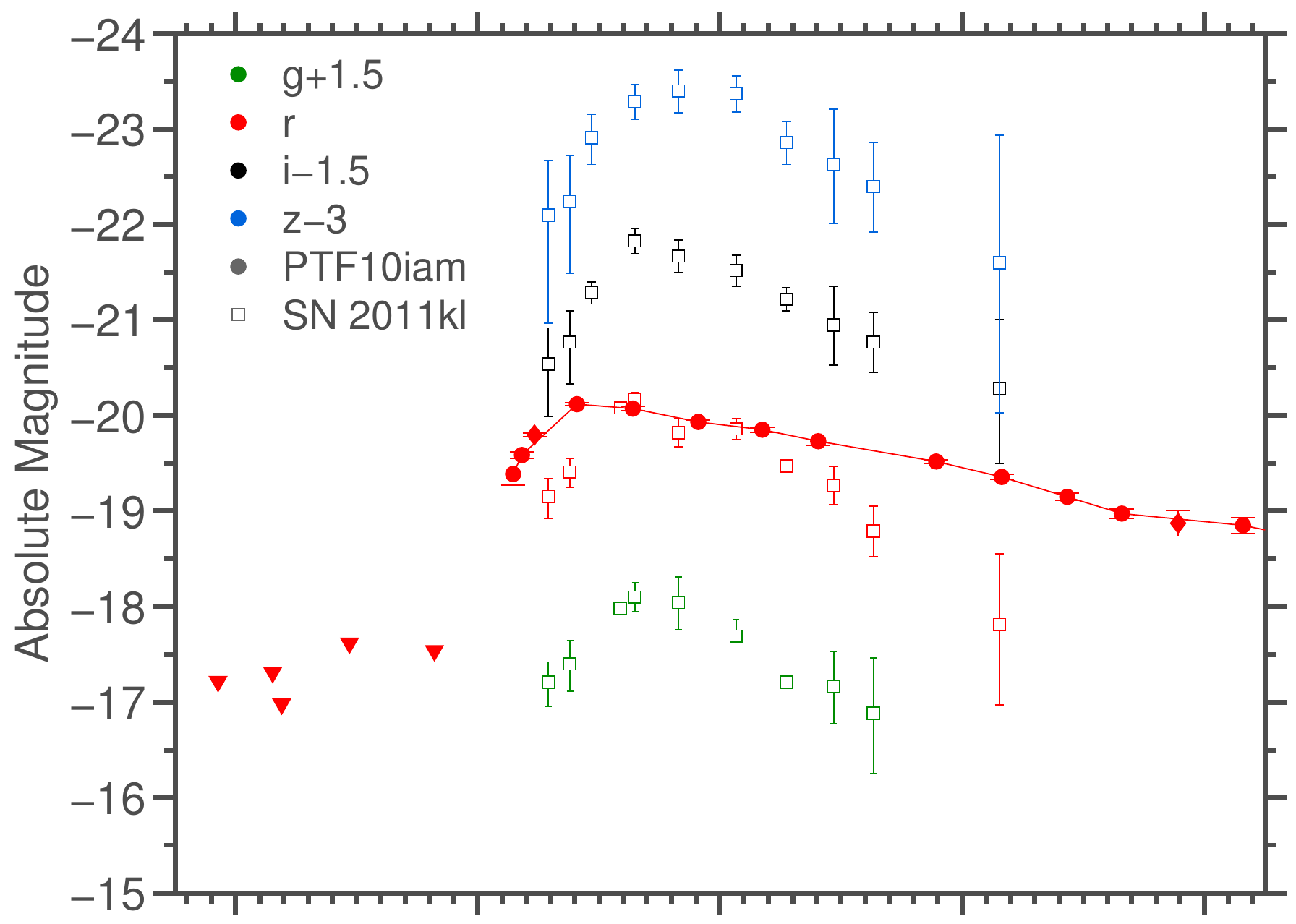}}
\settowidth{\rightimagewidth}{\includegraphics[height=6.7cm]{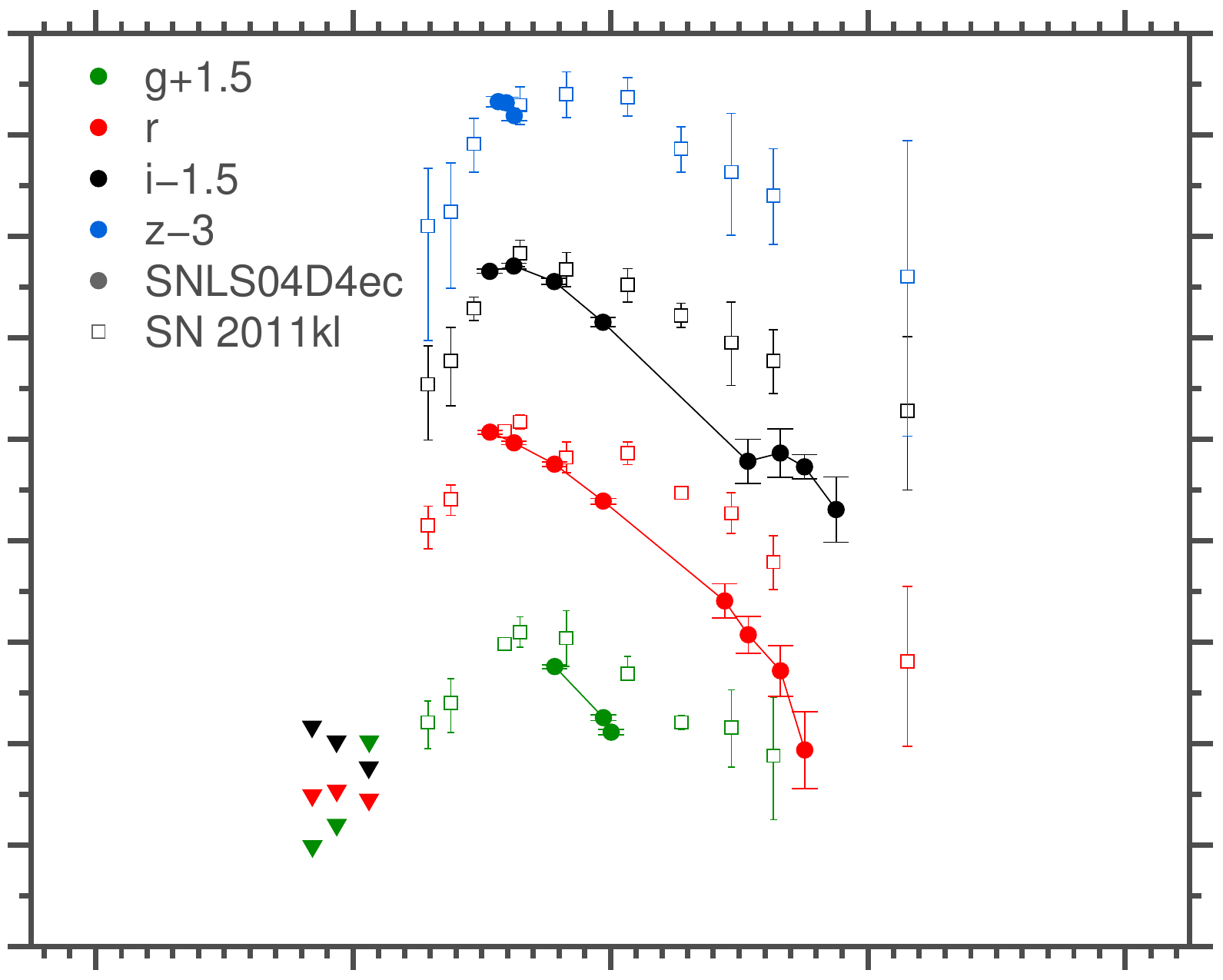}}

\begin{figure*}[t]
\includegraphics[height=6.7cm]{11kl_vs_ptf10iam.eps}\,\,\includegraphics[height=6.7cm]{11kl_vs_04d4ec.eps}
\includegraphics[width=\leftimagewidth]{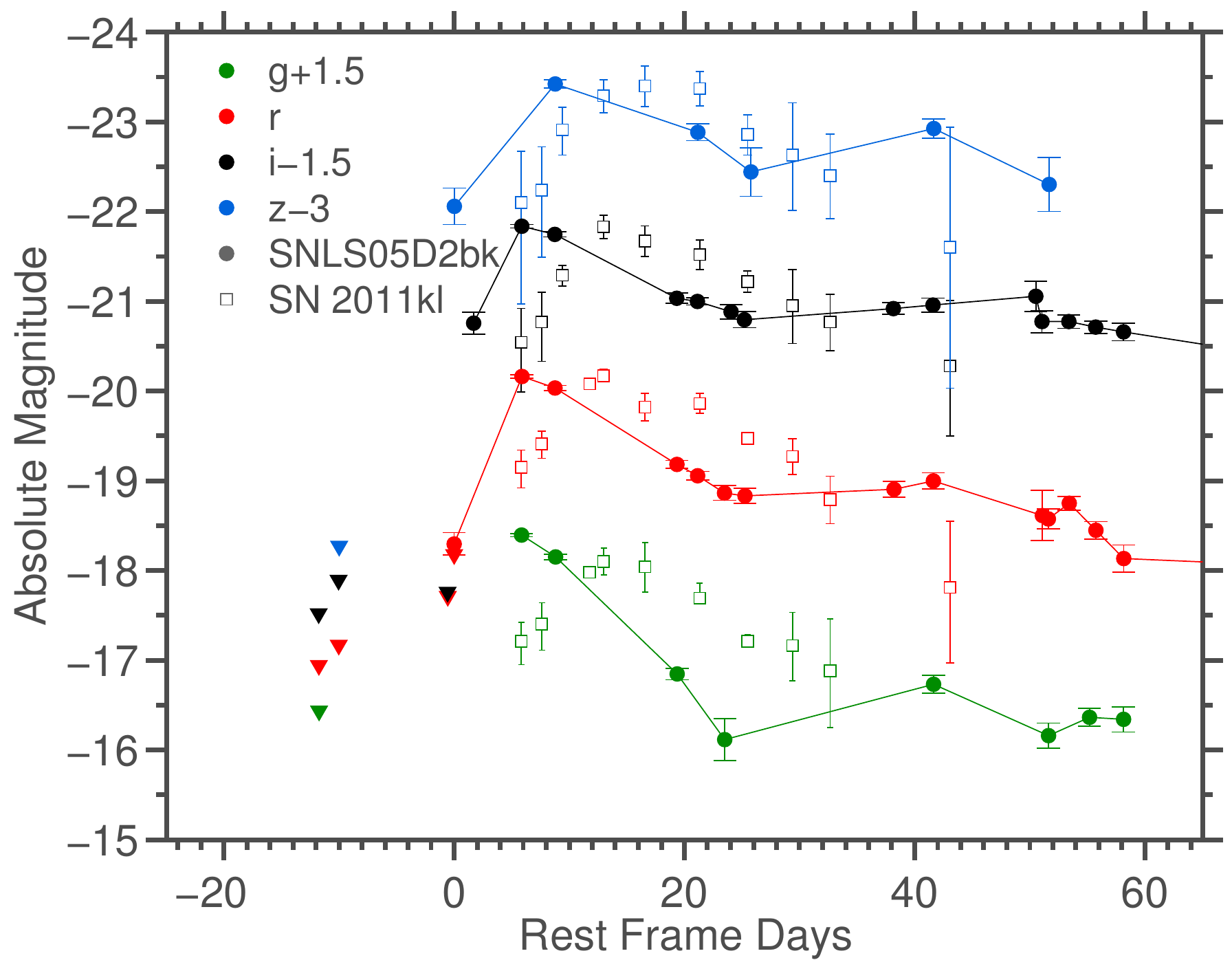}\,\,\includegraphics[width=\rightimagewidth]{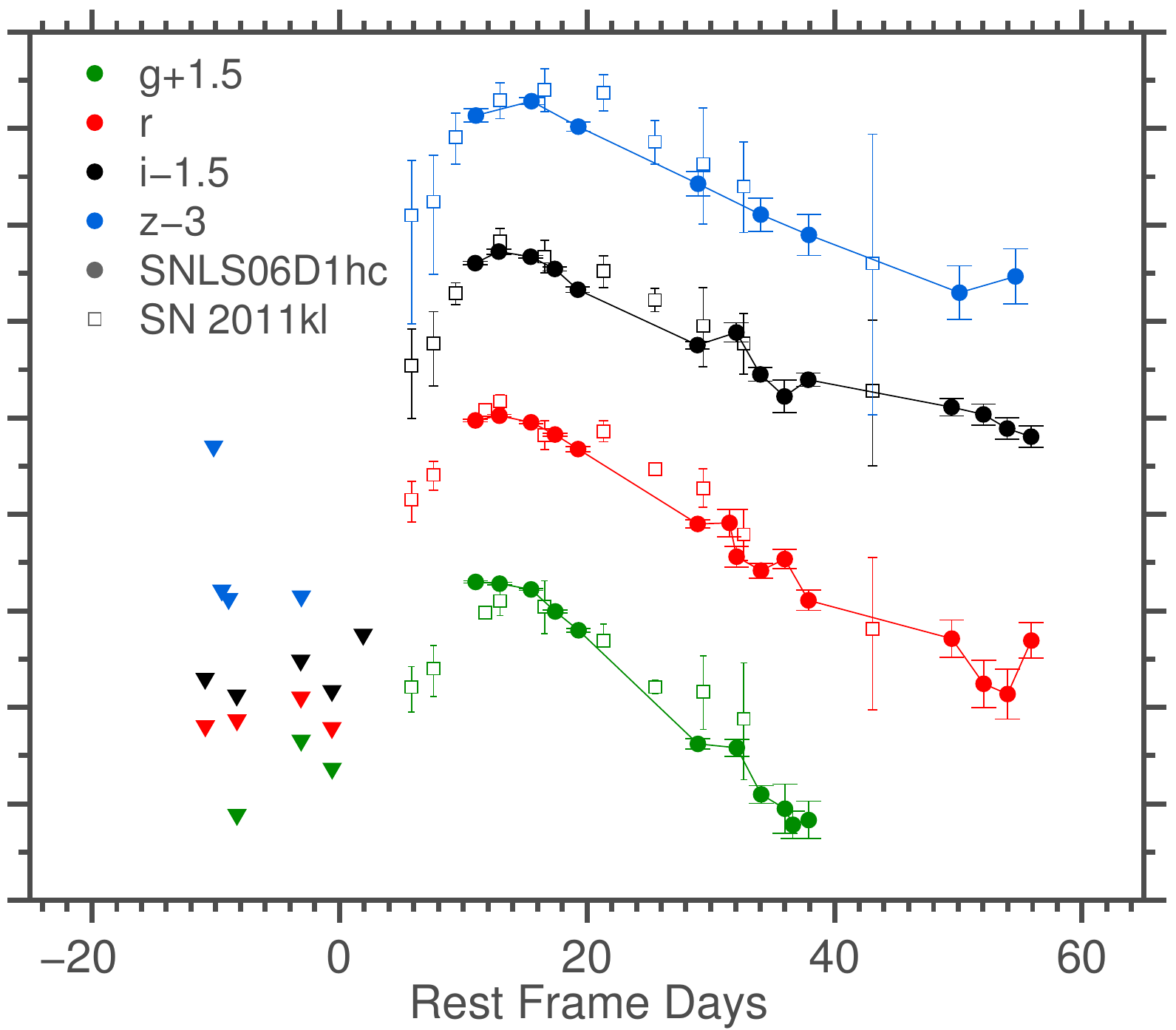}
\caption{\label{fig:comp_to_11kl}Comparison of the light curves of our events (filled symbols and lines) to SN\,2011kl (empty symbols; Greiner et al. 2015), a SN that accompanied an ultra-long-duration GRB. Time zero for SN\,2011kl is set to the time of the GRB trigger. For our events it is set to the estimated time of explosion (with an offset of $2$ days for SNLS04D4ec and $3$ days for SNLS06D1hc, to improve the match). No brightness matching was applied. PTF10iam is at a substantially different redshift than SN\,2011kl (z=$0.109$ vs. z=$0.677$), so the observed wavelength coverages do not match. SNLS06D1hc, on the other hand, is at a very similar redshift as SN\,2011kl, and the two events appear almost identical in their light curve shape (in each filter), indicating that they may be members of the same class of explosions.}
\end{figure*}

\subsection{PTF10iam Spectral Features}

\begin{figure*}[t]
\includegraphics[width=\textwidth]{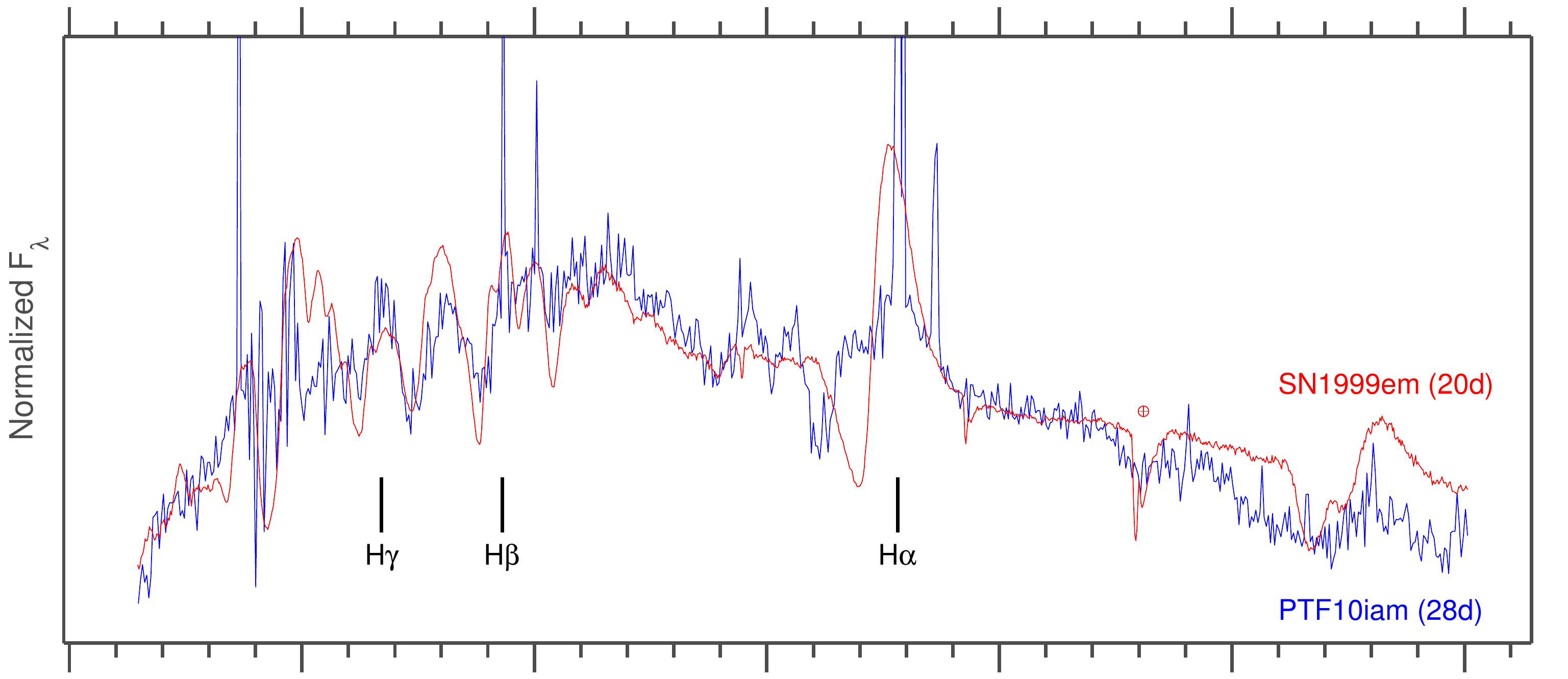}
\includegraphics[width=\textwidth]{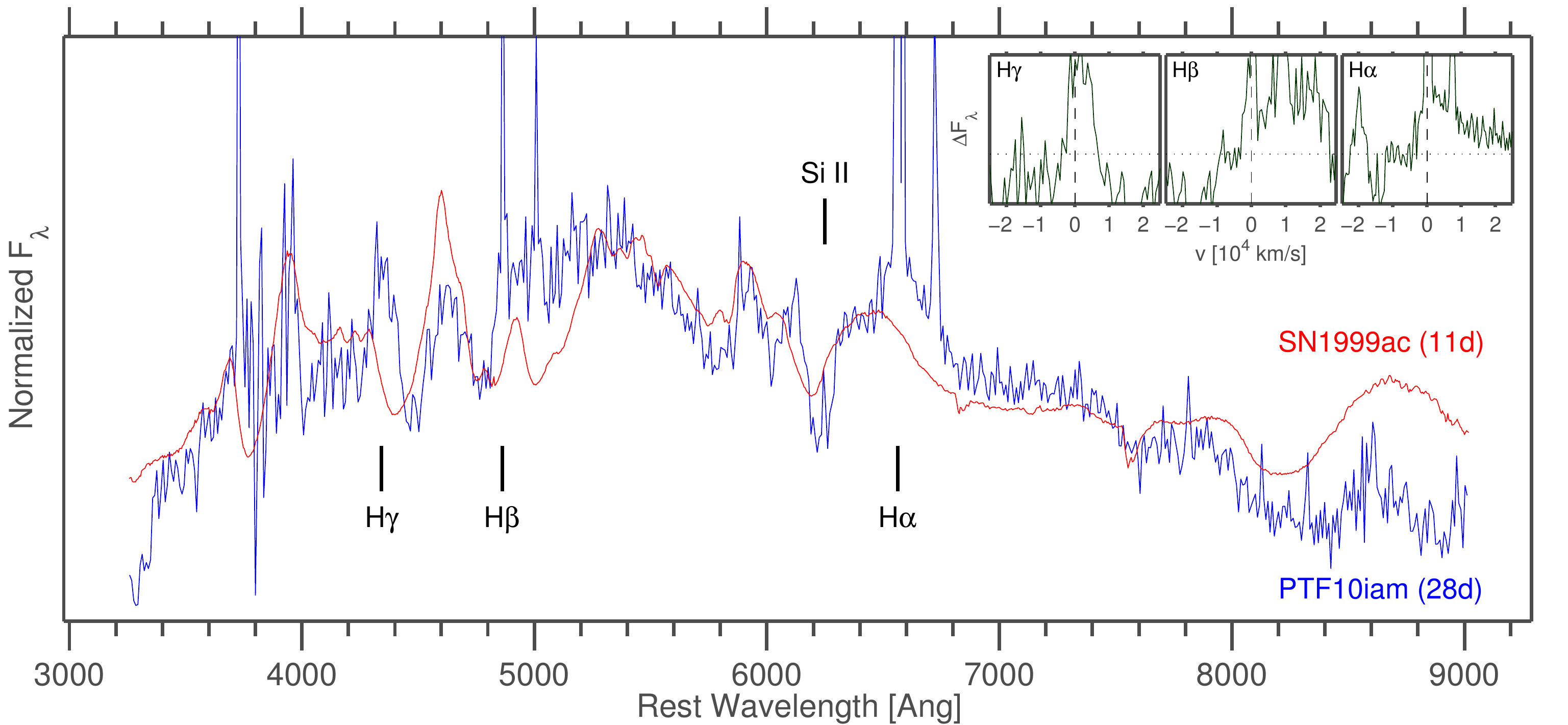}
\caption{\label{fig:10iam_vs_99ac}Superfit comparisons of PTF10iam with the Type~II SN 1999em (from Hamuy et al. 2001; top) and the Type~Ia SN 1999ac (from Garavini et al. 2005; bottom) with best fit extinction and host-contamination corrections applied. The fit to SN 1999em matches most of the spectral features, except for the absorption feature near $6200\,\textrm{\AA}$. The SN~Ia fit is able to match this feature as Si II, as well as other features in the spectrum, with the major difference being that PTF10iam has additional broad hydrogen emission lines (insets show the difference between the superfit host and extinction-corrected spectral fluxes of PTF10iam and SN 1999ac around the denoted hydrogen lines; zero flux is marked by the horizontal dotted line; the narrow features are from the not-fully subtracted host of PTF10iam).}
\end{figure*}

The spectrum of PTF10iam obtained $28$ (rest frame) days after peak is the first to show significant broad features. Most notable is broad H$\alpha$ emission and a broad absorption feature just bluewards of H$\alpha$. We present Superfit (Howell et al. 2005) results, comparing this spectrum to that of the Type~IIP SN 1999em (from Hamuy et al. 2011) and to the peculiar Type~Ia SN 1999ac (from Garavini et al. 2005) in Figure \ref{fig:10iam_vs_99ac}.

\begin{figure}
\includegraphics[width=\columnwidth]{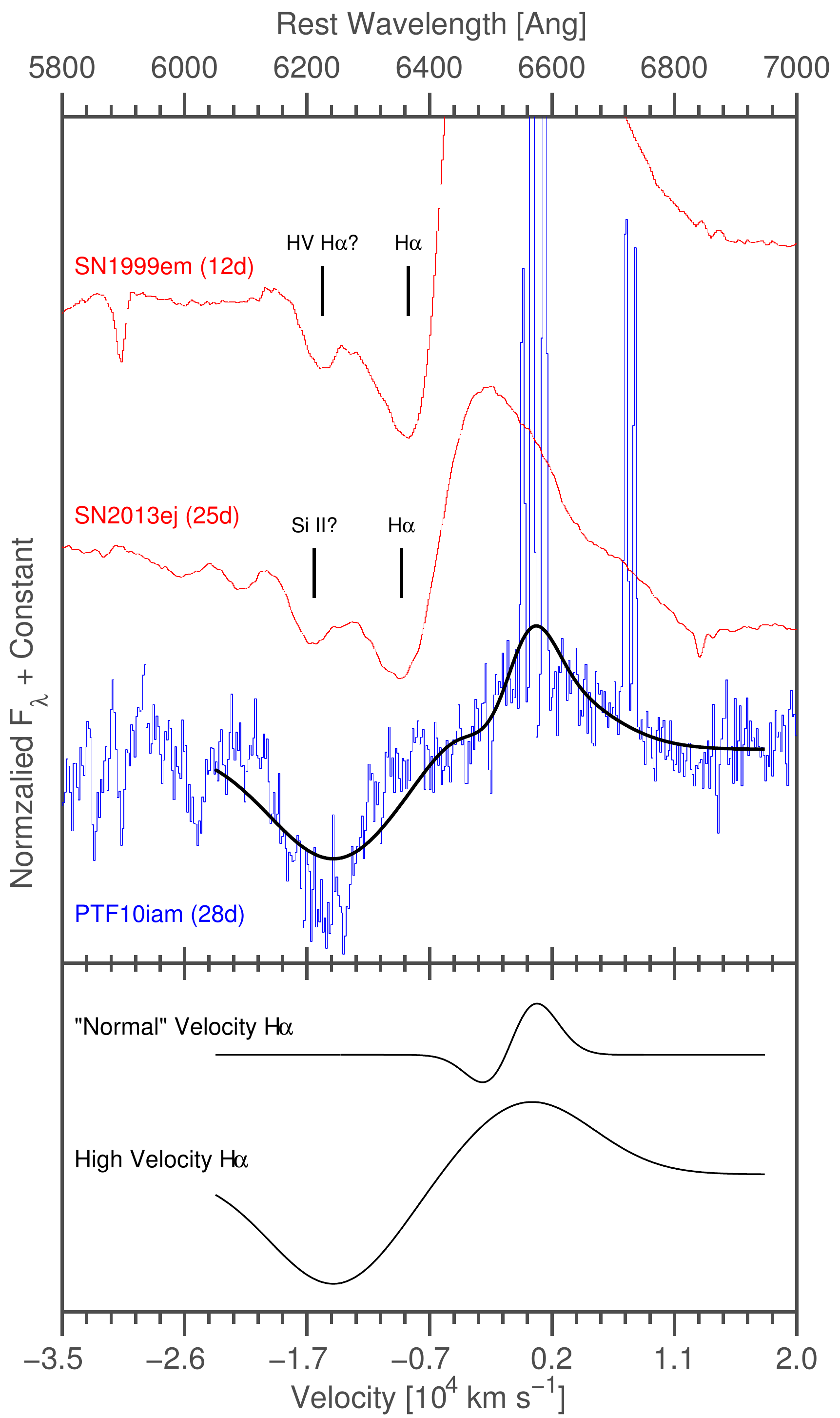}
\caption{\label{fig:10iam_halpha}The H${\alpha}$ region in PTF10iam (blue; spectrum taken $28$ rest frame days after peak) compared to spectra of SN1999em (Leonard et al. 2002) and SN2013ej (Valenti et al. 2013; red) after removing a low order polynomial from each spectrum. The narrow emission lines in the PTF10iam spectrum are from the host galaxy. The absorption feature at $\approx6200\textrm{\AA}$ could be related to high velocity hydrogen, a sign of possible CSM interaction (as interpreted by Chugai et al. 2007 for a later appearance of a similar feature in SN1999em), or to Si II (as interpreted by Valenti et al. 2013 for SN2013ej). In PTF10iam, however, this features is much deeper. We plot the best fit to a sum of four Gaussian functions (black), two representing a ``normal'' H${\alpha}$ P-Cygni profile and two representing a high velocity P-Cygni profile. The profiles provide a reasonable fit to the features, consistent with the high velocity H$\alpha$ interpretation, but the absorption depth would be greater than any previously observed high velocity hydrogen feature.}
\end{figure}

The spectrum of SN 1999em fits most of the features of PTF10iam quite well, confirming the initial SN~II classification of this event. However the H$\alpha$ P-Cygni profile is very different. The absorption feature in PTF10iam is notably more blueshifted, meaning it could be the result of a high-velocity hydrogen component. To test this possibility, we remove a low order polynomial from the spectrum of PTF10iam and plot the area around H$\alpha$ in Figure \ref{fig:10iam_halpha}. We do the same for an earlier spectrum of SN 1999em (Leonard et al. 2002) and for a spectrum of the Type~II SN 2013ej (Valenti et al. 2013). Unlike PTF10iam, these are both SNe~IIP (with a plateau in their light curve), but they also show an absorption feature blueshifted from the ``main'' P-Cygni component of H$\alpha$. Chugai et al. (2007) interpret a similar feature (appearing at $\approx50$ days post explosion) as high velocity H${\alpha}$ for SN 1999em. Valenti et al. (2013) interpret the early-phase appearance of this feature as Si II for SN 2013ej (see also Parrent et al. 2015 for a discussion on similar features in Type I SNe). It may be possible that this feature is related to Si II at early phases and high velocity H$\alpha$ at later phases. In either case, however, the feature is much more pronounced in PTF10iam than in the comparison Type II SNe mentioned above. 

We fit the full profile from the spectrum of PTF10iam with two H${\alpha}$ P-Cygni components, one at ``normal'' velocities and one at high velocities. Each P-Cygni profile is made of two equal-width but shifted and inverted Gaussians (the best fit Gaussian parameters are presented in Table \ref{tab:hvmodel_params}). The high velocity hydrogen interpretation has the advantage that it fits both the blueshifted absorption feature and the possible high velocity emission tail redwards of H${\alpha}$, which would not be explained by Si II. 

Chugai et al. (2007) consider high velocity hydrogen as a sign of interaction of the SN ejecta with the CSM. Such interaction could also power the light curve of PTF10iam, and would explain its extended blackbody radius. 

\renewcommand{\arraystretch}{1.4}
\begin{table}[h]
\caption{\label{tab:hvmodel_params}Parameters of the four best fit Gaussian functions used to reproduce the H${\alpha}$ emission profile and bluer absorption feature of PTF10iam presented in Figure \ref{fig:10iam_halpha}. The mean offset is shown in $10^3$\,km\,s$^{-1}$ relative to rest-frame H${\alpha}$, the $1\sigma$ width is shown in $10^3$\,km\,s$^{-1}$ and the normalization is shown in relative units. Bounds indicate 67\% confidence intervals.}
\begin{tabular}{lcc}
\hline
\hline
{Parameter} & {``Normal'' Velocity} & {High Velocity} \tabularnewline
{} & {Component} & {Component} \tabularnewline
\hline
{Emission:} & {} & {} \tabularnewline
{\hspace{1em} Mean Offset} & {$0$ (fixed)} & {$0$ (fixed)} \tabularnewline
{\hspace{1em} Width} & {$1.9$ ($1.5$,$2.3$)} & {$4.8$ ($4.8$,$4.9$)} \tabularnewline
{\hspace{1em} Normalization} & {$1$} & {$1.11$} \tabularnewline
\hline
{Absorption:} & {} & {} \tabularnewline
{\hspace{1em} Mean Offset} & {$-2.5$ ($-1.5$,$-3.4$)} & {$-14.5$} \tabularnewline
{\hspace{1em} Width} & \multicolumn{2}{c}{Fixed to the same values as in emission} \tabularnewline
{\hspace{1em} Normalization} & {$0.68$} & {$1.66$} \tabularnewline
\hline
\end{tabular}
\end{table}
\renewcommand{\arraystretch}{1}

\begin{figure*}
\includegraphics[trim=0 0 0 0,clip,width=\textwidth]{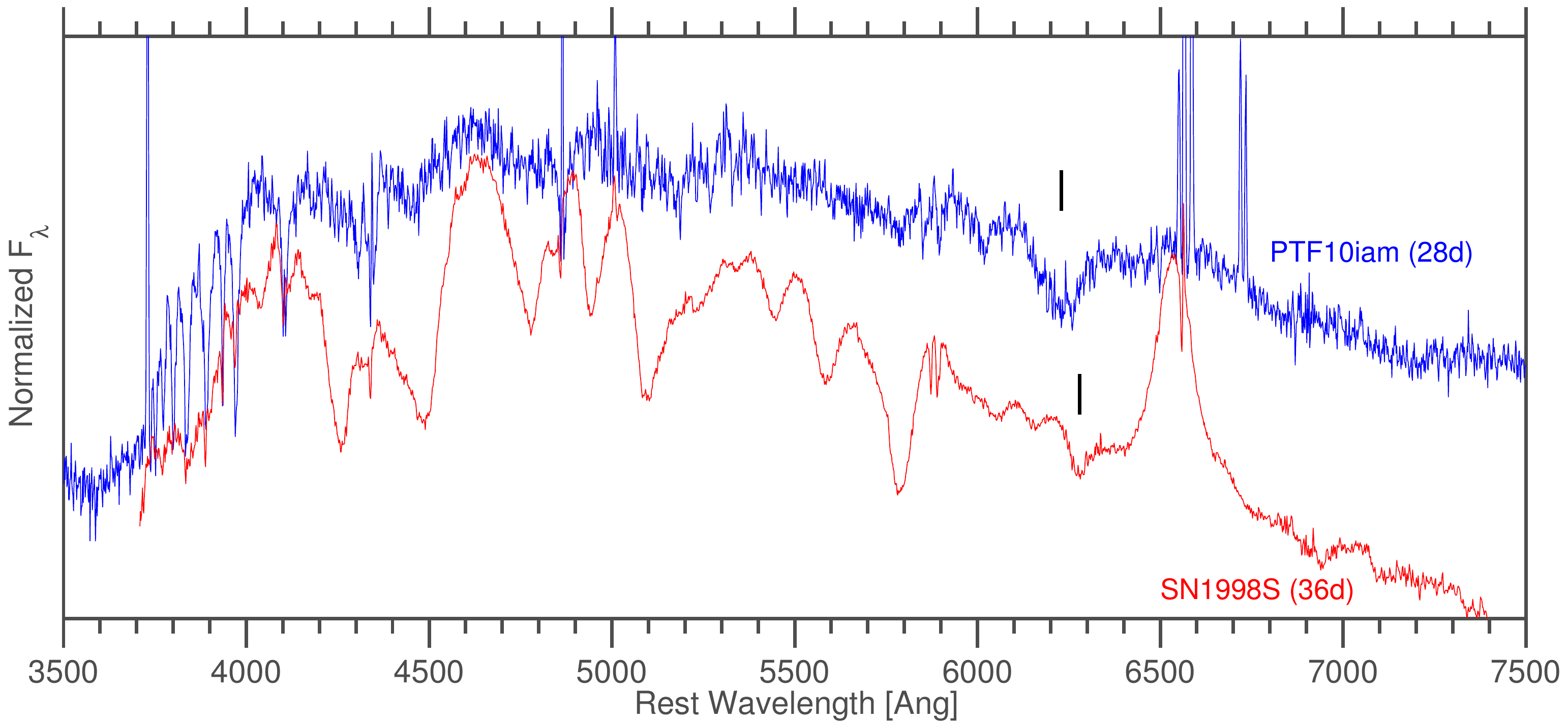}
\caption{\label{fig:10iam_vs_98s}A comparison of a spectrum of PTF10iam and SN 19998S (from Lentz et al. 2001). The broad absorption feature blueward of H${\alpha}$ interpreted as either high velocity H${\alpha}$ or Si II, is marked in both spectra. The feature is stronger and bluer for PTF10iam.}
\end{figure*}

\begin{figure}
\includegraphics[width=\columnwidth]{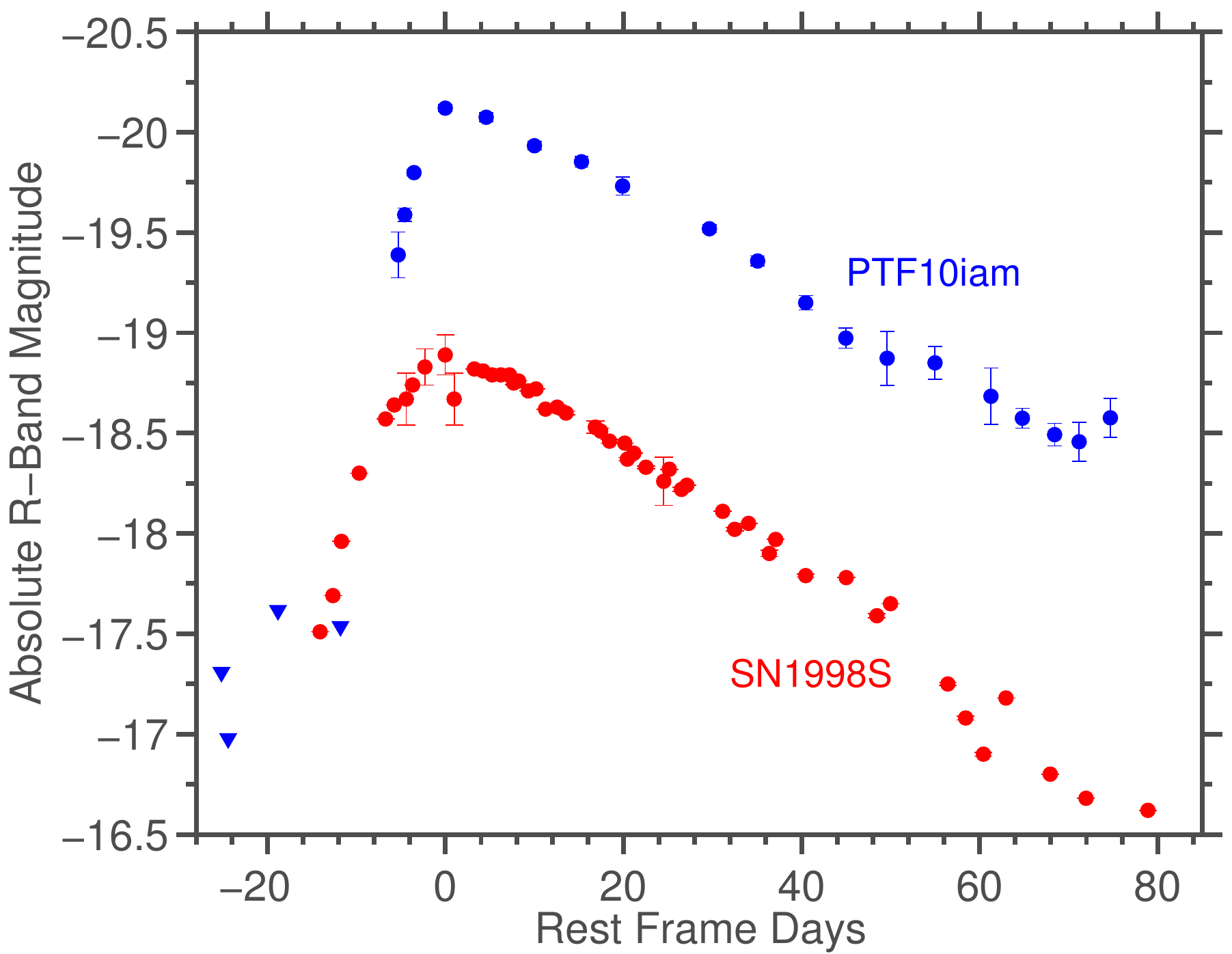}
\caption{\label{fig:10iam_vs_98s_lc}Light curve comparison between PTF10iam and SN 1998S (data from Li et al. 1998, Nakamura et al. 1998, Bignotti et al. 1998, Leonard et al. 2000, Liu et al. 2000 and Fassia et al. 2000). PTF10iam is faster to rise and more luminous, but the decline rates of both events are very similar.}
\end{figure}

A similar spectral feature was seen in a spectrum of SN 1998S (Li et al. 1998), though still weaker and not as blue as that of PTF10iam (Fig. \ref{fig:10iam_vs_98s}). Lentz et al. (2001) model this feature as a blend of Fe II and Si II, and rule out a high velocity H${\alpha}$ origin. We compare the light curves of PTF10iam and SN 1998S and find that SN 1998S was not as fast to rise nor as luminous at peak as PTF10iam, though both events do have similar post-peak decline rates (Fig. \ref{fig:10iam_vs_98s_lc}).

Given that we are not able to find a Type~II SN with a similar absorption feature and light curve behavior, we turn to the Si II interpretation. The similarities with the Type~Ia SN 1999ac (Fig. \ref{fig:10iam_vs_99ac}) is intriguing\footnote{SN 1999ac is a peculiar Type~Ia of the 99aa-like class (Li et al. 2001, Garavini et al. 2004).}. Not only does the Si II in the spectrum of SN 1999ac align well with the broad absorption feature of PTF10iam, but many other features (except for the hydrogen lines) fit quite well. This similarity suggests that the spectrum of PTF10iam may be explained as that of a (peculiar) Type~Ia SN with added broad hydrogen features (see insets in Figure \ref{fig:10iam_vs_99ac}, which present the difference between PTF10iam and SN 1999ac around each of the hydrogen lines). 

\begin{figure}
\includegraphics[width=\columnwidth]{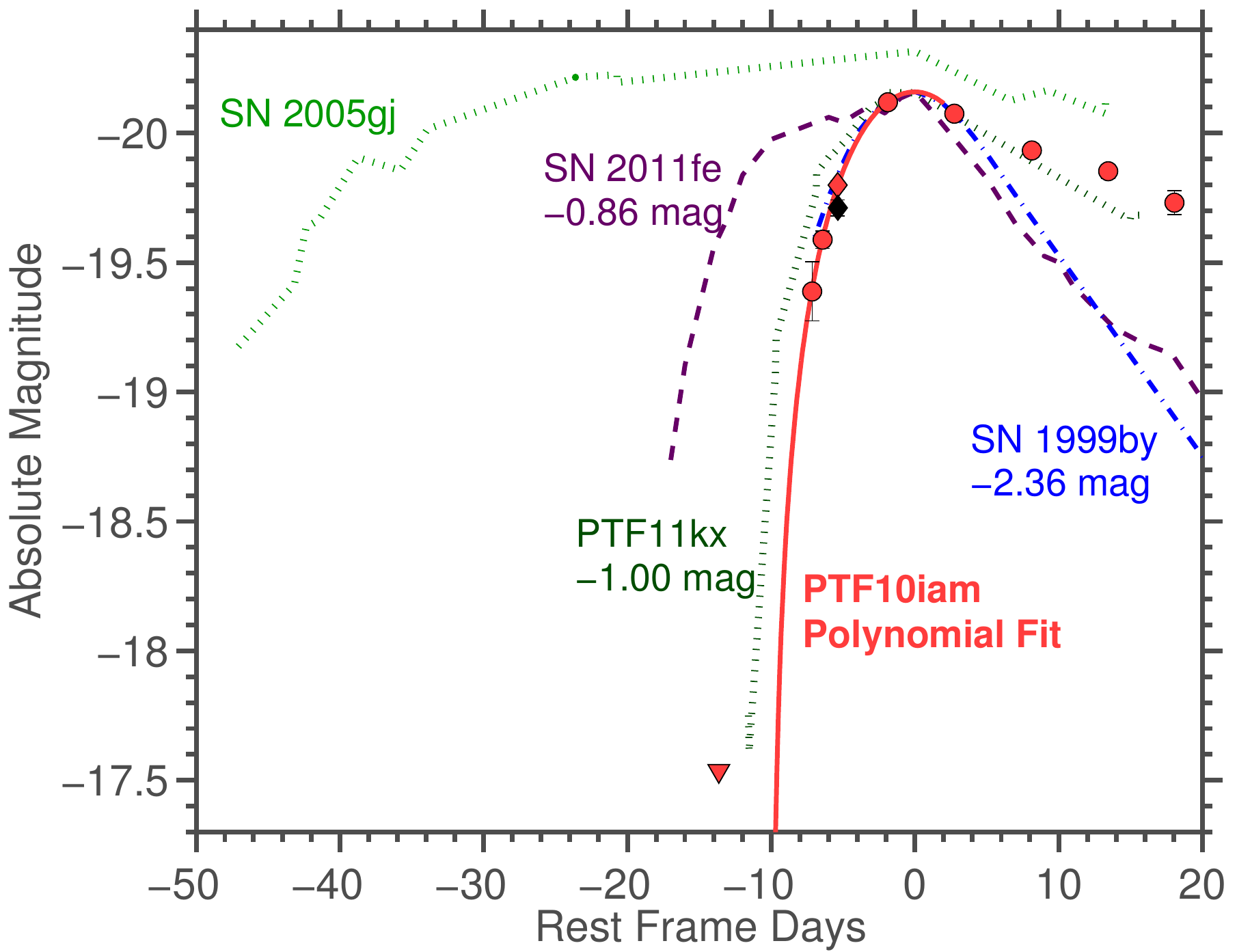}
\caption{\label{fig:10iam_vs_ias} Early light curve of PTF10iam (red circles are detections and the triangle is an upper limit) and the parabolic fit to the pre-peak data used to infer the rise time (solid red line). We compare the rise of PTF10iam to those of the normal Type~Ia SN\,2011fe (dashed purple line; data from Vinko et al. 2012; distance modulus from Lee \& Jang 2012), the 91bg-like SN\,1999by (dot-dashed blue line; data from Garnavich et al. 2004; distance modulus from NED), the Ia-CSM PTF11kx (dotted dark green line; Firth et al. 2015) - all shifted in brightness to match the peak of PTF10iam - and the Ia-CSM SN\,2005gj (dotted light green line; Aldering et al. 2006). PTF10iam has a faster rise compared to SN\,2011fe and SN\,2005gj. The rise of PTF10iam, SN\.1999by and PTF11kx are similar, but their peak magnitudes are very different. If this difference were due to interaction power it should have imprinted strong CSM signatures in the spectrum.}
\end{figure}

Some Type~Ia SNe have been observed to interact with a H-rich CSM (these are known as Type~Ia-CSM events or ``02ic-likes''; Hamuy et al. 2003; Livio \& Riess 2003; Dilday et al. 2012; Silverman et al. 2013; Leloudas et al. 2015a). To test whether PTF10iam could be such an event, we compare its early light curve to those of the normal Type~Ia SN\,2011fe, the Ia-CSM SNe\,2005gj and PTF11kx, and the 91bg-like SN\,1999by (Fig. \ref{fig:10iam_vs_ias}). PTF10iam clearly rises more rapidly than SN\,2011fe, ruling out an additional interaction component on top of a normal SN Ia, and much faster than the interacting SN\,2005gj. The rise of PTF10iam is similar to those of SN\,1999by and PTF11kx, but PTF10iam declines much more slowly than SN\,1999by, is more than $2$ magnitudes brighter at peak than SN\,1999by and is a magnitude brighter than PTF11kx. If PTF10iam were a 91bg-like event with added interaction, or a PTF11kx-like with even stronger interaction, the additional luminosity would require interaction power to dominate the emission. However the spectra of PTF10iam show no signs of interaction. Specifically they do not display the narrow or intermediate-width H$\alpha$ in their spectra that are prominent in Ia-CSM events.

We conclude that the spectrum of PTF10iam is inconsistent with a strong interaction power source, but that it may be interpreted as that of either a peculiar Type~II SN, or a hybrid Type~Ia - Type~II event. These interpretations will be discussed below in the context of the extreme light curve behavior of PTF10iam.

\subsection{Host Galaxies}

\subsubsection{Photometric Analysis}

We fit the host-galaxy $ugriz$ magnitudes with spectral energy distributions (SEDs) computed using PEGASE2 (Fioc \& Rocca-Volmerange 1997, 1999) stellar population synthesis models. We use the eight star-forming scenarios described in Table 1 of Le Borgne \& Rocca-Volmerange (2002) and the default modeling of internal dust presented therein, together with the initial mass function of Rana \& Basu (1992) to compute stellar masses and recent (averaged over the last $5\cdot10^8$\,yrs) specific star formation rate (sSFR). Uncertainties are evaluated through a Monte Carlo propagation of the host-galaxy magnitude uncertainties.

\subsubsection{Spectroscopic Analysis}

We scale the host galaxy spectra (Fig. \ref{fig:spec_hosts}) to the host galaxy photometry (Table \ref{tab:hostmags}) and correct for foreground Galactic extinction (Schlafly \& Finkbeiner 2011). The flux of each emission line was measured by fitting a Gaussian. We fixed the FWHM of the weakest lines to those of the lines that were significantly detected.

The host of PTF10iam is the most nearby one of our sample and has the highest signal to noise ratio. This allows for an estimation of the host extinction. Based on the Balmer decrement (Osterbrock 1989), we find $\textrm{E(B-V)}=0.52\pm0.13$. We adopt this redenning for deriving SFRs, but we caution that it should be considered an upper limit due to the presence of stellar absorption (which affects H$\beta$ more than H$\alpha$). This host-integrated extinction does not necessarily affect the line of sight to PTF10iam and may originate in a dusty region behind the SN (indeed, we rule out significant extinction for PTF10iam in Section \ref{sec:photometry}). After correcting for this extinction, we derive SFRs from the luminosity of the H$\alpha$ line and (separately) from the luminosity of the [O II] line. For both we use the relations in Kennicutt et al. (1998), corrected to a Chabrier IMF by dividing by a factor of 1.7. Both SFR estimates agree within the uncertainties (Table \ref{tab:host_params}), which contain the measurement error, the host reddening uncertainty and the systematic uncertainties of the conversion relations. We compute metallicities based on line flux ratios and the calibrations presented in Pettini \& Pagel (2004) and Kewley \& Ellison (2008). The results are presented in Table \ref{tab:host_params}. For metallicities that are based on the R23 scale (McGaugh 1991, Kobulnicky \& Kewley 2004; denoted M91 and KK04 respectively) the upper branch solution is selected based on criteria in Kewley \& Ellison (2008).

The host galaxies of the SNLS objects are more distant, causing the H$\alpha$ region to be redshifted outside the observed wavelength. Although the H$\beta$ line is detected in the hosts of SNLS04D4ec and SNLS05D2bk, the non-detection of higher order Balmer lines prevents us from deriving accurate estimates of the host extinction. A nominal value of $\textrm{E(B-V)}>1.1$\,mag can be derived from the upper limit of the H$\gamma$ flux. However this derivation is further complicated by signs of stellar absorption that affect the regions of the higher order Balmer lines, and that are difficult to correct for with the signal to noise ratio of these spectra. For SNLS06D1hc we do not detect any Balmer lines. 

Because of this uncertainty, we do not apply any host reddening correction to the hosts of the SNLS events. SFRs are calculated based on the luminosity of only the [O II] line, and we compute metallicities only in the R23 scale. The upper branch solution was selected for all the galaxies based on the low [O III]/[O II] ratio (e.g. Nagao et al. 2006). The uncertainty in host reddening does not affect this choice as this ratio would become even lower if a non-negligible extinction is assumed. Due to the non-detection of any Balmer lines in the spectrum of SNLS06D1hc we can not provide any metallicity measurements for its host. Our results are present in Table \ref{tab:host_params} (the Dougie host galaxy does not display any emission lines, and is therefore excluded from this analysis).

The metallicities of all hosts are close to solar or super-solar and the galaxies show clear signs of an evolved stellar population, such as stellar absorption. These galaxies are markedly different from the hosts of H-poor SLSNe that have been shown to be preferentially star bursting dwarf galaxies (e.g. Neill et al. 2011, Lunnan et al. 2014, Leloudas et al. 2015b).

\renewcommand{\arraystretch}{1.4}
\begin{table*}
\caption{\label{tab:host_params}Properties of the host galaxies of our events and of Dougie obtained using fits to host $ugriz$ photometry from Table \ref{tab:hostmags} and analysis of the host emission lines (when available) from the spectra presented in Figure \ref{fig:spec_hosts}. Errors denote $1\sigma$ uncertainties. For Dougie we find zero SFR from the photometric analysis (also when varying the input magnitudes in the Monte Carlo simulation; formally this gives a limit of $\log\left(SFR\right)<-3$).}
\begin{tabular*}{\textwidth}{lccp{0mm}cccccc}
\hline
\hline
{Object} & \multicolumn{2}{c}{Photometric Analysis} & {} & \multicolumn{6}{c}{Spectroscopic Analysis}\tabularnewline
\cline{2-3}\cline{5-10}
{} & {$\log\left(M\right)$} & {$\log(sSFR)$} & {} & {H$\alpha$ SFR} & {OII SFR} & \multicolumn{4}{c}{$12+{\log}$(O/H)} \tabularnewline
{} & {[$\log\left(M_{\odot}\right)$]} & {[$\log\left(\textrm{yr}^{-1}\right)$]} & {} & {[$M_{\odot}\textrm{yr}^{-1}$]} & {[$M_{\odot}\textrm{yr}^{-1}$]} & {(M91)} & {(KK04)} & {(N2)} & {(O3N2)} \tabularnewline
\hline
{PTF10iam Host} & {$10.40\pm0.08$} & {$-9.70\pm0.07$} & {} & {$4.86\pm1.58$} & {$11.77\pm7.65$} & {$8.67\pm0.06$} & {$8.87$} & {$8.57\pm0.05$} & {$8.62\pm0.04$} \tabularnewline
{SNLS04D4ec Host} & {$9.77\pm0.07$} & {$-9.36\pm0.09$} & {} & {n/a} & {$2.87\pm0.82$} & {$8.59\pm0.08$} & {$8.75$} & {n/a} & {n/a} \tabularnewline
{SNLS05D2bk Host} & {$10.27\pm0.08$} & {$-9.46\pm0.08$} & {} & {n/a} & {$5.06\pm1.43$} & {$8.73\pm0.03$} & {$8.93$} & {n/a} & {n/a} \tabularnewline
{SNLS06D1hc Host} & {$9.39\pm0.09$} & {$-9.44\pm0.09$} & {} & {n/a} & {$0.21\pm0.07$} & {n/a} & {n/a} & {n/a} & {n/a} \tabularnewline
\hline
{Dougie} & {$10.35\pm0.04$} & {no SF} & {} & {n/a} & {n/a} & {n/a} & {n/a} & {n/a} & {n/a} \tabularnewline
\hline
\end{tabular*}
\end{table*}
\renewcommand{\arraystretch}{1}

\section{Possible Light Curve Power Sources} \label{sec:powersources}

The rapid rise and luminous peak of our events challenge traditional SN power sources. Nickel-decay power would require very high nickel to total mass ratios which are seen in models of pure and double detonations of carbon-oxygen white dwarfs (but not observed in normal Type~Ia SNe). We compare our data to models of such detonations and to general energy conservation considerations for high nickel mass explosions. We then turn to massive stars and consider three other possible power sources: interaction with the CSM, shock breakout in an optically thick wind and magnetar spindown. 

\subsection{White Dwarf Detonation} \label{sec:powersources_wds}

Sim et al. (2010) investigated pure detonations of sub-Chandrasekhar carbon-oxygen white dwarfs by artificially igniting them in the center. For their most massive white dwarf ($M_{WD}=1.15M_{\odot}$) they find high nickel to total mass ratios and consequently rapidly rising luminous light curves. The same behavior is seen by Kromer et al. (2010) who investigate double detonations of white dwarfs (detonating the base of the helium shell, causing a second detonation inside the carbon-oxygen core).

\begin{figure}
\includegraphics[trim=0 0 0 0,clip,width=1.0\columnwidth]{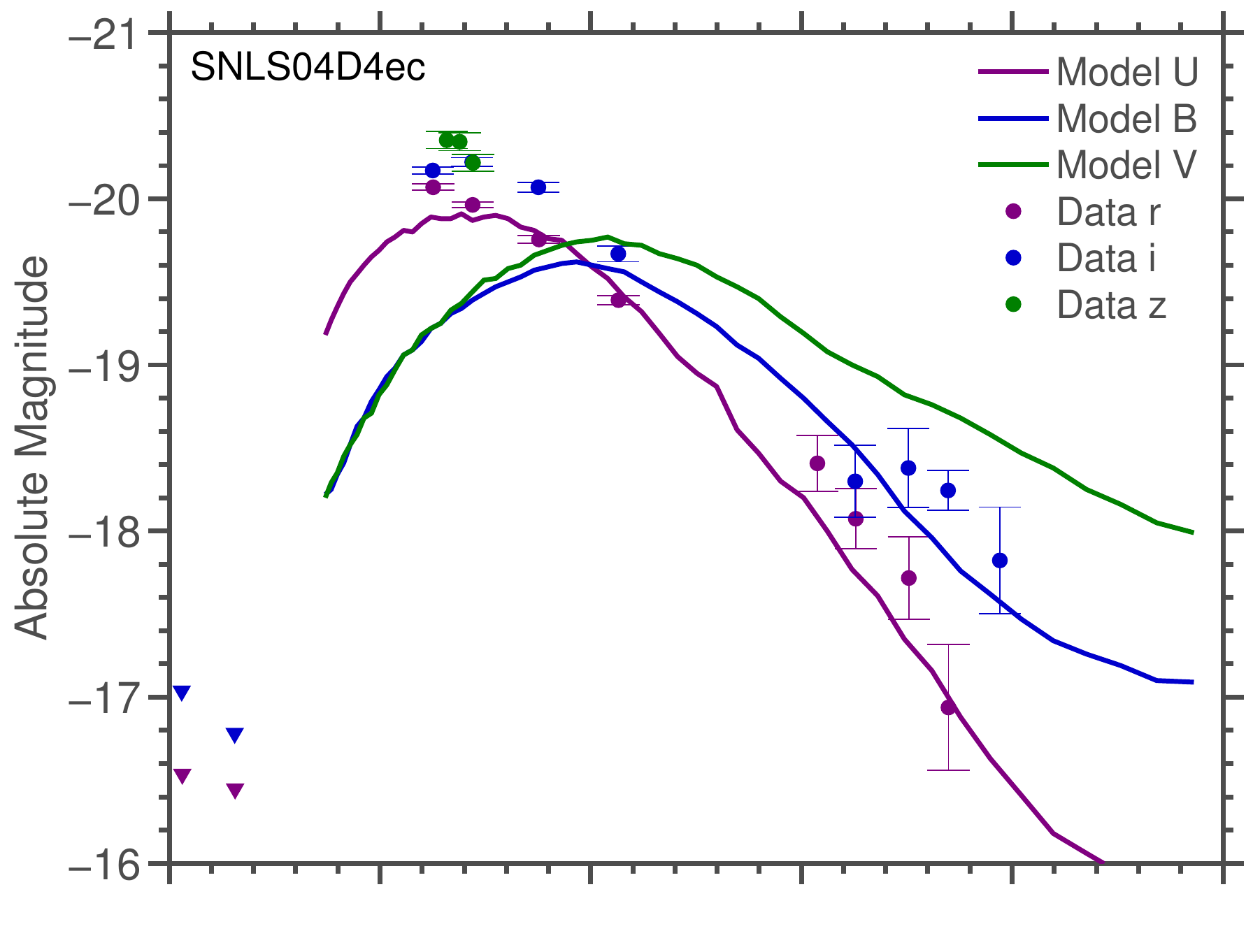}
\includegraphics[trim=0 0 2 0,clip,width=1.0\columnwidth]{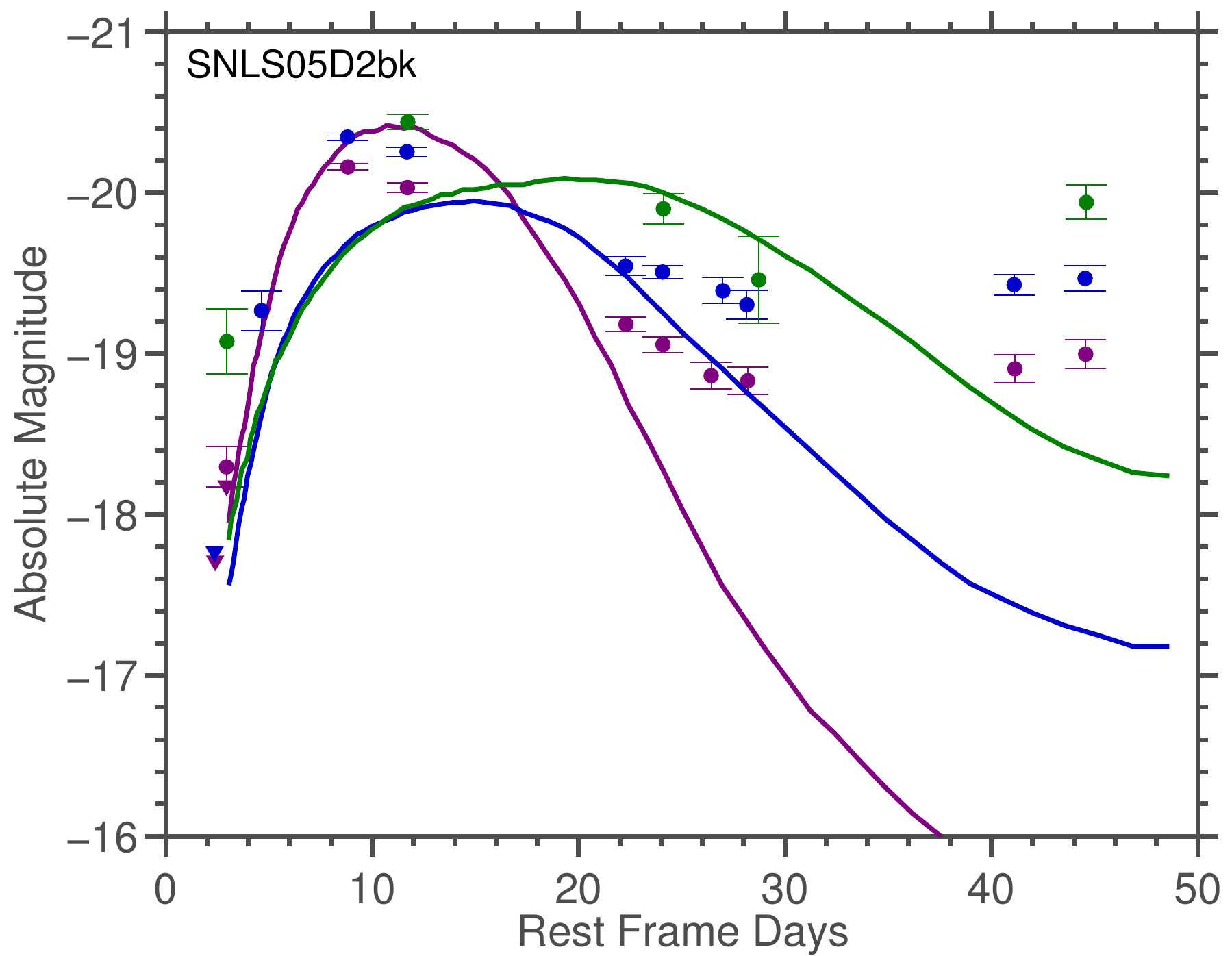}
\caption{\label{fig:wd_det}Two of the closest fits between white dwarf detonation models and our light curves. Top: SNLS04D4ec compared to the Sim et al. (2010) $1.15M_{\odot}$ model. Bottom: SNLS05D2bk compared to the Kromer et al. (2010) model number $6$. Both SNe do not trace the model post-peak declines nor their color evolution. The other models from the Sim et al. (2010) and Kromer et al. (2010) sets were even further from the data, and none were able to reasonably fit SNLS06D1hc at any phase.}
\end{figure}

Since hydrogen is not expected to show up in the spectra of exploding white dwarfs, we focus on the SNLS events for now. We compare our observed light curves to those modeled by Sim et al. (2010) and Kromer et al. (2010), and find that none match the models well in all filters simultaneously. Given the redshift of our events ($z\approx0.6$), we compare observed $r$-band with model $U$-band, observed $i$-band with model $B$-band and observed $z$-band with model $V$-band. We present two of the closest matches between the data and the models in Figure \ref{fig:wd_det}. As can be seen, even for these cases, the match is unsatisfactory.

An important caveat to these comparisons is that Sim et al. (2010) and Kromer et al. (2010) do not consider iron group elements in their models. Such elements could introduce additional blue-band opacities, and for the Kromer et al. (2010) models could also influence the nucleosynthesis yields in the helium shell, further affecting the light curves.

The poor match of the post-peak light curve between the models and our data disfavor this interpretation for the SNLS events. The hydrogen seen in the spectrum of PTF10iam disfavors this scenario also for that event.

However, there have been suggestions of an explosion channel that would involve white dwarfs detonating inside hydrogen-rich envelopes - so-called ``Type 1.5'' SNe (Arnett 1969; Iben \& Renzini 1983; Lau et al. 2008). Such SNe are expected to occur when carbon is explosively ignited in the core of an intermediate-mass star during its AGB phase. These explosions could be similar to Type~Ia SNe with the additiona of hydrogen-rich ejecta. It is therefore reasonable to expect that they would show signs of hydrogen in their spectra (coming from the envelope), possibly in addition to deep Si II absorption (as seen in SNe Ia), and would synthesize large amounts of nickel, generating a luminous light curve peak.

The progenitors of Type 1.5 SNe are expected to be metal poor, while the derived metallicity for the host galaxies for our events is close to solar (Table \ref{tab:host_params}; though these are global values and not specific to the SN sites). Sparks \& Stecher (1974) suggested that a white dwarf spiraling in to the core of a non-degenerate companion and merging with it (the so-called ``core-degenerate'' scenario) could give rise to a similar explosion, without obvious metallicity constraints. In that case, however, some or all of the envelope of the companion might be ejected during the in-spiral. This scenario has thus been used as a possible progenitor channel for Type Ia-CSM events or for events with fast-moving carbon spectral components (Soker et al. 2014).

Since it's not clear exactly what to expect for true Type 1.5 SNe (assuming they even exist in nature), we now relax many of the assumptions used to derive nickel-powered light curve properties and test only global energy conservation. Katz et al. (2013) present a method for calculating the nickel mass required to power a given bolometric light curve, which relies only on the assumptions of homologous expansion, radiation-dominated internal energy and nickel decay being the sole power source. Their argument is that at late enough times (when the internal energy becomes negligible), the total radiated luminosity equals the total energy deposited by nickel plus the energy lost to expansion. This translates to:
\begin{equation} \label{eq:total_l}
\int_{t_{exp}}^{t_{late}}Q\left(t\right){\cdot}t\,dt=\int_{t_{exp}}^{t_{late}}L\left(t\right){\cdot}t\,dt
\end{equation}
where $Q\left(t\right)$ is the energy deposition from nickel decay, $L\left(t\right)$ is the total radiated luminosity, and $t_{late}$ is a late enough time when the internal energy is negligible (Katz et al. 2013 indicate $t_{late}\gtrsim40$ days). 

We assume full trapping of the positrons, and take the $\gamma$-ray optical depth to be:
\begin{equation}
\tau_{\gamma}\left(t\right)=\left(\frac{T_0}{t}\right)^{-2}
\end{equation}
where $T_0$ is left as a free parameter\footnote{Under certain additional assumptions, $T_0$ can be connected to the explosion energy and to the mass and density profile of the ejecta (Clocchiatti \& Wheeler 1997)}.

We use the $t>25$\,days bolometric data points of SNLS05D2bk and SNLS06D1hc from section \ref{sec:bb} to calculate the right hand side of Equation \eqref{eq:total_l}, performing a linear interpolation at $0.1$\,day intervals for the numerical integration, and fit for the nickel mass $M_{Ni}$ and $\gamma$ escape timescale $T_0$ used to calculate the left hand side (while keeping the explosion time $t_{exp}$ constant at the values listed in Table \ref{tab:events_params}). We find that for SNLS05D2bk, $T_0>200$\,days (essentially full $\gamma$-trapping) and $M_{Ni}=1.88\pm0.17M_{\odot}$. For SNLS06D1hc we find $T_0=48_{-15}^{+162}$\,days (i.e. almost full trapping) and a nickel mass very close to the carbon ignition mass $M_{Ni}=1.38_{-0.11}^{+0.07}M_{\odot}$ ($1\sigma$ confidence bounds). The errors on $M_{Ni}$ and $T_0$ were estimated using $500$ Monte Carlo simulated fits. The bolometric light curve, total radiated energy and best fits are plotted in Figure \ref{fig:ni_fits}.

\settowidth{\leftimagewidth}{\includegraphics[height=6.7cm]{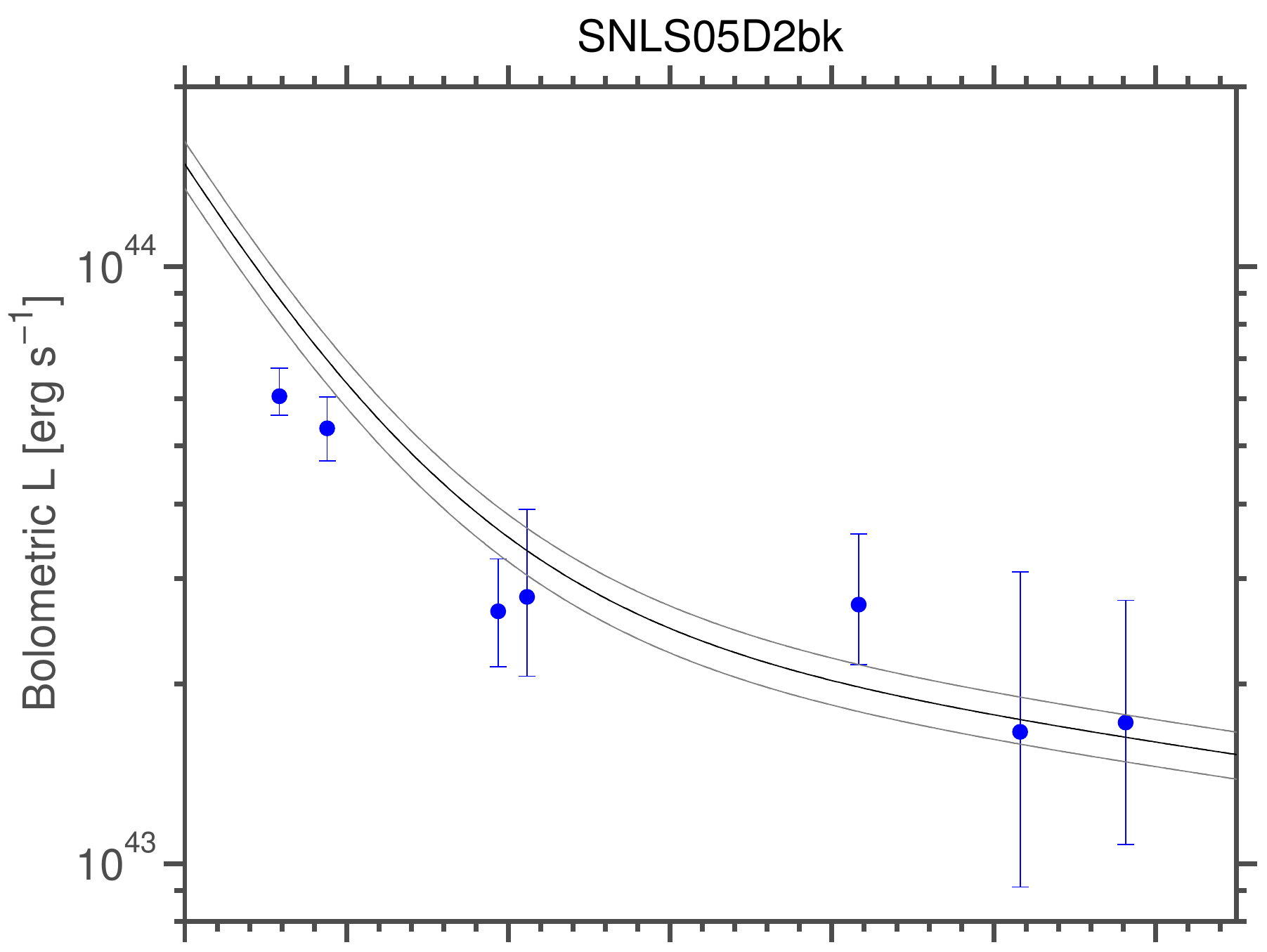}}
\settowidth{\rightimagewidth}{\includegraphics[height=6.7cm]{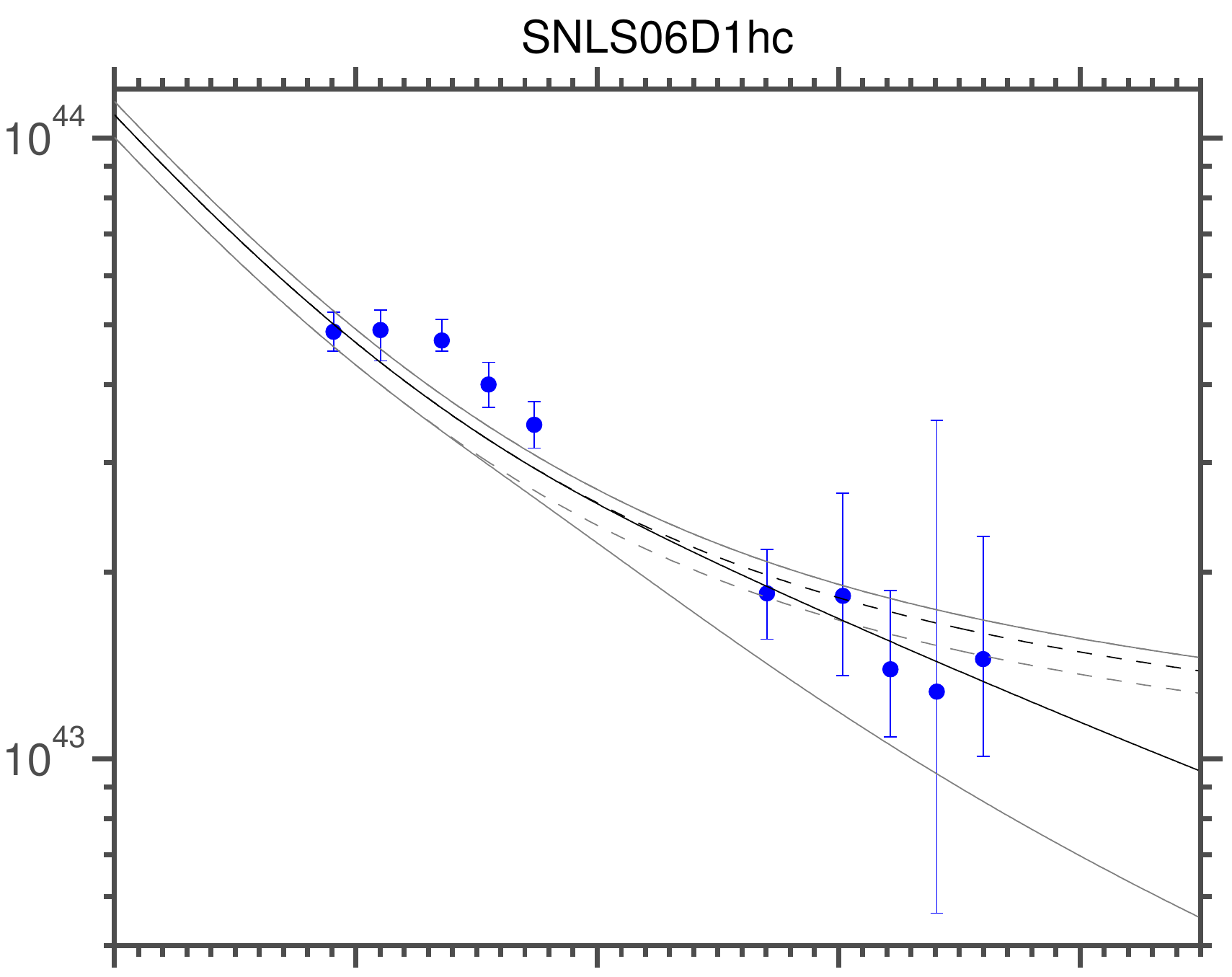}}

\begin{figure*}[t]
\includegraphics[height=6.7cm]{ni_05d2bk_l.eps}\,\,\,\,\,\,\,\,\includegraphics[height=6.7cm]{ni_06d1hc_l.eps}
\includegraphics[width=\leftimagewidth]{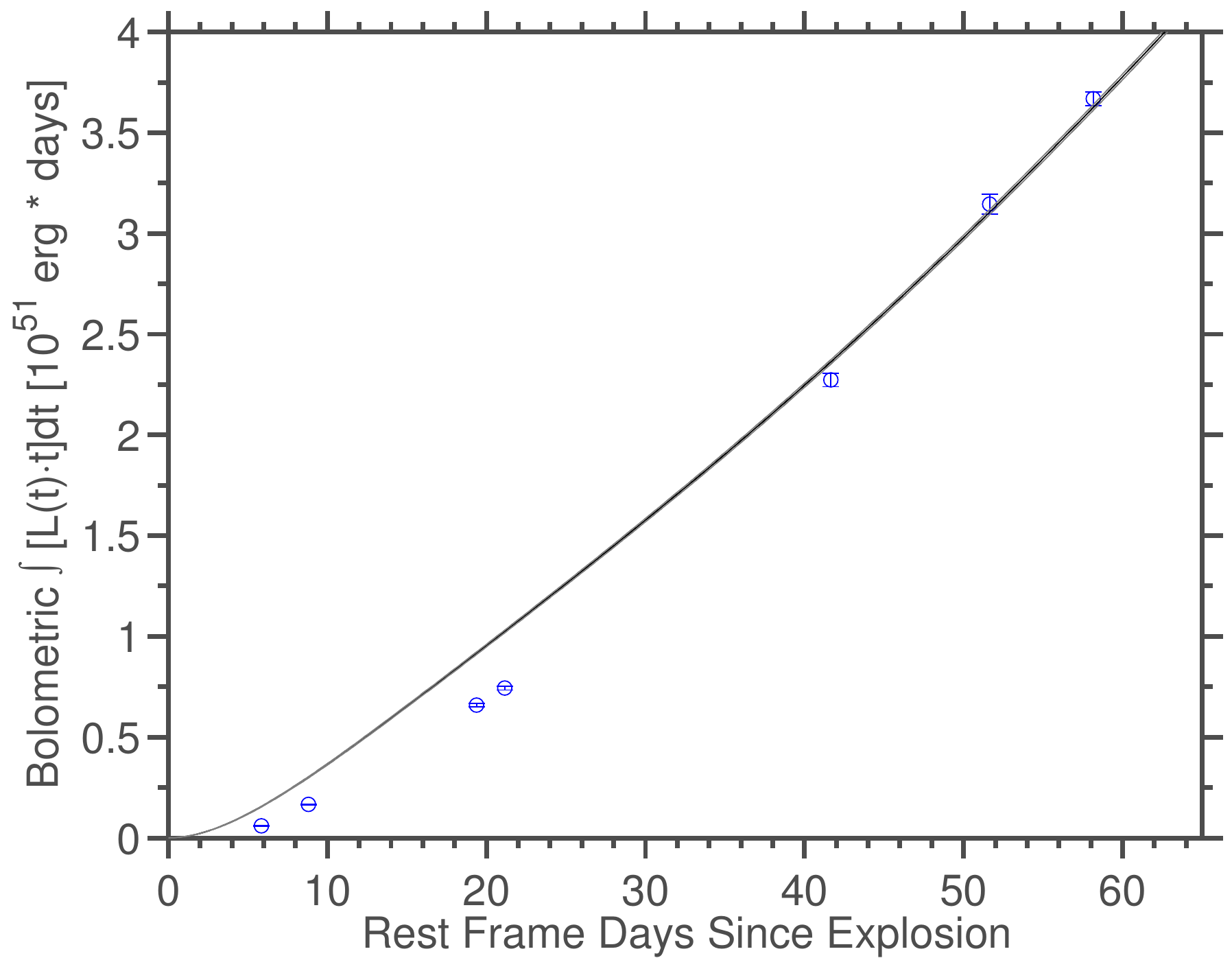}\,\,\,\,\,\,\,\,\includegraphics[width=\rightimagewidth]{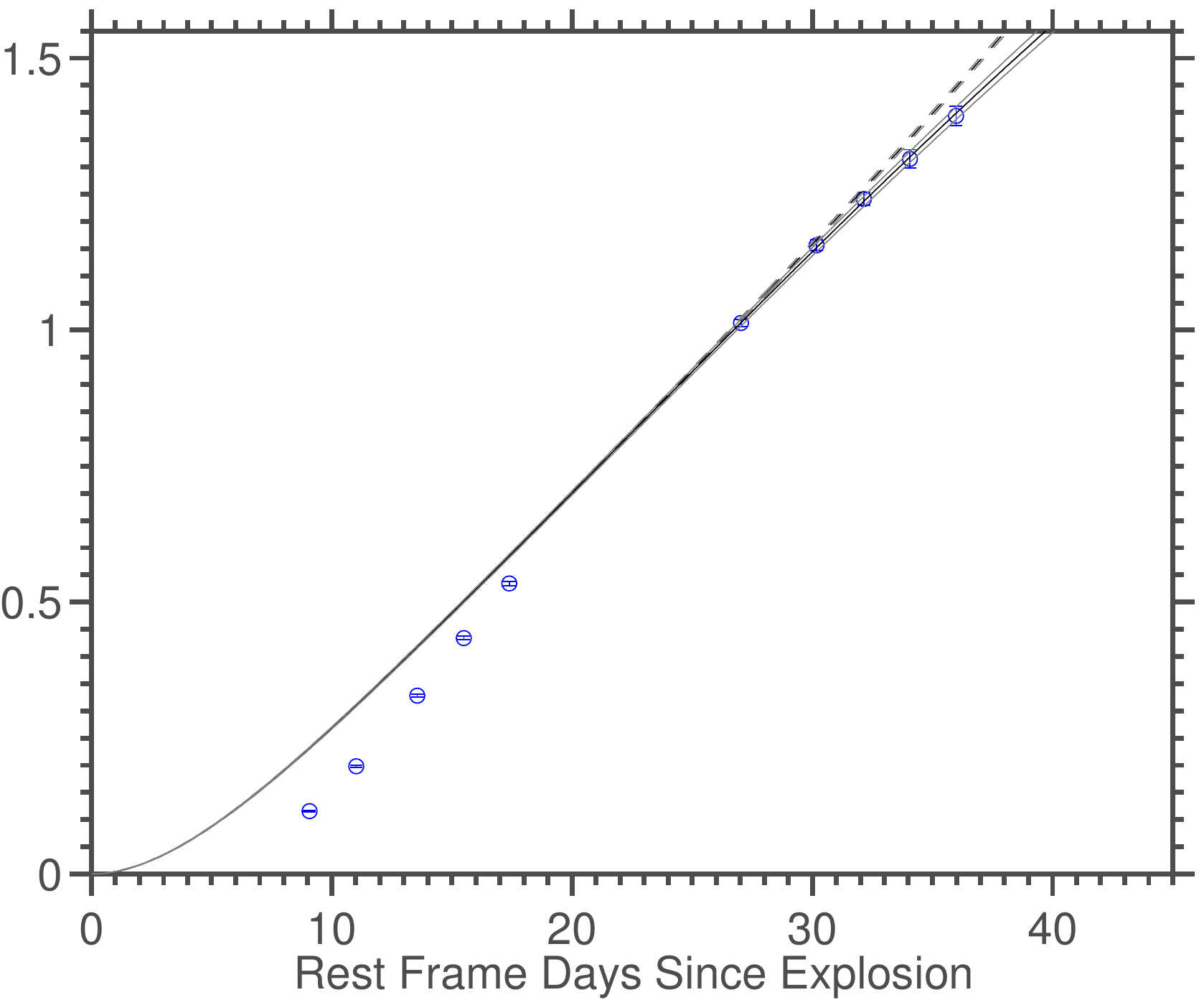}
\caption{\label{fig:ni_fits}Fits to the $t>25$\,days time-integrated bolometric luminosity (bottom) of SNLS05D2bk and SNLS06D1hc with a nickel-powered light curve (dashed lines are for full $\gamma$-trapping, visibly different from the solid lines only for SNLS06D1hc). The integrated luminosity is consistent with a nickel-decay power source for the light curves, though there are only few data points and they do not extend to late enough times to make this determination secure. Top plots show the instantaneous luminosity.}
\end{figure*}

The integrated bolometric luminosities of SNLS05D2bk and SNLS06D1hc at late times are consistent with a high-mass nickel decay power source. However, the number of late-time data points for the fit is small and no constraints are provided for the ejecta mass without additional assumptions. While this method does not provide a strong argument for high mass nickel decay indeed being the main power source of the light curves, it is an indication that this possibility is not completely ruled out by the data. More detailed models could perhaps test whether the addition of a hydrogen envelope, as expected for Type 1.5 SNe, can account for the differences in light curves between the detonation models and our observations.

We conclude that a Type 1.5 SN origin for our events is an intriguing possibility, but not a conclusive interpretation given the lack of detailed model predictions for this scenario and the lack of strong constraints on the long-term bolometric light curves of our events. We now turn to explosion scenarios involving the core collapse of massive stars.

\subsection{CSM Interaction} \label{sec:powersources_csm}

\begin{figure}
\includegraphics[trim=0 0 0 0 clip=true,width=1.0\columnwidth]{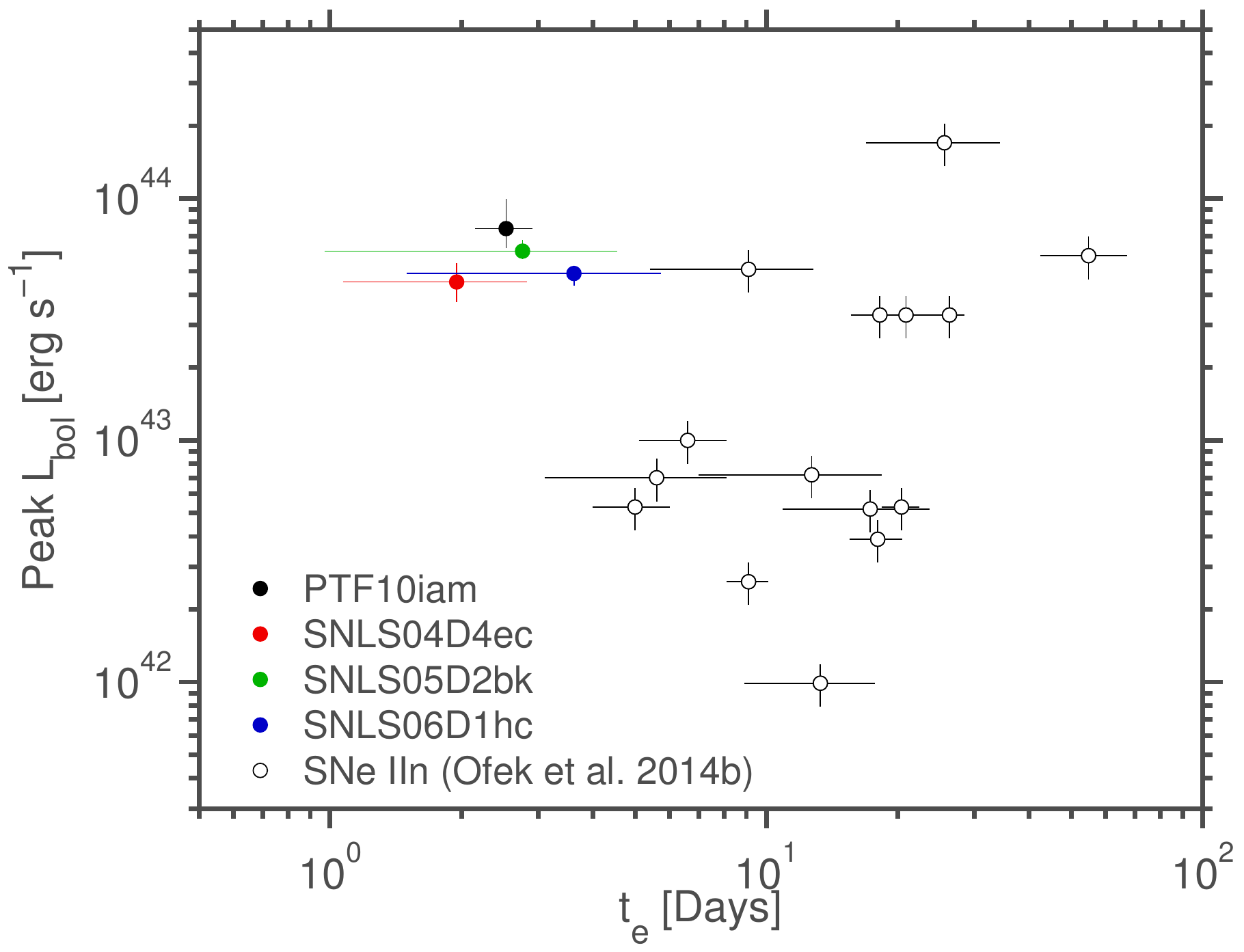}
\caption{\label{fig:peak_vs_te}Estimated peak bolometric luminosities vs. the exponential rise fit parameter $t_e$ of our events compared to Type~IIn SNe from Ofek et al. (2014b). Our events rise more rapidly than interaction-powered SNe.}
\end{figure}

If a massive star explodes in a dense CSM, the collision of the ejecta with that CSM can produce strong emission and be a major light-curve power source. Depending on the distribution of the CSM, the light curve can be made luminous and either rapidly or slowly evolving. This is the common interpretation of Type~IIn SNe (Schlegel 1990), which indeed exhibit luminous light curves with varying rise times (e.g. Kiewe et al. 2012; Ofek et al. 2014b). In addition, if the peculiar absorption feature in the spectrum of PTF10iam is high velocity H$\alpha$, then it may be evidence of CSM interaction (Chugai et al. 2007). To compare the peak magnitudes and rise times of our events to those of SNe IIn, we follow Ofek et al. (2014b) and fit an exponential rise of the form:
\begin{equation}
L=L_{0}\cdot\left\{1-\exp\left[\left(t_{exp}-t\right)/t_{e}\right]\right\}
\end{equation}
to each light curve (between discovery and peak). The free parameter here is $t_e$ which is treated as a characteristic timescale describing the rise. We plot the best fit $t_e$ for our events, compared to the Ofek et al (2014b) Type~IIn sample in Figure \ref{fig:peak_vs_te}.

We find that our events are comparable in peak luminosity to the brightest SNe IIn from the Ofek et al. (2014b) sample, but that ours rise faster. Another discrepancy with the CSM model is that the spectra of PTF10iam and of SNLS06D1hc do not display the strong intermediate-width and narrow Balmer-series emission features characteristic of SNe IIn (e.g. Kiewe et al. 2012). Moriya \& Tominaga (2012) suggest that a shallow density profile ($\rho_{CSM}{\propto}r^{-w}$ with $w\lesssim1$) of the CSM could allow interaction to power the light curve while not creating IIn-like features in the spectra. However we show below that the light curves of our events imply a much steeper density profile of $w>2$. Recently, Smith et al. (2015) suggested a new model to account for events with interaction-powered light curves but showing no narrow emission lines in their spectra. The model involves a rather complex non-symmetrical CSM distribution, and can explain the observations of SNe 1998S and PTF11iqb as weakly interacting events. Our events are much more luminous, requiring strong interaction, and are thus not readily explained by this model.

In summary, our events have shorter rise times and higher peak luminosities compared to other SNe IIn and they lack IIn-like spectral features (while displaying indications of a steep density profile, see below). If our events are powered by CSM interaction, then the initial conditions must be different than for most CSM-powered SNe. One possibility is brief interaction with a CSM clump or shell (formed for e.g. in the scenario suggested by Quataert et al. 2015).

\subsection{Shock Breakout in a Wind (SBW)} \label{sec:powersources_sbw}

First light from a propagating shock in a SN will emerge when the optical depth $\tau$ is approximately equal to $c/v$ (with $c$ the speed of light and $v$ the speed of the SN shock). This is known as the shock breakout (e.g. Colgate 1974, Weaver 1976, and more recently Nakar \& Sari 2010, Rabinak \& Waxman 2011). The duration of such a signal will be smeared by the light crossing time of the radius at shock breakout (i.e. the radius of the star), typically seconds (for compact stars) to hours (for supergiants). However, if the star is surrounded by an optically thick wind, the shock will continue to propagate in the wind and will break out at a much larger radius (and lower effective temperature). The emission leading up to shock breakout in such cases (known as shock breakout in a wind; hereafter SBW) has been studied extensively in recent years (e.g. Ofek et al. 2010, Balberg \& Loeb 2011, Chevalier \& Irwin 2011 and Ginzburg \& Balberg 2014). Svirski et al. (2012) further studied the emission following the shock breakout. We now compare our observations to these models.

\subsubsection{Up to Breakout: Rise and Peak Luminosity}

Following Drout et al. (2014), we use the Margutti et al. (2014) solutions (see their Appendix A) to the Chevalier \& Irwin (2011) equations, which relate the SN rise time ($t_{rise}$), the shock breakout radius ($R_{bo}$) and the energy radiated during the light curve rise ($E_{rise}$) with the total ejected mass ($M_{ej}$), the pre-explosion mass-loss parameter ($D_{*}$) and the opacity ($\kappa$), where
\begin{equation}
D_{*}=\left(\frac{\dot{M}}{M_{\odot}\textrm{\,yr}^{-1}}\right)\cdot\left(\frac{v_{w}}{1000\,\textrm{km\,s}^{-1}}\right)^{-1}
\end{equation}
with $\dot{M}$ the mass loss rate and $v_w$ the mass loss wind speed. We approximate $R_{bo}$ with the the first blackbody radius we measure for each event, and $E_{rise}$ as $t_{rise}{\cdot}L_{peak}$, where $t_{rise}$ is the rise time calculated in section \ref{sec:rise_times} and $L_{peak}$ is the peak bolometric luminosity deduced in section \ref{sec:bb}. 

These approximations introduce an uncertainty of a factor of a few to each of the derived quantities. We list the derived values of $D_{*}$ and $M_{ej}/E_{51}^2$ (where $M_{ej}$ is in units of $M_{\odot}$ and $E_{51}$ is the explosion energy in units of $10^{51}$\,erg) in Table \ref{tab:sbw_params}. Since the rise times for the SNLS events are limits, so are their derived parameters. The derived mass loss rates are high, similar to what was found by Drout et al. (2014) for their sample, and may imply an enhanced mass loss episode before explosion.

We also calculate the expected $v_{bo}$ (the velocity of the material at shock breakout) from Svirski et al. (2012) and find that for PTF10iam it is $\approx5,300$\,km\,s$^{-1}$, which is lower than the value measured for the possible high velocity H${\alpha}$ component seen in the spectrum (Fig. \ref{fig:10iam_halpha}).

\subsubsection{After Breakout: Post-Peak Decline Rate}

Svirski et al. (2012) show that the expected decline of the light curve following a SBW peak is a power law: 
\begin{equation}L{\propto}t^{\alpha}
\end{equation}
with ${\alpha}=-0.3$. More generally,
\begin{equation}
\alpha=\frac{\left(2-w\right)\left(m-3\right)+3\left(w-3\right)}{m-w}
\end{equation}
where $w$ is the power law index of the CSM density profile ($\rho_{CSM}{\propto}r^{-w}$) and $m$ (denoted as $n$ by Svirski et al. 2012) is the index of the velocity distribution of the ejecta ($\rho_{ej}{\propto}v^{-m}$; see Ofek et al. 2014a and references therein for more details). The value $\alpha=-0.3$ comes from the standard wind index $w=2$ and $m=12$ assumed for stars with a convective envelope ($m=10$ is used for radiative envelopes; Ofek et al. 2014a).

\begin{figure*}
\includegraphics[trim=0 0 0 0,clip,width=0.52\columnwidth]{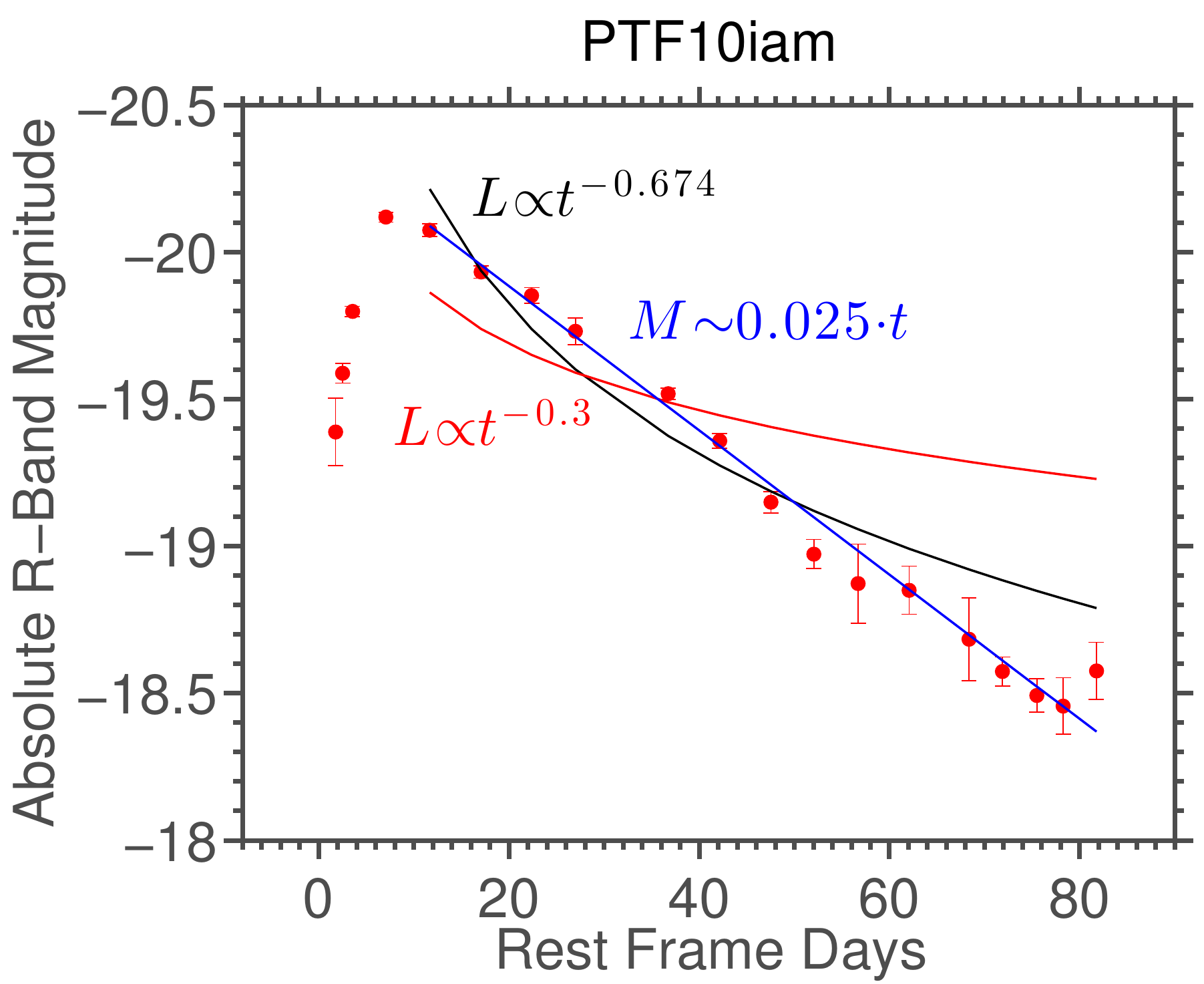}\includegraphics[trim=0 0 0 0,clip,width=0.52\columnwidth]{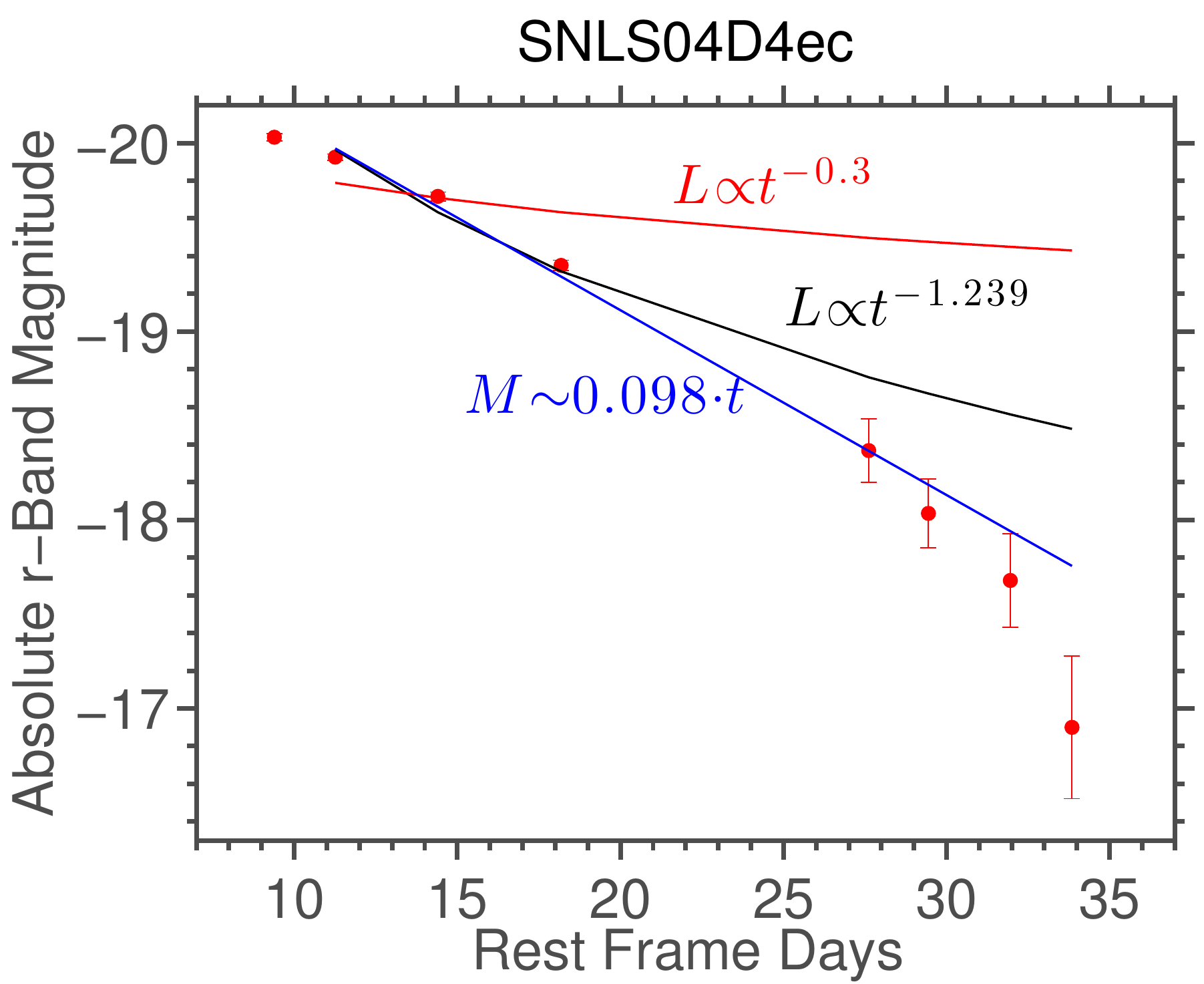}\includegraphics[trim=0 0 0 0,clip,width=0.52\columnwidth]{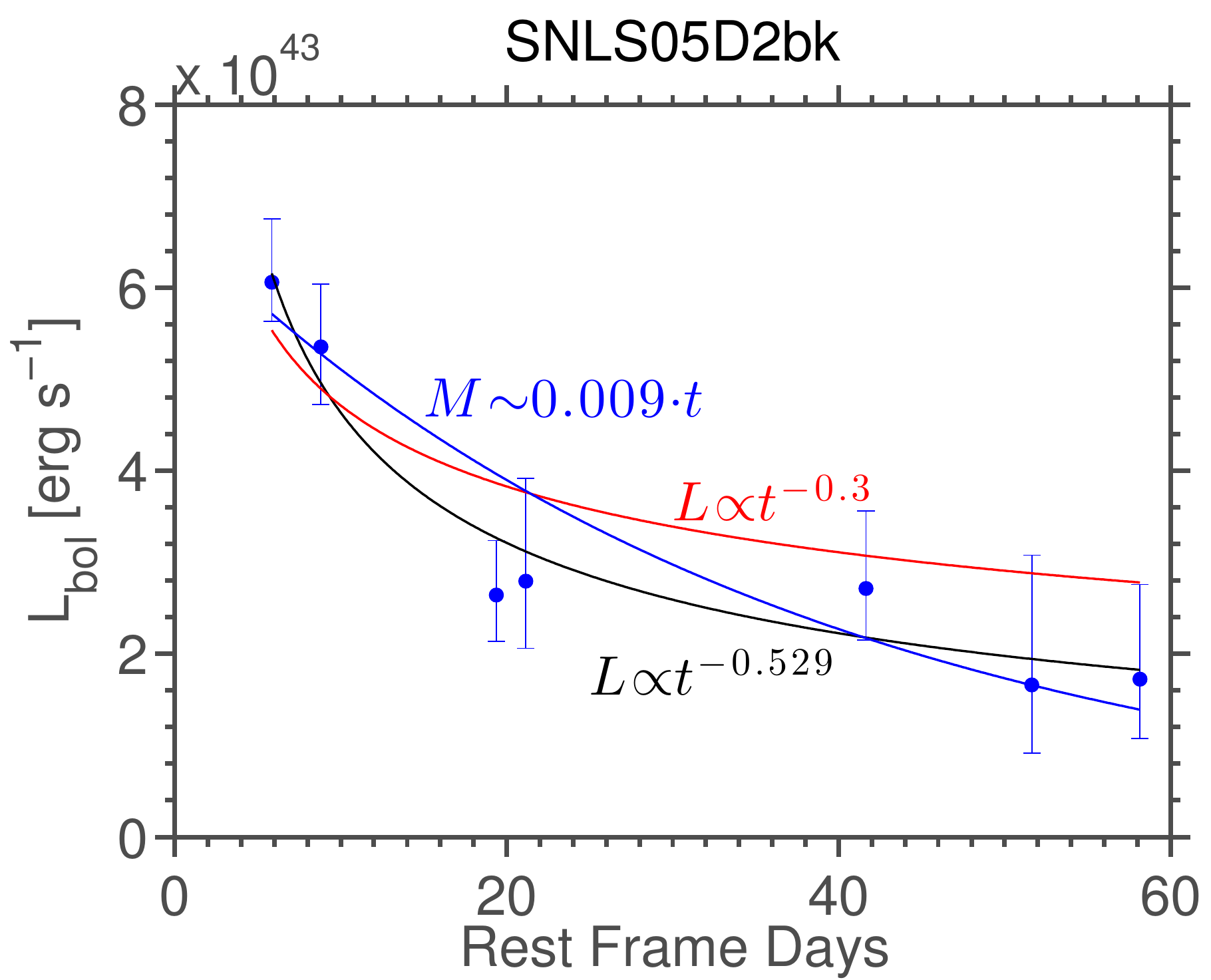}\includegraphics[trim=0 0 0 0,clip,width=0.52\columnwidth]{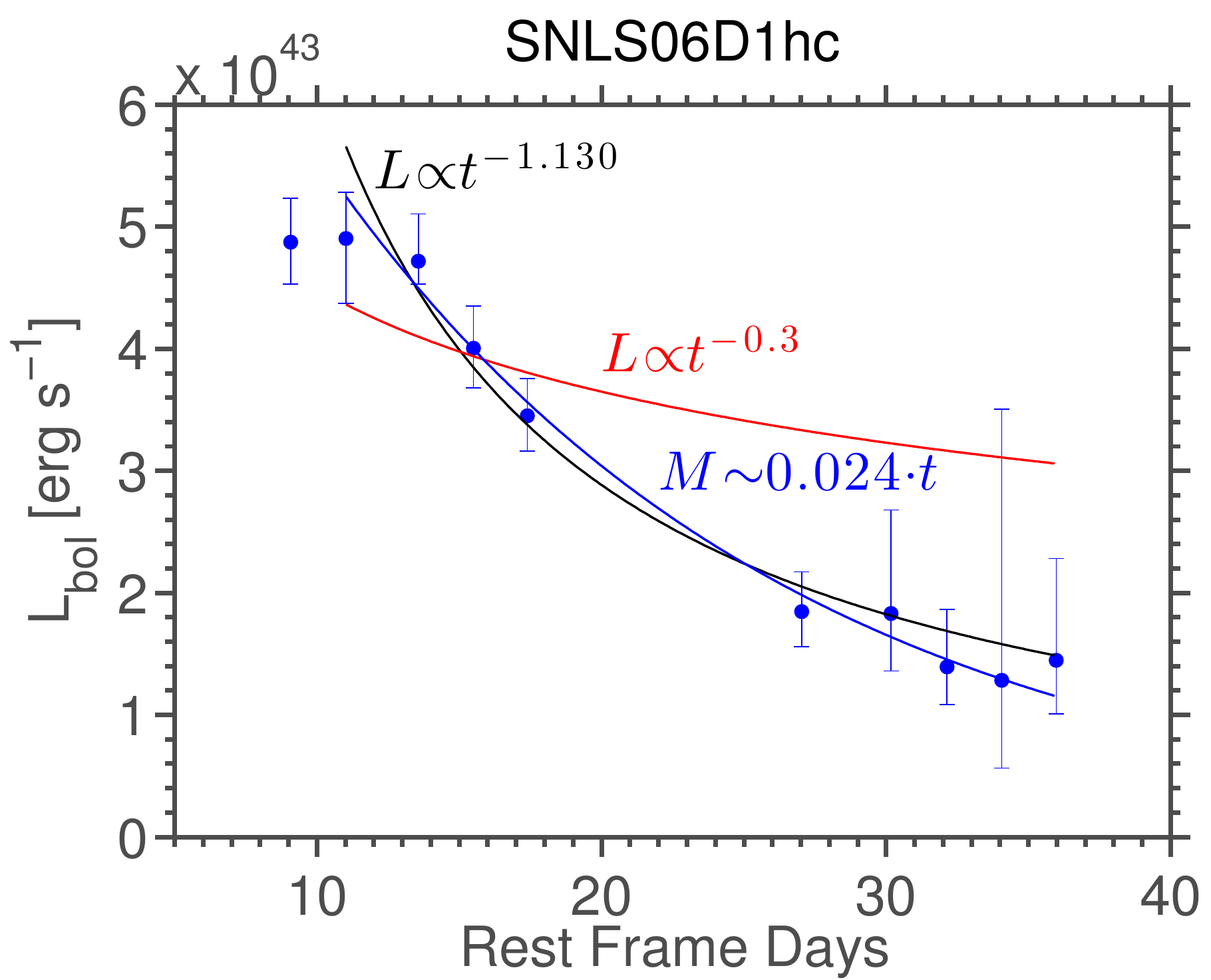}
\caption{\label{fig:declines}Left to right: Fits to the decline of the $R$-band light curve of PTF10iam, the $r$-band light curve of SNLS04D4ec, and the bolometric light curves of SNLS05D2bk and SNLS06D1hc. All light curves decline faster than $L{\propto}t^{-0.3}$ (red), expected for shock breakout in a wind, and are consistent with either a more rapid power law decline (black) or an exponential decline (blue), typical of light curves powered by radioactive decay.}
\end{figure*}

We fit a power law decline to the $R$-band light curve of PTF10iam and find a best fit to $\alpha=-0.667\pm0.063$, requiring a steep CSM profile (e.g. $w=2.55\pm0.14$ for $m=12$). We find a good fit also to an exponential decline, typical of radioactive decay (and not expected for SBW), with a decline rate of $2.45\pm0.07$ mag/$100$ days (left panel of Figure \ref{fig:declines}).

\begin{figure*}
\includegraphics[height=7.7cm]{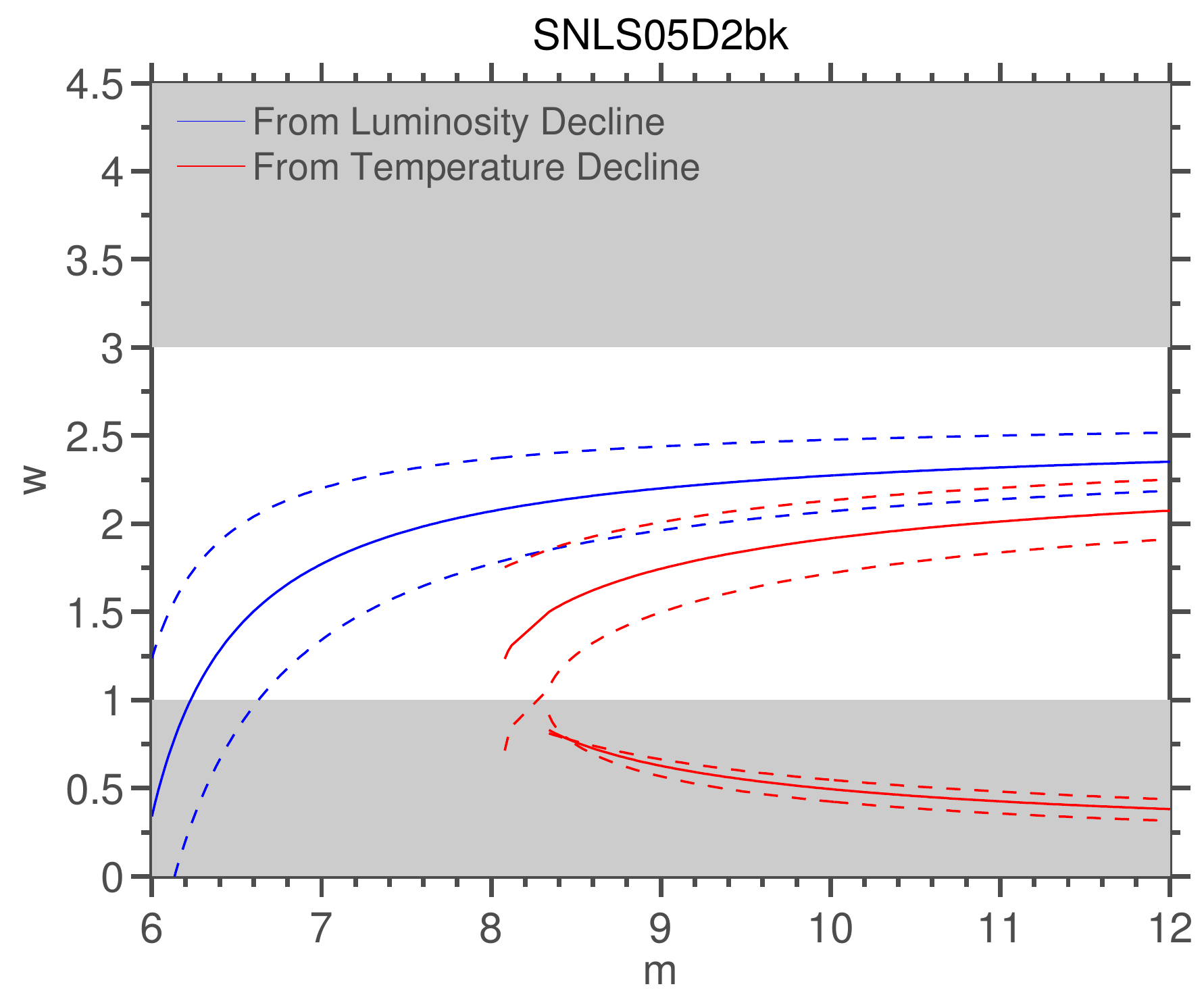}\includegraphics[height=7.7cm]{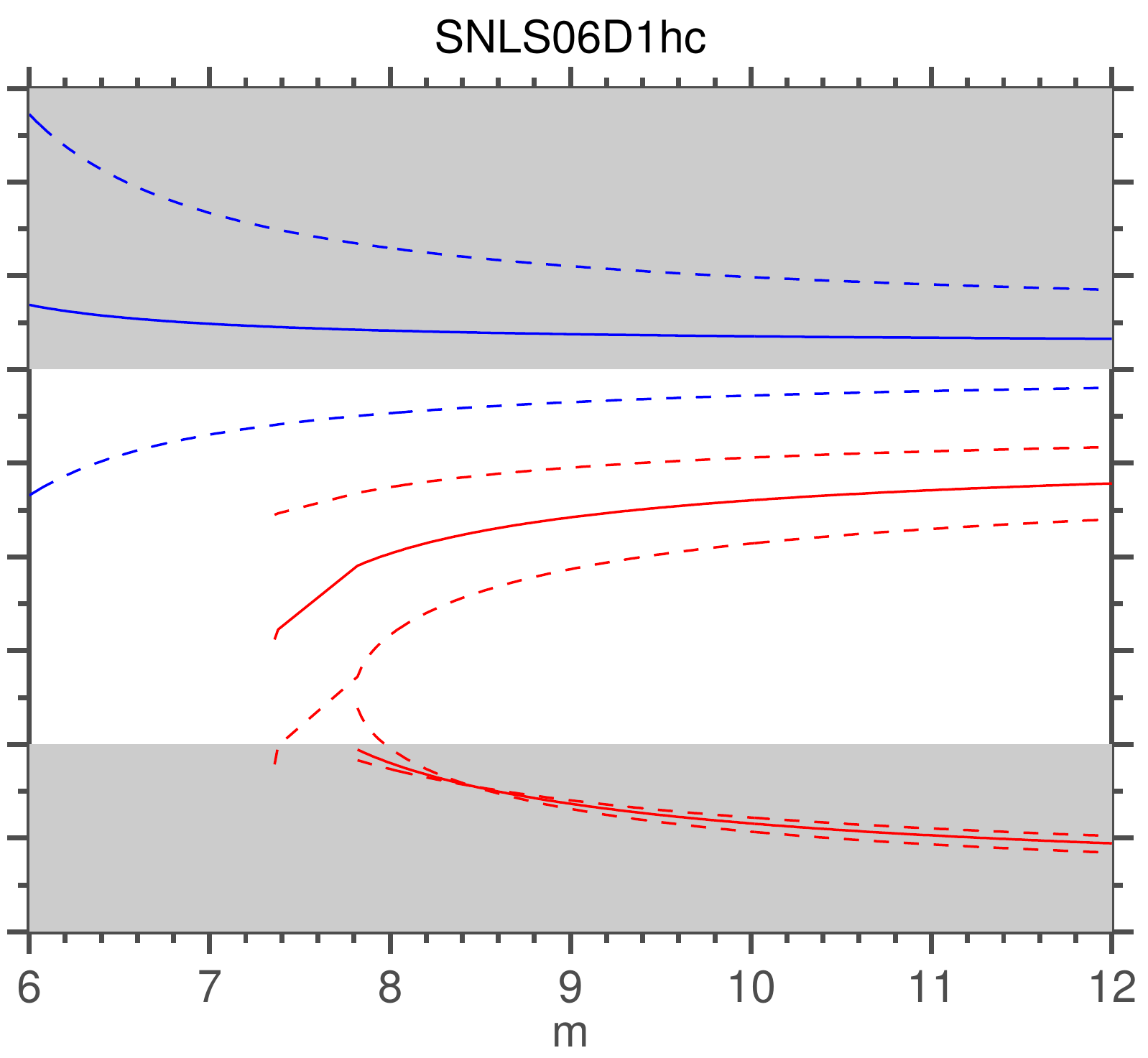}
\caption{\label{fig:w_vs_m}Constraints on the values of $w$ and $m$ for SNLS05D2bk (left) and SNLS06D1hc (right) derived from the best fit power law decline rates of their bolometric luminosity (blue) and temperature (red). One sigma errors are denoted by corresponding blue or red dashed lines. Two solutions for $w$ exist given a value for $m$ from the temperature decline slope due to the quadratic nature of Equation \ref{eq:w_vs_m_beta}. The gray areas mark the regions outside the allowed values of $w$ by the Svirski et al. (2012) model. For SNLS06D1hc, the values of $w$ are not consistent between the luminosity and temperature decline rates, for any $m$. For SNLS05D2bk, the values are marginally consistent for large $m$, and are close to the constant-wind value $w=2$.}
\end{figure*}

We fit the decline of the bolometric light curves of SNLS05D2bk and SNLS06D1hc and find that they are also steeper than $\alpha=-0.3$. For SNLS06D1hc, the corresponding values of $w$ (for $7<m<12$) are all in the $w\gtrsim3$ regime which is not allowed by the Svirski et al. (2012) model. The fit values are presented in Figure \ref{fig:declines} and Table \ref{tab:sbw_params} together with the calculated values of $w$ for selected values of $m$. Additional values of $m$ are considered in Figure \ref{fig:w_vs_m}.

\renewcommand{\arraystretch}{1.4}
\begin{table*}
{\caption{\label{tab:sbw_params}Best fit parameters to the light curves and temperatures of our events using models of shock breakout in a wind (see text for details). The luminosity decline rates are from fits to the post-peak $R$-band light curve of PTF10iam, the $r$-band light curve of SNLS04D4ec and the bolometric light curves of SNLS05D2bk and SNLS06D1hc. All are steeper than the expected $\alpha=-0.3$, are not fully consistent between the different derivations, and some values are beyond the validity range of the model. This disfavors shock breakout as the power source for most of these events (SNLS05D2bk is marginally consistent with the model). The best fit to an exponential decline rate for the light curves (indicative of radioactive decay power rather than shock breakout in a wind) is also shown.}}
\begin{center}
\begin{tabular*}{\textwidth}{lccccccp{0mm}ccc}
\hline
\hline
{Object} & \multicolumn{6}{c}{Luminosity} & {} & \multicolumn{3}{c}{Temperature}\tabularnewline
\cline{2-7}\cline{9-11}
{} & \multicolumn{2}{c}{Rise and Peak} & \multicolumn{3}{c}{Power Law Decline} & {Exp. Decline} & {} & \multicolumn{3}{c}{Power Law Decline}\tabularnewline
{} & {$M_{ej}/E_{51}^2$} & {$D_{*}$} & {$\alpha$} & {$m$} & {$w$} & {[mag/$100$d]} & {} & {$\beta$} & {$m$} & {$w$}\tabularnewline
%{} & {[$M_{\odot}$]} & {[$M_{\odot}$/yr]} & {} & {} & {} & {} & {[mag/$100$d]}\tabularnewline
\hline
{PTF10iam} & {$0.35$}& {$1.13\cdot10^{-2}$} & {$-0.667\pm0.063$} & {$10$} & {$2.50\pm0.19$} & {$2.45\pm0.07$} & {} & {} & {} & {}\tabularnewline
{} & {}& {} & {} & {$12$} & {$2.55\pm0.14$}& {} & {} & {} & {} & {}\tabularnewline
{SNLS04D4ec} & {$<0.66$} & {$<8.3\cdot10^{-3}$} & {$-1.188\pm0.190$} & {$10$} & {$3.25\pm0.49$} & {$8.80\pm0.74$} & {} & {} & {} & {}\tabularnewline
{} & {} & {} & {} & {$12$} & {$3.24\pm0.40$} & {} & {} & {} & {} & {}\tabularnewline
{SNLS05D2bk} & {$<0.46$} & {$<4.6\cdot10^{-3}$} & {$-0.530\pm0.075$} & {$10$} & {$2.27\pm0.20$} & {$1.17\pm0.28$} & {} & {$-0.247\pm0.067$} & {10} & {$1.92\pm0.21$}\tabularnewline
{} & {}& {} & {} & {$12$} & {$2.35\pm0.16$} & {} & {} & {} & {12} & {$2.08\pm0.17$}\tabularnewline
{SNLS06D1hc} & {$<1.47$} & {$<3.88\cdot10^{-2}$} & {$-1.130\pm0.123$} & {$10$} & {$3.18\pm0.32$} & {$2.64\pm0.18$} & {} & {$-0.367\pm0.073$} & {10} & {$2.30\pm0.23$}\tabularnewline
{} & {}& {} & {} & {$12$} & {$3.16\pm0.26$} & {} & {} & {} & {12} & {$2.39\pm0.19$}\tabularnewline
\hline

\end{tabular*}
\end{center}
\end{table*}
\renewcommand{\arraystretch}{1}

For $m=10$ or $m=12$, a decay in the optical bands steeper than $t^{-0.3}$ is not generally expected in SBW-powered light curves as long as the radiation is in thermal equilibrium with the emitting plasma. In such equilibrium, the blackbody temperature indeed decreases slowly, but the flux in the soft optical bands should not decay rapidly (it may even rise) since it is always in the Rayleigh-Jeans tail of the spectrum (as enforced by recombination).

If a breakout pulse in equilibrium is followed by a deviation from equilibrium after some (measurable) time, one expects an initially rather constant (or slightly rising) optical luminosity, when in equilibrium, followed by a steep decay when out of equilibrium. When the radiation is out of thermal equilibrium, two processes act simultaneously to decrease the optical luminosity: First, the temperature at which most free-free or bound-free photons are produced rises with time. Second, the fraction of energy that photons at their emission temperature carry, compared to the total energy, declines since when out of equilibrium, much of the energy in the system is carried by photons that scatter multiple times with the hot shocked electrons (and thus have temperatures higher than their emission ones). However, our events don't show any break in the optical light curves which would indicate a transition from equilibrium to non-equilibrium.

A third option, of a breakout pulse in thermal equilibrium, {\it promptly} followed by a deviation from equilibrium, is harder to rule out at the single event level. The decay rate of optical luminosity out of equilibrium may vary widely between events, depending on, e.g., the shock velocity, the ratio between free-free and bound-free emission and the possible effects of line absorption.

However, given a sample of SNe which all have high optical peak luminosities, indicating breakout radiation in or near thermal equilibrium, and which all have a rapid decay due to a prompt post-breakout deviation from thermal equilibrium, is statistically unlikely. One would expect, within a sample, a division of shock breakout events into three categories: (1) The breakout pulse is already out of thermal equilibrium (e.g., due to a high shock velocity), and thus rather faint in the optical; (2) The breakout pulse is in thermal equilibrium but the following radiation promptly deviates from equilibrium - a combination that could potentially explain an observed luminous optical peak and a prompt rapid decay; (3) The radiation remains in equilibrium for some period after the breakout, such that an initial phase of a rather constant optical luminosity is expected. It is unlikely that all SNe in a sample selected by peak optical luminosity, would fall into option (2). 

The light curves of PTF09uj (Ofek et al. 2010) and of the Drout et al. (2014) events, all considered likely cases of SBW, display even steeper decline rates compared to ours. Considering them together with our events, we therefore disfavor the interpretation SBW radiation with a prompt departure from thermal equilibrium for explaining all of these events. To the best of our knowledge, the only reported event with an $\alpha\approx-0.3$ decline rate and which also fits other SBW characteristics is SN 2010jl (Ofek et al. 2014a). 

As an additional check we fit the temperature evolution of SNLS05D2bk and SNLS06D1hc with a power law
\begin{equation}
T_{BB}{\propto}t^{\beta}
\end{equation}
The generalization of the Svirski et al. (2012) power-law index $\beta=-0.2$ is:
\begin{equation}
\label{eq:w_vs_m_beta}
\beta=\frac{4w\left[\left(m-3\right)\left(5-w\right)-3\left(m-w\right)\right]}{\left(m-w\right)\left(9w-7\right)}
 \end{equation} 
Using the best fit $\beta$ to the data we find values of $w$ which are not consistent with those found from the luminosity decline (in Table \ref{tab:sbw_params} we ignore the $w<1$ solutions, which are outside the validity range of the model, but they are plotted in Figure \ref{fig:w_vs_m}). For SNLS05D2bk, however, the inconsistency is not strong ($\lesssim2\sigma$ for the larger values of $m$), and the temperature decline rate does imply a wind profile of $w=2$. 

We conclude that SNLS05D2bk may be marginally consistent with shock breakout in a wind, though the second peak is not predicted by such models. For the other events in our sample, due to their steep post-peak decline, shock cooling in a wind is more strongly disfavored.

\subsection{Magnetar Spindown} \label{sec:powersources_magnetar}

\begin{figure}
\includegraphics[trim=0 0 0 0,clip,width=\columnwidth]{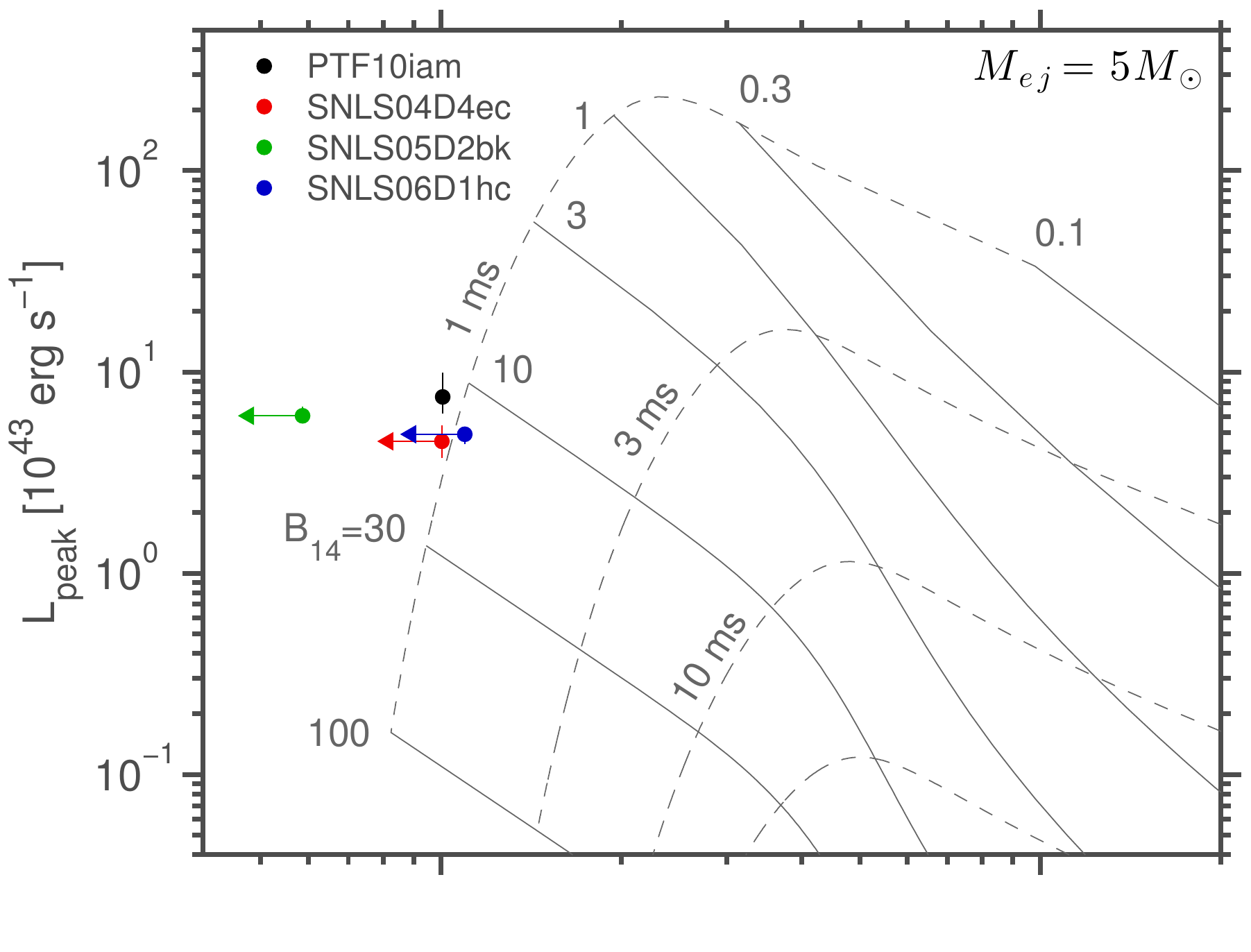}
\includegraphics[trim=0 0 0 0,clip,width=\columnwidth]{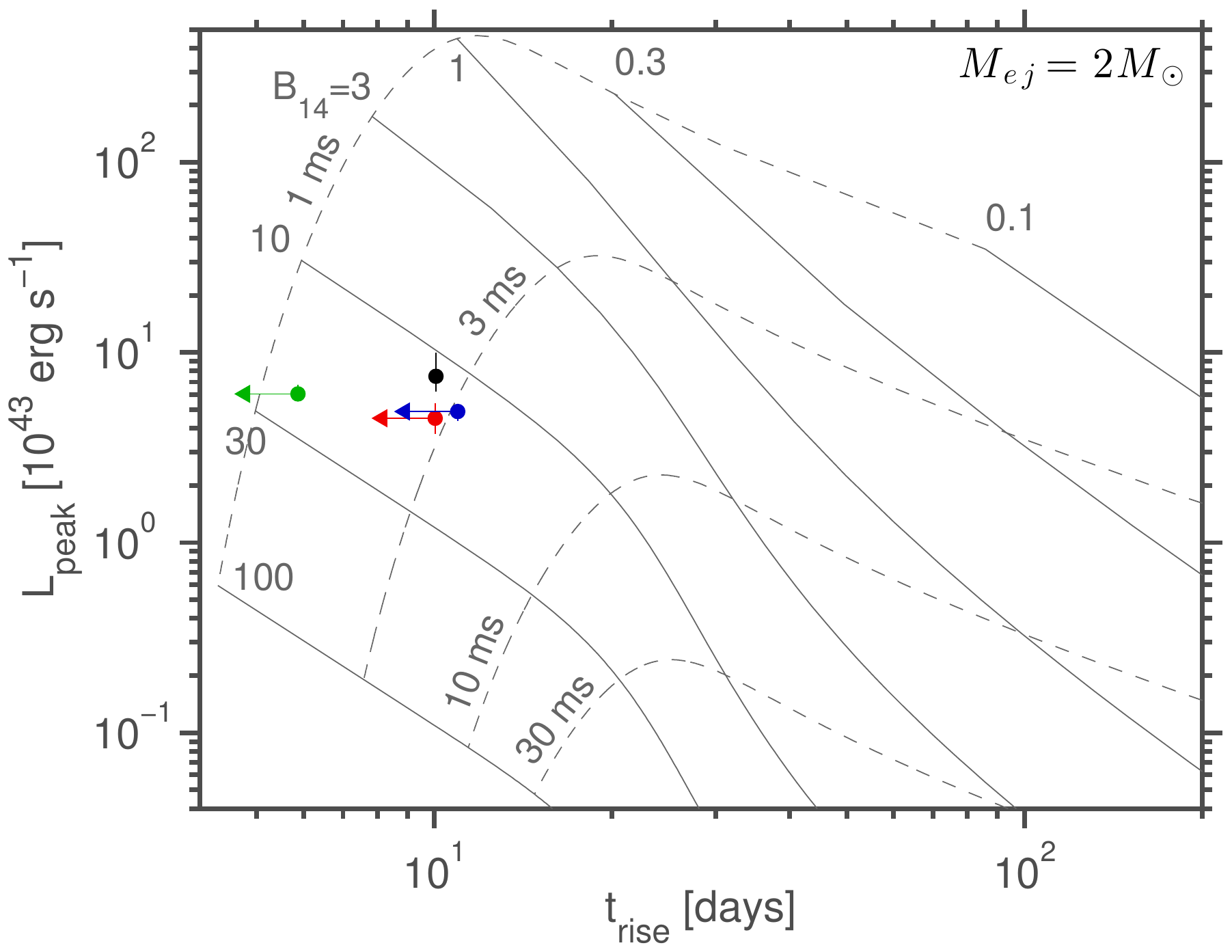}
\caption{\label{fig:magnetar_phasespace}Magnetar initial spin period and magnetic field contours (in units of $10^{14}$\,G) from Kasen \& Bildsten (2010; assuming an explosion energy of $10^{51}$\,erg and an opacity $\kappa=0.2$\,cm$^2$\,gr$^{-1}$) for an ejecta mass of $5M_{\odot}$ (top) and $2M_{\odot}$ (bottom). The magnetar models can reproduce the rise time and peak luminosity of our events only with extremely rapid initial spin periods and low ejecta masses.}
\end{figure}

The spindown of a highly magnetized ($B\sim10^{14}-10^{15}$\,G) neutron star (known as a ``magnetar'') formed during the core collapse of a massive star can influence the light curve of the ensuing SN (Kasen \& Bildsten 2010; Woosley 2010). Following Kasen \& Bildsten (2010), we plot the magnetar parameter contours on a peak luminosity vs. rise time plot (Fig. \ref{fig:magnetar_phasespace}) for two different ejecta masses assuming an explosion energy of $10^{51}$\,erg and an opacity $\kappa=0.2$\,cm$^2$\,gr$^{-1}$. 

The magnetar models can reproduce the rapid rise times of our events for an ejecta mass as high as $5M_{\odot}$, but only assuming an initial spin period close to breakup ($P_i=1$\,ms). This period can be increased slightly for $M_{ej}=2M_{\odot}$, but is still required to be extreme ($1-3$\,ms; Fig. \ref{fig:magnetar_phasespace}). Increasing the initial spin period would require further decreasing the ejecta mass. 

This problem is also apparent when considering the full light curve shapes. We fit the magnetar model from Inserra et al. (2013) to the photometry of our events. We assume the same opacity ($\kappa=0.2$\,cm$^2$\,gr$^{-1}$) and explosion energy ($10^{51}$\,erg) as above, but allow for spindown energy that is never ultimately radiated to contribute as well. We add half of the integrated un-radiated spindown energy to the original explosion energy and re-run the fits, iterating this process until the total explosion energy (original plus spindown contributions) changes by no more than one percent. We fix the explosion dates to the values from Table \ref{tab:events_params} and fit for the ejecta mass, initial spin period and magnetic field on all bands simultaneously, assuming a blackbody spectrum (the Inserra et al. 2013 prescription produces a bolometric luminosity and radius, which allows for an effective temperature to be deduced). We restrict the initial spin periods to be $>1$\,ms. 

\begin{figure*}
\includegraphics[trim=0 0 0 0,clip,width=0.52\columnwidth]{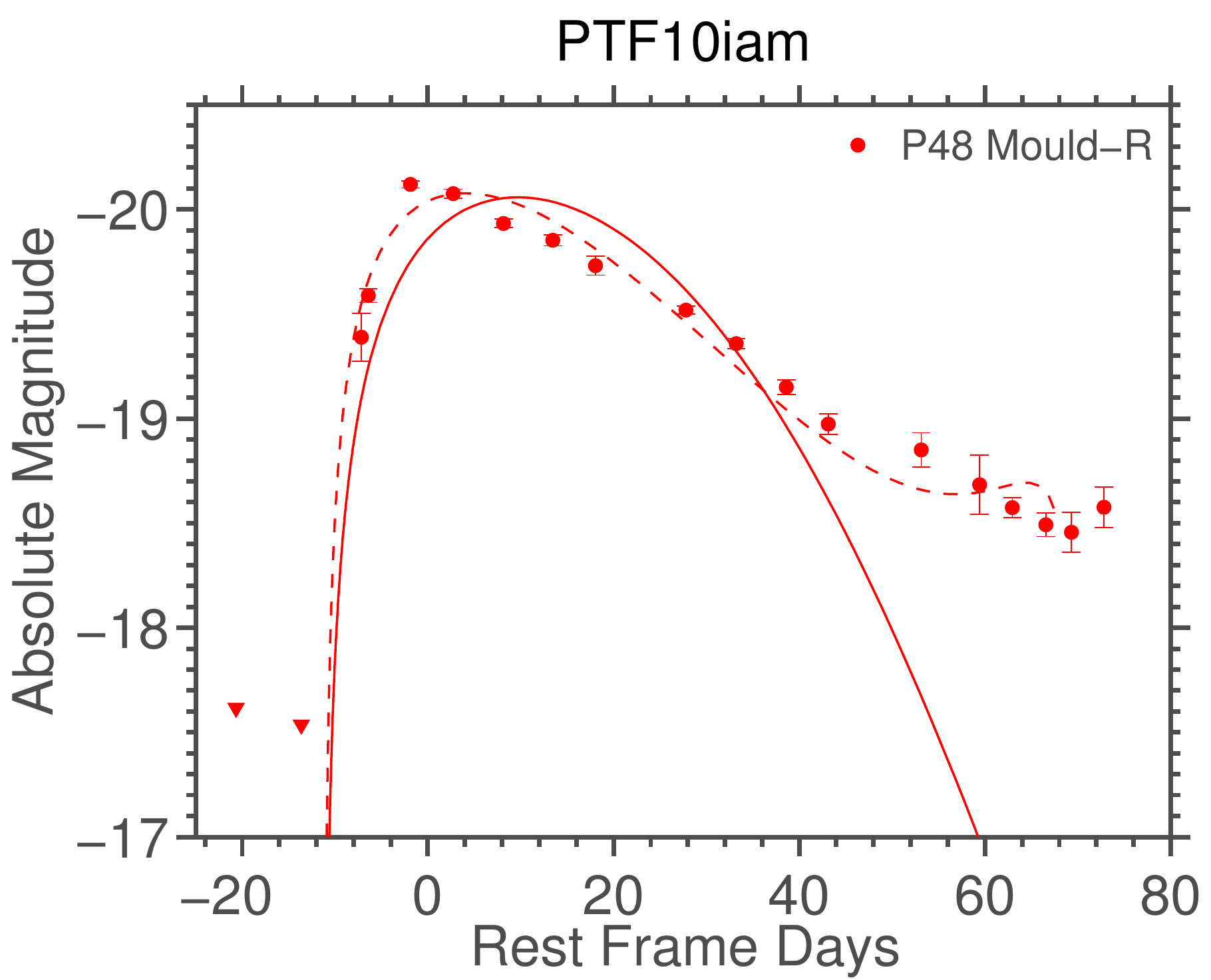}\includegraphics[trim=0 0 0 0,clip,width=0.52\columnwidth]{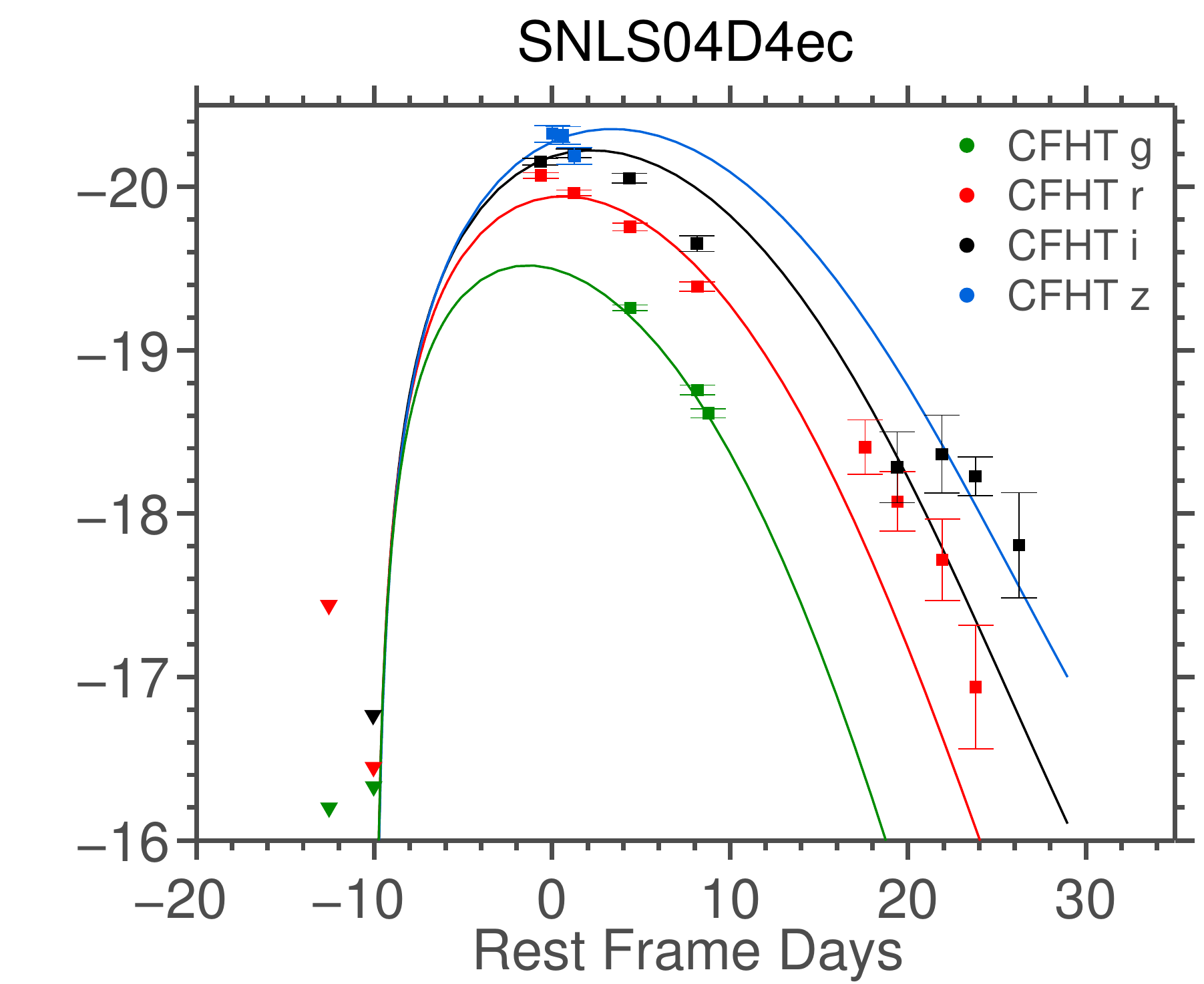}\includegraphics[trim=0 0 0 0,clip,width=0.52\columnwidth]{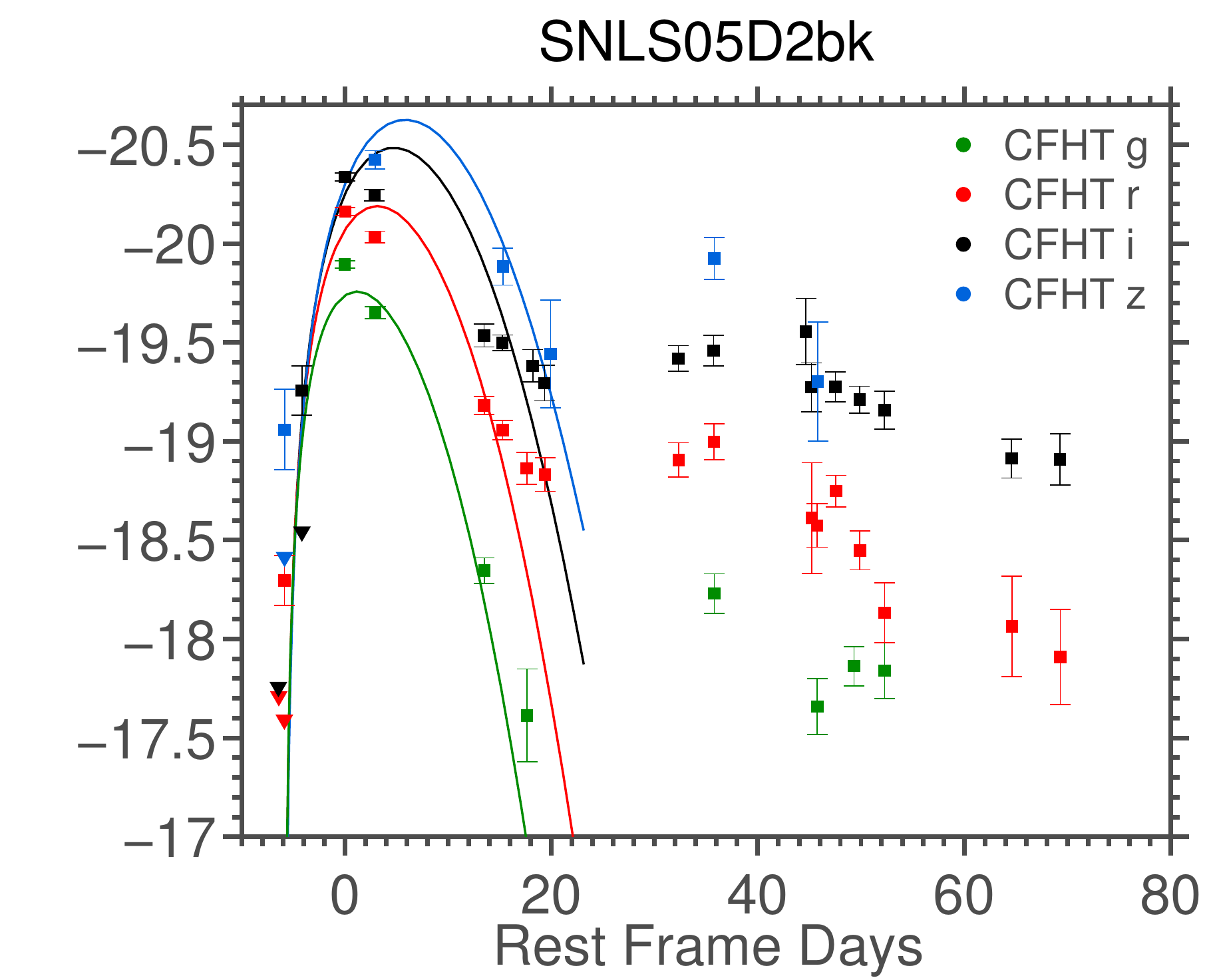}\includegraphics[trim=0 0 0 0,clip,width=0.52\columnwidth]{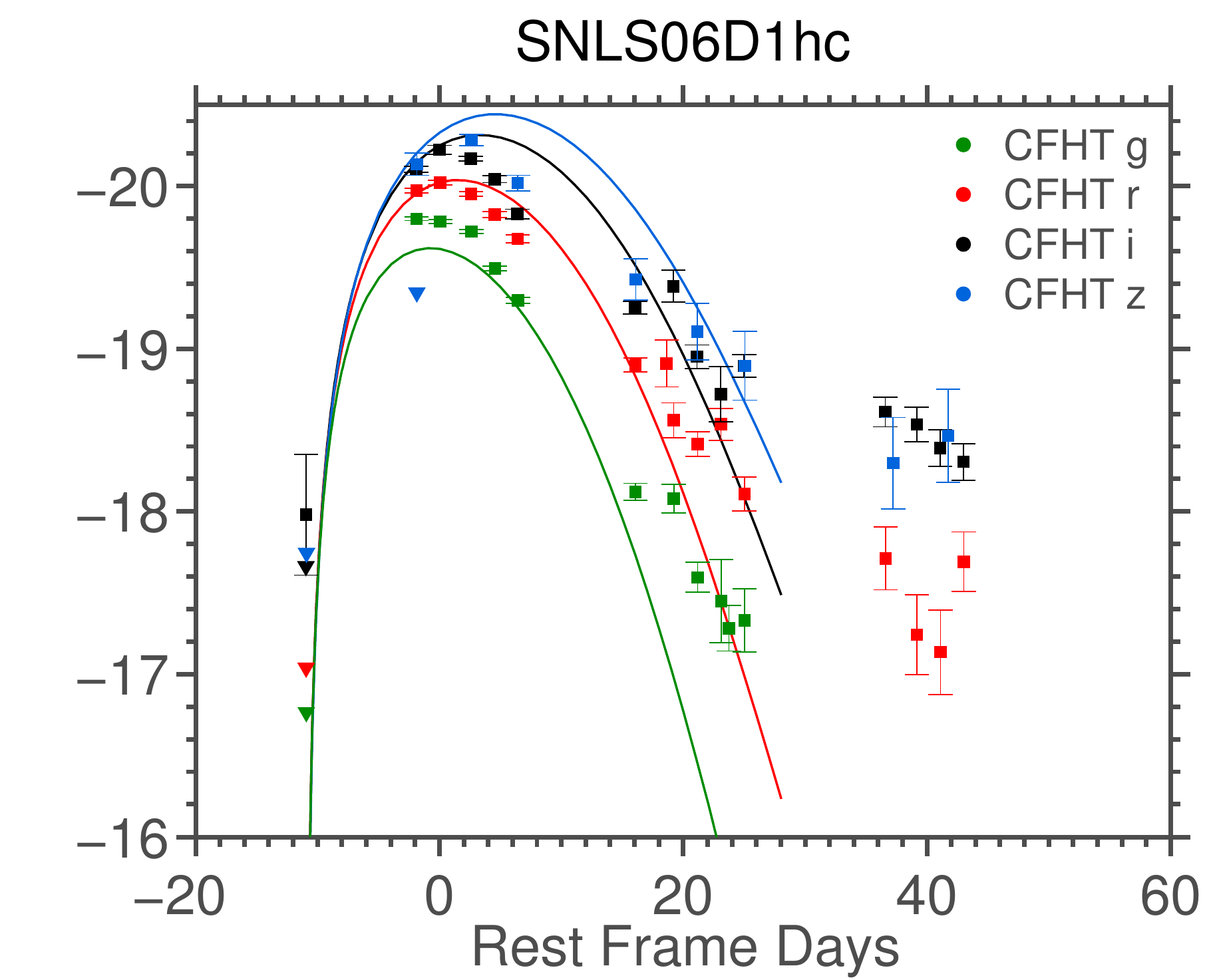}
\caption{\label{fig:lcs_magnetars}Best-fit magnetar models (solid lines) to the light curves of our events. Triangles denote $3{\sigma}$ non-detection upper limits. We use the Inserra et al. (2013) models with an explosion energy of $10^{51}$\,erg and an opacity $\kappa=0.2$\,cm$^2$\,gr$^{-1}$, we fix the explosion dates to the values listed in Table \ref{tab:events_params}, and we restrict the initial spin period $P_{i}$ to be $>1$\,ms. For PTF10iam the model is not a good fit and requires the lowest allowable initial spin period ($1$\,ms). We perform a fit with no restrictions on the initial period and present it in the dashed line. The SNLS observations are better fit by the models, but all require very low ejecta masses (the best-fit model parameters are listed in Table \ref{tab:magnetar}), and are unable to reproduce the late time light curve behavior.}
\end{figure*}

Our best-fit results are presented in Table \ref{tab:magnetar} and plotted in Figure \ref{fig:lcs_magnetars}. For PTF10iam, the best fit was given by the lowest allowed initial spin period ($1$\,ms) so we re-ran the fit with no restrictions. We present the results of this unrestricted fit as well. We ignore the apparent re-brightening of SNLS05D2bk and the flattening of the late light curve of SNLS06D1hc in the fits, since the models are not able to reproduce these features. Such features may be reproduced by fallback accretion on to a black hole (Dexter \& Kasen 2014), but we do not explore that scenario further here.

\renewcommand{\arraystretch}{1.4}
\begin{table}
\begin{center}
\caption{\label{tab:magnetar}Best fit parameters to the light curves of our events assuming they are powered by magnetar spindown (see text for details). The extremely low ejecta masses disfavor the magnetar interpretation. For PTF10iam the best fit is at the minimal allowed spin period, and we present the fit parameters also with no restrictions on the spin period.}
\begin{tabular}{lccc}
\hline
\hline
{Object} & {$M_{ej}$ [$M_{\odot}$]} & {$P_i$ [ms]} & {$B$ [$10^{14}$\,G]} \tabularnewline
\hline
{PTF10iam} & {$4.75\pm0.06$} & {$1.00$} & {$12.41\pm0.11$} \tabularnewline
{PTF10iam} & {$6.63\pm0.10$} & {$0.43\pm0.01$} & {$11.45\pm0.12$} \tabularnewline
{(unrestricted)} & {} & {} & {} \tabularnewline
{SNLS04D4ec} & {$1.27\pm0.12$} & {$6.13\pm1.92$} & {$22.61\pm2.69$} \tabularnewline 
{SNLS05D2bk} & {$1.23\pm0.05$} & {$3.58\pm0.43$} & {$25.96\pm0.84$} \tabularnewline
{SNLS06D1hc} & {$1.50\pm0.06$} & {$5.83\pm0.69$} & {$18.29\pm0.92$} \tabularnewline 
\hline

\end{tabular}
\end{center}
\end{table}
\renewcommand{\arraystretch}{1}

The required ejecta masses are very low for explosions of massive stars. For hydrogen-rich progenitors (as expected for PTF10iam from the broad hydrogen features in its spectrum), it would be more accurate to use an opacity of $\kappa=0.34$\,cm$^2$\,gr$^{-1}$. This would increase the ejecta mass by a factor of 1.7, but it would still be too small compared to an expected ejecta mass of $\approx10M_{\odot}$ (mainly from the hydrogen envelope). The SNLS events, for which the hydrogen content is unknown, could come from stripped-envelope progenitors. Even in that case, the ejecta masses given by the magnetar fits are on the very low bounds of what is measured for stripped-envelope SNe (see e.g. Perets et al. 2010, supplementary information; Tauris et al. 2010; Lyman et al. 2014; Taddia et al. 2015).

We conclude that magnetar spindown is disfavored as the power source of our events due to the poor light curve fit to PTF10iam and the low ejecta masses required to fit the SNLS events. Greiner et al. (2015), on the other hand, prefer the magnetar scenario for SN\,2011kl (see also Metzger et al. 2015). Their ejecta mass estimates are also strangely low (as they note), though higher than ours since they assume higher expansion velocities, and a slightly longer rise time. 

\section{Summary}

We present observations of four transients with light curves showing a rapid rise ($\approx10$ days) to a luminous peak ($\approx5\cdot10^{43}$ erg\,s$^{-1}$). These properties put our events in a unique part of SN phase space (even when compared to the diverse class of Type~IIn explosions). To the best of our knowledge, the only published event with similar light curve features is SN\,2011kl, which was accompanied by an ultra-long-duration GRB (Greiner et al. 2015). No GRBs were associated with our events, but available data can not rule-out a GRB association for any of our SNe.

For the only event in our sample with detected broad spectroscopic features, we see broad H emission and a deep absorption feature near $6200\,\textrm{\AA}$, which can be interpreted as either high velocity H$\alpha$ or as Si II. 

Due to the lack of spectral coverage or detection of obvious SN features for the SNLS events, it is not possible to determine if all four of our transients have the same origin. However, given the similarity in the light curve shapes, luminosities and color temperatures of all our events, we consider them, tentatively, as belonging to one class (perhaps including also SN\,2011kl).

We discuss several possible power sources for the light curves: white dwarf detonation (perhaps inside a hydrogen envelope), CSM interaction, shock breakout in a wind and magnetar spindown. Each interpretation has its strengths and weaknesses, summarized in Table \ref{tab:summary}, and we do not favor any particular explanation over the others.

\renewcommand{\arraystretch}{1.4}
\begin{table*}
{\caption{\label{tab:summary}Summary of the power sources considered for our events, of which neither is capable of fully explaining the observations. More detailed models and more constraining observations of future events may help distinguish between these possibilities, or suggest new ones.}}
\begin{center}
\begin{tabular*}{\textwidth}{lllll}
\hline
\hline
{Light Curve} & {Feature at} & {Strengths} & {Weaknesses} & {Implications}\tabularnewline
{Power Source} & {$\approx6200\textrm{\AA}$} & {} & {} & {if True}\tabularnewline
\hline
{Nickel decay} & {Si II} & {\begin{minipage}[t]{0.5\columnwidth}
Explains the possible spectral similarity of PTF10iam to a SN Ia and of all light curves to the rapid rise and luminous peaks of pure and double detonation models.
\end{minipage}}
& {\begin{minipage}[t]{0.5\columnwidth}
Post-peak light curve behavior is not consistent with models.
\end{minipage}}
& {\begin{minipage}[t]{0.3\columnwidth}
Possible first identification of Type~1.5 SNe.
\end{minipage}}\tabularnewline
\hline
{CSM interaction} & {HV H$\alpha$?} & {\begin{minipage}[t]{0.5\columnwidth}
Similar absorption feature, lack of narrow emission features and light curve decline rate as 98S, explained as an interaction-powered SN.
\end{minipage}}
& {\begin{minipage}[t]{0.5\columnwidth}
Light curves are much more luminous than 98S, absorption feature is deeper and bluer, light curve shapes are different than interaction-powered IIn's and no intermediate-width or narrow Balmer emission lines are seen in the spectra.
\end{minipage}} 
& {\begin{minipage}[t]{0.31\columnwidth}
A new type of strongly interacting SN.
\end{minipage}}\tabularnewline
\hline
{Shock breakout in a wind} & {HV H$\alpha$?} & {\begin{minipage}[t]{0.5\columnwidth}
Can reproduce the rapid rise and high peak luminosity.
\end{minipage}}
& {\begin{minipage}[t]{0.5\columnwidth}
Light curve decline is too rapid for a standard wind profile (and in one case too rapid for the model validity regime), as well as inconsistent with the temperature decline rate.
\end{minipage}}
& {\begin{minipage}[t]{0.31\columnwidth}
Most SBW events deviate from thermal equilibrium promptly after peak luminosity.
\end{minipage}}\tabularnewline
\hline
{Magnetar Spindown} & {?} & {\begin{minipage}[t]{0.5\columnwidth}
Can reproduce the rapid rise and high peak luminosity of the SNLS events.
\end{minipage}}
& {\begin{minipage}[t]{0.5\columnwidth}
Requires very low ejecta masses, not consistent with a massive star collapse (especially if a H envelope is present).
\end{minipage}} & {}\tabularnewline
\hline

\end{tabular*}
\end{center}
\end{table*}
\renewcommand{\arraystretch}{1}

Recently, Kashiyama \& Quataert (2015) suggested that outflows from a fallback accretion disk around a newly formed black hole could produce rapidly evolving blue transients. The predicted peak luminosity, however, is lower, while the predicted pre- and post-peak evolution is much faster than in our events. Gilkis et al. (2015) suggest inefficient jets from accretion onto a newly formed neutron star are responsible for most or all luminous SNe. However, it remains to be seen if this mechanism can reproduce the low rise times of our events.

The origin of the new class of rapidly rising luminous transients identified here remains a mystery. We encourage more detailed models (especially of white dwarf detonations inside hydrogen envelopes), as well as more complete observational coverage of future such events, in order to better constrain their nature.

~\\
We thank J. Silverman and J. Johansson for helpful discussions, and S. Sim and M. Kromer for sharing their white dwarf detonation models. This paper is based on observations obtained at the Cerro Paranal Observatory (ESO program 176.A-0589) and with the Samuel Oschin Telescope as part of the Palomar Transient Factory project. We are grateful for the assistance of the staffs at the various observatories where data were obtained. This work made use of the astronomy \& astrophysics package for Matlab (Ofek 2014). Some of the work presented here is supported by the National Science Foundation under Grant No. 1313484. I.A. and A.G. acknowledge support by the Israeli Science Foundation, and an EU/FP7/ERC grant. A.G. further acknowledges grants from the BSF, GIF and Minerva, as well as the ``Quantum Universe'' I-Core program of the planning and budgeting committee and the ISF, and a Kimmel Investigator award. The work of W.M.W. and L.B. was supported by the National Science Foundation under grants PHY 11-25915 and AST 11-09174. The Dark Cosmology Centre is funded by the Danish National Research Foundation. Some of the data presented herein were obtained at the W. M. Keck Observatory, which is operated as a scientific partnership among the California Institute of Technology, the University of California and the National Aeronautics and Space Administration. The Observatory was made possible by the generous financial support of the W. M. Keck Foundation. Some data is based on observations obtained at the Gemini Observatory processed using the Gemini IRAF package, which is operated by the Association of Universities for Research in Astronomy, Inc., under a cooperative agreement with the NSF on behalf of the Gemini partnership: the National Science Foundation (United States), the National Research Council (Canada), CONICYT (Chile), the Australian Research Council (Australia), Ministério da Ciência, Tecnologia e Inovação (Brazil) and Ministerio de Ciencia, Tecnología e Innovación Productiva (Argentina). This work made use of the NASA/IPAC Extragalactic Database (NED) which is operated by the Jet Propulsion Laboratory, California Institute of Technology, under contract with the National Aeronautics and Space Administration. Funding for SDSS-III has been provided by the Alfred P. Sloan Foundation, the Participating Institutions, the National Science Foundation, and the U.S. Department of Energy Office of Science. SDSS-III is managed by the Astrophysical Research Consortium for the Participating Institutions of the SDSS-III Collaboration including the University of Arizona, the Brazilian Participation Group, Brookhaven National Laboratory, Carnegie Mellon University, University of Florida, the French Participation Group, the German Participation Group, Harvard University, the Instituto de Astrofisica de Canarias, the Michigan State/Notre Dame/JINA Participation Group, Johns Hopkins University, Lawrence Berkeley National Laboratory, Max Planck Institute for Astrophysics, Max Planck Institute for Extraterrestrial Physics, New Mexico State University, New York University, Ohio State University, Pennsylvania State University, University of Portsmouth, Princeton University, the Spanish Participation Group, University of Tokyo, University of Utah, Vanderbilt University, University of Virginia, University of Washington, and Yale University.

\end{document}